\documentclass[fleqn,usenatbib]{mnras}

\usepackage{newtxtext,newtxmath}

\usepackage[T1]{fontenc}
\usepackage{ae,aecompl}

\usepackage{graphicx}	
\usepackage{amsmath}	
\usepackage{amssymb}	
\usepackage{booktabs}

\newcommand{\hii}     {H\,{\sc ii}}
\newcommand{\CII}     {[C\,{\sc ii}]}
\newcommand{\OI}      {[O\,{\sc i}]}
\newcommand{\thirtCII}{[$^{13}$C\,{\sc ii}]} 
\newcommand{\kms}     {km\,s$^{-1}$}
\newcommand{\nel}     {$n_{\rm e}$}
\newcommand{\tgas}     {$T_{\rm gas}$}
\newcommand{\tdust}     {$T_{\rm dust}$}
\newcommand{\teff}    {$T_{\rm eff}$}
\newcommand{\tmb}     {$T_{\rm mb}$}
\newcommand{\tex}     {$T_{\rm ex}$} 
\newcommand{\tauCII}   {$\tau_{\rm C^+}$}  
\newcommand{\HCOp}    {HCO$^+$(3--2)}
\newcommand{\offsets} {($\Delta \alpha, \Delta \delta$)}
\newcommand{\vlsr}    {$V_{\rm lsr}$}
\newcommand{\NCO}     {$N{\rm (CO)}$}
\newcommand{\NHCOp}   {$N{\rm (HCO^+)}$}
\newcommand{\NCp}     {$N{\rm (C^+)}$}

\newcommand{\Vexp}     {$V_{\rm exp}$}

\title[PDRs around H\,{\normalsize \textit{II}} regions S235~A and S235~C]{The PDR structure and kinematics around the compact H\,{\Large \textbf{II}} regions S235~A and S235~C with \CII, \thirtCII, \OI{} and HCO$^+$ line profiles}

\author[M. S. Kirsanova et al.]{M. S. Kirsanova,$^{1,2}$\thanks{E-mail: kirsanova@inasan.ru}
V. Ossenkopf-Okada,$^{3}$
L. D. Anderson,$^{4,5,6}$
P. A. Boley,$^{2,7}$
\newauthor
J. H. Bieging,$^{8}$
Ya. N. Pavlyuchenkov,$^{1}$
M. Luisi,$^{5,6}$
N.~Schneider,$^{3}$
M.~Andersen,$^{9}$
\newauthor
M.~R.~Samal,$^{10}$
A.~M.~Sobolev$^{7}$
C.~Buchbender$^{3}$
R.~Aladro$^{11}$
Y.~Okada$^{3}$
\\
$^{1}$Institute of Astronomy, Russian Academy of Sciences, 119017, 48 Pyatnitskaya Str., Moscow, Russia\\
$^{2}$Moscow Institute of Physics and Technology, 141701, 9 Institutskiy per., Dolgoprudny, Moscow Region, Russia\\
$^{3}$I. Physikalisches Institut, Universit{\"a}t zu K{\"o}ln, Z{\"u}lpicher Strasse 77, D-50937 K{\"o}ln, Germany\\
$^{4}$Department of Physics and Astronomy, West Virginia University, Morgantown WV 26506\\
$^{5}$Center for Gravitational Waves and Cosmology, West Virginia University, Chestnut Ridge Research Building, Morgantown, WV 26505\\
$^{6}$Adjunct Astronomer at the Green Bank Observatory, P.O. Box 2, Green Bank WV 24944\\
$^{7}$Ural Federal University, 620075, 19 Mira Str., Ekaterinburg, Russia\\
$^{8}$Steward Observatory, The University of Arizona, Tucson, AZ 85721, USA\\
$^{9}$Gemini South, Casilla 603, La Serena, Chile\\
$^{10}$Physical Research Laboratory, Navrangpura, Ahmedabad, Gujarat 380009, India\\
$^{11}$Max-Planck-Institut f{\"u}r Radioastronomie, Auf dem H{\"u}gel 69, 53121 Bonn, Germany
}

\date{Accepted 2020 July 11. Received 2020 June 22; in original form 2020 February 28.}

\pubyear{2019}

\begin{document}
\label{firstpage}
\pagerange{\pageref{firstpage}--\pageref{lastpage}}
\maketitle

\begin{abstract}
The aim of the present work is to study structure and gas kinematics in the photodissociation regions (PDRs) around the compact \hii{} regions S235~A and S235~C. We observe the \CII{}, \thirtCII{} and \OI{} line emission, using SOFIA/upGREAT and complement them by data of HCO$^+$ and CO. We use the \thirtCII{} line to measure the optical depth of the \CII{} emission, and find that the \CII{} line profiles are influenced by self-absorption, while the \thirtCII{} line remains unaffected by these effects. Hence, for dense PDRs, \thirtCII{} emission is a better tracer of gas kinematics. The optical depth of the \CII{} line is up to 10 in S235~A. We find an expanding motion of the \CII-emitting layer of the PDRs into the front molecular layer in both regions. Comparison of the gas and dust columns shows that gas components visible neither in the \CII{} nor in low-J CO lines may contribute to the total column across S235~A. We test whether the observed properties of the PDRs match the predictions of spherical models of expanding \hii{} region + PDR + molecular cloud. Integrated intensities of the \thirtCII, \CII{} and \OI{} lines are well-represented by the model, but the models do not reproduce the double-peaked \CII{} line profiles due to an insufficient column density of C$^+$. The model predicts that the \OI{} line could be a more reliable tracer of gas kinematics, but the foreground self-absorbing material does not allow using it in the considered regions.
\end{abstract}



\section{Introduction}\label{sec:intro}

Massive O and B-stars reveal the places of their birth by powerful UV radiation. The stellar UV photons illuminate and heat the surrounding gas up to thousands of Kelvin. UV photons from the massive stars ionize hydrogen atoms and lead to the formation of \hii{} regions \citep[e.g.][and many others]{Stromgren1939, Kahn1954}. The \hii{} regions continue to grow in size, driven by the difference of the thermal pressure between the hot ionized and the cold neutral gas~\citep[see][and many others]{Mathews1965, Spitzer1978}. Shock waves accompany the expansion and move ahead of the ionization front into the neutral surrounding material. The shock compresses the neutral gas and collects it in a dense neutral shell \citep[see simulations by][]{Tenorio-Tagle1979, Beltrametti1982, Franco1990, Garcia-Segura1996, Hosokawa2006, Krumholz2007, Kirsanova2009, Raga2012, Tremblin2014, Bisbas2015, Williams2018}. A photodissociation region (PDR) is formed between the hot ionized and cold molecular material compressed by the shock \citep[e.g.][]{Tielens1985, Tielens1985ii, Tielens1993, Sternberg1995, Hollenbach1999, LePetit2006, Rollig2007, Bron2018}. In spite of the long research history, direct observations of the expansion of the compressed PDR material are still missing, while it has been suggested in works by \citet{Gomez2010, Deharveng2011, Pilleri2012, Torres-Flores2013, Schneider2018, Mookerjea2019, Pabst2019, Sitnik2019}.

The PDRs around \hii{} regions often appear as ring-like or arc structures in 8~\micron{} {\it Spitzer} and 12~\micron{} {\it WISE} images due to emission from bending/stretching modes of PAHs excited by the UV radiation, and also in {\it Herschel} images at longer wavelengths due to thermal emission from heated dust grains \citep[e.g.][]{Deharveng2005, Churchwell2006, Churchwell2007, Watson2009, Deharveng2010, Anderson2012, Topchieva_2017}. Several thousand PDRs have been identified thanks to the space-based surveys, but the PAH or dust emission does not carry any information about their velocity structure. Moreover, the dust is not directly coupled to the gas, and, in general, gas and dust temperatures differ. The gas cooling occurs predominantly through fine-structure lines of \OI{} at $63~\mu{}$m (hot and dense has) and \CII{} at $158~\mu{}$m (warm and less dense gas with temperature $\leq 200$~K and density of the order 100-1000~cm$^{-3}$) \citep[see e.~g.][]{Tielens1985, Hollenbach1999, Roellig2006}. Only observations of these spectral lines, obtained with high spectral resolution, allow us to trace gas kinematics and the physical conditions of the gas in the emitting region.

The neutral molecular shells around extended \hii{} regions are non-uniform and clumpy, so it is difficult, if ever possible, to detect the expansion of \hii{} regions as a whole using position-velocity (pv) diagrams \citep[e. g.][]{Dirienzo2012, Anderson2015, Deharveng2015, Trevino-Morales2016, Zhang2016, Anderson2019, KirsanovaPavyar2019}. Common manifestations of inhomogeneous structure near \hii{} regions are `pillars', which result when the clumpy medium is carved by the UV radiation of hot stars \citep[see e.~g.][]{Smith2010, Flagey2011, Benaglia2013, Panwar2019}. Moreover, according to theoretical expectations, the expansion velocity (\Vexp) of extended \hii{} regions embedded in molecular clouds is $\sim 1$~\kms{} with an age about 0.5--1~Myr \citep[see simulations in][]{Hosokawa2006, Kim2016, Zavagno2007, Kirsanova2009, Akimkin2015, Akimkin2017}, which is less than the typical velocity dispersions observed in molecular gas in high-mass star-forming regions.  In this case, implicit approaches may help. For example, multi-wavelength observations allowed \citet{Lopez2014} to estimate the gas pressure exerted on the surrounding neutral gas shells from the ionized gas and conclude that thermal pressure from the ionized gas dominates the expansion of the shells, rather than pressure from stellar wind or from dust grains accelerated by radiation pressure from the ionizing stars. Direct estimation of \Vexp{} from the pv diagrams can also support this conclusion, as was recently shown by \citet{Pabst2019} for a wind-blown bubble.

The best candidates to search for observational signatures of the expansion are embedded compact \hii{} regions.  The simulations mentioned above predict that \Vexp{} is higher for younger compact \hii{} regions. Moreover, small compact \hii{} regions might have geometries that are closer to spherical, because at small radii they have a lower chance of interacting with density fluctuations in the surrounding ISM, resulting in a more uniform expansion in all directions.

The aim of the present work is to study the geometry and gas kinematics in the PDRs around two compact \hii{} regions, and test whether their observed properties match the predictions of spherical models. We present observations of the \CII{} and \OI{} emission from PDRs, complemented by data of molecular line emission from the surrounding natal molecular cloud. We use the \thirtCII{} line to measure the optical depth of the \CII{} emission to study why the line profiles of the latter line are double-peaked: due to high optical depth or gas kinematics. Recent studies \citep[e.g.][]{Graf2012, Ossenkopf2015, Mookerjea2018, Mookerjea2019, Okada2019, Guevara_2020} have shown that the \CII{} lines can be optically thick in dense PDRs. 

\citet{Anderson2019} examined the radial velocities of the \CII{} line in the extended \hii{} region Sh2-235 (S235 hereafter). The velocities coincide with the $^{13}$CO(2--1) and carbon radio recombination lines (RRLs), but differ from the hydrogen RRLs and the fine-structure [N\,{\sc{ii}}] at 205~\micron{} by up to 4~\kms{}. The ionized gas in S235 escapes from the parent molecular cloud and surrounding PDR to the observer because of the absence of foreground dense material. There are two compact \hii{} regions, S235~A and S235~C, in the same molecular cloud, also mapped in \CII{} by \citet{Anderson2019}. The velocity difference between the hydrogen RRLs and the $^{13}$CO(2--1) lines appears to be insignificant there, so we expect that these compact \hii{} regions might have geometries that are closer to spherical.

\section{Target regions}\label{sec:target}

The compact \hii{} regions S235~A and S235~C are located 8\arcmin{} south of S235 (see Fig.~1 in the work by \citet{Anderson2019} to find a large-scale view S235 and the surrounding regions)]. All of the \hii{} regions in the area are embedded in the same giant molecular cloud, G174+2.5 \citep{Evans1981, Heyer1996, Kirsanova2014, Bieging2016, Ladeyschikov2016}. For the distance to the \hii{} regions, we adopt a value of 1.6~kpc from the maser parallax measurements by \citet{Burns2015}, who obtained $1.56^{+0.09}_{-0.08}$~kpc. Consequently, 10\arcsec{} in our observations translate into a physical scale of 0.08~pc. We also checked the possibility of using GAIA DR2 \citep{Gaia16,Gaia18} parallax measurements, in combination with the catalog of inferred distances to GAIA sources by \citet{Bailer-Jones2018} and found that the astrometric solutions for the ionizing stars in S235 A and C are poor. For example, the Renormalised Unit Weight Error (RUWE; see Gaia technical note GAIA-C3-TN-LU-LL-124-01), which characterizes the quality of the astrometric data, is greater than 1.3 for both sources. Due to the relative faintness of stars in the region ($G=18-20$\,mag), the corresponding large uncertainties in the GAIA parallaxes do not allow for a reliable distance determination.

Using infrared spectroscopy, \citet{Thompson1983} found that the ionizing star in S235~A \citep[which is also known as IRS~3 source in the work by][]{Evans1981} has spectral type of B0.5, and showed that the star is nearly on the Zero-Age Main Sequence. They measured a visual extinction $A_{\rm V} \approx 10$~mag in S235~A. \citet{Thompson1983}, and later \citet{Felli2004}, determined the luminosity of the ionizing star in S235~A to be $L_{*} = 1.1\times10^4$~$L_\odot$. For the ionizing star in S235~C \citet{Thompson1983} found $L_{*}=(0.8-1.2)\times10^4$~$L_\odot$ and a spectral type from B3 to B0.3. According to the calibration tables from \citet{Smith2002, Pecaut2013}, these luminosities correspond to stellar effective temperatures in the range \teff=25000--30000~K.

There are four bright point-source objects in the 3.6~\micron{} {\it Spitzer} image in the area around S235~A and S235~C. Their coordinates are given in Table~\ref{tab:ionizstars}. The sources S235~A$^{\star}$ and S235~C$^{\star}$ are the ionizing stars of the corresponding \hii{} regions \citep[see also][]{Felli1997, Chavarria2014, Bieging2016}. S235~B is a reflection nebula near a Herbig Be-type star S235B$^{\star}$ \citep{Boley2009}. S235~A-2$^{\star}$ is projected on a PDR to the south-east of S235~A$^{\star}$ and visible in K~band continuum emission \citep[e.g.][]{Felli1997, Klein2005}. 

\citet{Glushkov1975} determined the electron density (\nel) in S235~A using optical spectroscopy of [\ion{S}{II}] lines and found \nel{} at selected positions on the order of $10^3-10^4$~cm$^{-3}$, as well as a non-uniform distribution of $A_{\rm V}$ between 6 and 12~mag. \cite{Israel1978} determined size ($S$) and \nel{} of the \hii{} regions using aperture synthesis observations at 5.0 and 1.4~GHz. They found $S = 0.28$ and 0.34~pc for S235~A and S235~C, and \nel$=10^3$~cm$^{-3}$ and $4\times 10^2$~cm$^{-3}$ for S235~A and S235~C, respectively. \citet{Felli1997} obtained an image of Br$\gamma$ emission in S235~A and found a brightness distribution consisting of two spots, with the ionizing star situated between them. Visual inspection of the 3.6~\micron{} image, shown in Fig.~\ref{fig:CIImom0}, gives a projected distance of 0.17-0.20~pc from the ionizing stars in S235~A and S235~C, respectively, to the bright arcs delineating the PDRs from the  east and south-east side, and a maximum of about 0.3~pc for the west side of S235~C.

\begin{table}
\caption{Bright point sources on the {\it Spitzer} 3.6~\micron{} image in the area of S235~A and S235~C. References: $^1$\citet{Chavarria2014}, $^2$this work, $^3$\citet{Dewangan2011}, $^4$\citet{Thompson1983}, $^5$\citet{Boley2009}, $^6$\citet{Gyulbudaghian1978}.}
\begin{tabular}{ccc}
\hline
Source & $\alpha({\rm J2000.0})$,  $\delta({\rm J2000.0})$       & type\\
       & ($^h$ $^m$ $^s$),         ($^{\circ}$ \arcmin{} \arcsec) & \\
\hline
S235~A$^{\star}$   & 5 40 52.58, +35 42 18.6$^1$ & B0.5$^4$ \\
S235~A-2$^{\star}$ & 5 40 53.39, +35 42 07.1$^2$ & Class~I YSO$^3$\\
S235~B$^{\star}$   & 5 40 52.39, +35 41 29.4$^1$ & Herbig Be (B1Ve)$^5$\\
S235~C$^{\star}$   & 5 40 51.41, +35 38 30.0$^1$ & B3-B0.3$^4$\\
\hline
\end{tabular}
\label{tab:ionizstars}
\end{table}

\section{Observational data}\label{sec:obs}

\subsection{{\it SOFIA} C\,{\small II}, $^{13}$C\,{\small II} and O\,{\small I} observations}\label{sec:obs:cii}

SOFIA/upGREAT\footnote{GREAT is a development by the MPI f\"{u}r Radioastronomie and the KOSMA / Universit\"{a}t zu K\"{o}ln, in cooperation with the MPI f\"{u}r Sonnensystemforschung and the DLR Institut f\"{u}r Planetenforschung} mapping observations of the \CII{} $158~\mu{}$m line were done in 2016-2017 and are described by \citet{Anderson2019}. The typical noise level of that data is 5~K at 0.385~\kms{} resolution on a $T_{\rm mb}$ scale. This noise level only allowed for a tentative detection of the \thirtCII{} line. For a reliable detection, deep integrations were performed towards three positions in guaranteed time on November 21 and 23, 2018, Project-ID 83\_0624. For these integrations we chose positions where the \CII{} line showed some double peak structure, to find out whether that structure is reproduced in \thirtCII{}. The center of the upGREAT Low-Frequency Array (LFA) was pointed towards $\alpha_1({\rm J2000.0}) =
05^{\rm h}40^{\rm m}53\fs 17^{s}$, $\delta_1({\rm J2000.0}) =
+35^{\circ}42\arcmin11\farcs 03$, and $\alpha_2({\rm J2000.0}) = 05^{\rm h}40^{\rm m}42\fs 14^{s}$, $\delta_2({\rm J2000.0}) =
+35^{\circ}41\arcmin56\farcs 99$ in S235~A and towards the peak of the \CII{} emission at $\alpha({\rm J2000.0}) = 05^{\rm h}40^{\rm m}52\fs 09^{s}$ and $\delta({\rm J2000.0}) = +35^{\circ}38\arcmin37\farcs 03$ in S235~C. The signal was integrated for 115~s at each position. While the detection of \thirtCII{} was the main goal of these observations, the \OI~$63~\mu{}$m line was observed simultaneously with the High-Frequency Array (HFA) as a side-product. Because of the different footprints of the two arrays on the sky, only the central positions match, so that we can compare the \OI{} data with the \thirtCII{} data only for the central pixels.

The data were calibrated by the standard GREAT pipeline \citep{Guan2012} to convert the observed counts to the main beam temperature ($T_{\mathrm{mb}}$) using a main beam efficiency of $\eta_{\rm mb}=0.68$ for \thirtCII{} and $\eta_{\rm mb}=0.7$ for \OI{}. For \thirtCII{} both polarizations were averaged. We subtracted linear baselines and spectrally resampled the data to 0.192~\kms{} channel width for \thirtCII{} and 0.14~\kms{} for \OI{}. The spatial resolution of the \thirtCII{} data is 14.6\arcsec, while the spatial resolution of the \OI{} data is 5.9\arcsec. The sensitivity varies slightly between the individual pixels of the arrays, leading to noise levels of 0.25~K for the best pixels of the LFA and 0.45~K for the worst pixel and noise levels of 0.5~K for the central pixel of the HFA. The \CII{} line was always detected simultaneously with \thirtCII{} with the same noise level at the same positions.

From the \CII{} and \thirtCII{} lines we calculate the optical depth of the \CII{} line (\tauCII{}) under assumption of the same excitation temperature \tex{} for both transitions. With the carbon atomic abundance ratio $r={\rm ^{12} C}/{\rm ^{13} C}=80$ \citep{Wilson1999}, used in studies of the G174+2.5 giant molecular cloud, situated in the external Perseus spiral arm by e.g. \citet{Bieging2016, Kirsanova2017}, we obtain \tauCII{} from the line intensity ratio in the each velocity channel where the intensities of both lines are higher than $3\sigma$ of the noise:

\begin{equation}\label{eq:tau}
\frac{T_{\rm {mb} [^{13}CII]}}{T_{\rm {mb} [CII]}} = \frac{1-{\rm exp}(-\tau_{\rm C^+} / r) }{1-{\rm exp}(-\tau_{\rm C^+}) } \approx \frac{\tau_{\rm C^+} / r}{1-{\rm exp}(-\tau_{\rm C^+}) } 
\end{equation}

The value of $T_{\rm {mb}}{\rm [^{13}CII]} $ is a weighted sum over three hyper-fine components. Fractionation effects that would change the abundance ratio $r$ are not expected for the high densities of our sources \citep{RoelligOssenkopf2013}. To convert the optical depth into a column density of C$^+$, we need to know the excitation temperature. For those velocity channels where we can determine $\tau_{\rm C^+}$, we use the radiation transfer equation:

\begin{equation}\label{eq:tex}
T_{\rm {mb}} = ( J_\nu(T_{\rm ex})-J_\nu(T_{\rm bg}) ) ( 1-{\rm exp} ( -\tau_{\rm C^+}) ) \quad {\rm (K)},
\end{equation}

there $J_\nu(T)=({\rm h}\nu/{\rm k})/({\rm exp}(\frac{{\rm h}\nu}{{\rm k}T})-1)$. The background can be ignored at 1.9~THz, so that $J_\nu(T_{\rm {bg}})\approx 0$. With \tex{} the value of $N_{\rm C^+}$ can be calculated for each velocity channel according to Eq.~3 from \citet{Ossenkopf2013}:

\begin{equation}\label{eq:thick}
N_{\rm C^+} = 1.4\times 10^{17} \frac{1+2{\rm exp}(-91.2 {\rm K}/T_{\rm ex})}{1-{\rm exp}(-91.2 {\rm K}/T_{\rm ex})} \int \tau_{\rm C^+} {\rm d}v \quad ({\rm cm^{-2}}),
\end{equation}
for ${\rm d}v$ in \kms.

For the velocity channels where \thirtCII{} was not detected above the $3\sigma$ level, we treat the \CII{} emission as optically thin and apply Eqs.~1 and 2 from \citet{Ossenkopf2013}:

\begin{equation}\label{eq:thin}
N_{\rm C^+} = \frac{8 \pi {\rm k \nu^2}}{{\rm h c^3}A} \frac{1+2{\rm exp}(-91.2 {\rm K}/ \langle T_{\rm ex} \rangle)}{2{\rm exp}(-91.2 {\rm K}/ \langle T_{\rm ex} \rangle )} \int T_{\rm {mb} [CII]} {\rm d}v \quad ({\rm cm^{-2}}),
\end{equation}
assuming that the average of the excitation temperatures $ \langle T_{\rm ex} \rangle$ calculated over the channels with a \thirtCII{} detection also applies to the optically thin channels. $A=2.3\times 10^{-6}$ is the Einstein-A coefficient and ${\rm d}v$ is in cm~s$^{-1}$. Summation over optically thin and thick channels gives the total $\rm C^+$ column density in the every observed position. The uncertainties of $N_{\rm C^+}$ were estimated using standard error propagation.

\subsection{{\it SMT} \HCOp{} observations}\label{sec:HCOpdata}

We use previously unpublished observations of the \HCOp{} emission at 267~GHz made with the 10m Heinrich Hertz Submillimeter Telescope telescope taken on April 21, 2009 
(S235~A and B region) and May 5, 2009 (S235C). The observations were made in the on-the-fly (OTF) observing mode, where the telescope scans continuously in a raster pattern with spectra sampled every 0.1~second. For these maps, the scan rate was 5\arcsec{} per second, and the spectra were binned by 4 samples, for an effective sample spacing of 2\arcsec. The scan direction was along Right Ascension and the row spacing was 5\arcsec. The telescope resolution is 28\arcsec{} (FWHM), so the maps are well-sampled in both coordinates. The map size for the S235AB region was 3.5\arcmin$\times$3.5\arcmin{} centered at $\alpha({\rm J2000.0}) = 05^{\rm h}40^{\rm m}51.94\fs^{}$ and $\delta({\rm J2000.0}) = +35^{\circ}41\arcmin{}35\farcs 49$.  The map of the S235C region was 3\arcmin{}$\times$3\arcmin{} centered at $\alpha({\rm J2000.0}) = 05^{\rm h}40^{\rm m}52.5\fs^{}$ and $\delta({\rm J2000.0}) = +35^{\circ}38\arcmin24\farcs$.  The receiver was the sideband-separating mixer system as described in \citet{Bieging2016}, with sideband separations of about 15~dB, which provides excellent intensity calibration using the standard method \citep[see ][]{Bieging2016}. Typical system temperatures were 280 -- 300 K (SSB). The spectrometer was a filterbank with 128 channels of 250~kHz filter width and channel spacing, giving a velocity resolution for the HCO$^+$~$J=3-2$ line of 0.281~\kms{} covering 36~\kms{} total bandwidth. Typical rms noise levels in each spectrometer channel were about 0.2~K.

As we do not have H$^{13}$CO$^+$(3--2) line observations in the same region and can not estimate the optical depth of the HCO$^+$(3--2) line, we stick to an LTE approach and the optically thin approximation to calculate the HCO$^+$ column density (\NHCOp) in ${\rm cm^{-2}}$ with Eq.~80 from \citet{Mangum2015} assuming a beam filling factor $f=1$: 

\begin{equation}\label{eq:thinhcop}
N = \frac{3 {\rm h} Q_{\rm rot}}{8 \pi^3 \mu^2 S g_{\rm u}} \times \frac{{\rm exp}\left( \frac{E_{\rm u}}{{\rm k}T_{\rm ex}} \right) } { {\rm exp}\left( \frac{{\rm h} \nu}{{\rm k}T_{\rm ex}} \right) - 1} \times \frac{W}{ J_\nu(T_{\rm {ex} }) -  J_\nu(T_{\rm {bg} }) } 
 \quad ({\rm cm^{-2}}),
\end{equation}
where $W = \int T_{\rm mb} {\rm d}v / f$ is the integrated intensity of the emission in K~\kms. The rotational partition function $Q_{\rm rot}$ can be approximated as $Q_{\rm rot} \sim \frac{kT_{\rm ex}}{\rm hB}+\frac{1}{3}$. For \HCOp{} the rotational angular momentum constant is $B_0=44590$~MHz, the dipole moment is $\mu=3.9$~Debye, the energy of the upper state $J_{\rm u}=3$ divided by Boltzmann constant is $k$: $E_{\rm u}/{\rm k}=25.7$~K, $S=\frac{J_{\rm u}^2-K_{\rm u}^2}{J_{\rm u}(2J_{\rm u}+1)}$, $g_{\rm u} = 2J_{\rm u}+1$, $K_{\rm u}=0$. \citet{Bieging2016} showed that the bulk of the molecular CO gas in the S235 complex has \tex{} of about 20~K, so we use the same value to calculate \NHCOp{}. We note that the actual excitation temperature of the \HCOp{} line is probably higher as this line most likely arises partially in the warm PDR (see below our model predictions for S235~A and S235~C in Sec.~\ref{sec:modelpred}). Therefore, the provided map of the \NHCOp{} distribution gives probably a lower limit only. 

\subsection{Archival CO and $^{13}$CO data}\label{sec:COdata}

We use CO(2--1) and $^{13}$CO(2--1) data from \citet{Bieging2016} to calculate the CO column density in the same directions where we have \CII{} and \thirtCII{} data. The spatial resolution of the CO data is 38\arcsec{} and the spectral resolution of the data is 0.3~\kms.

The column density of CO molecules (\NCO) is calculated with the same channel-by-channel procedure as for the \CII{} and \thirtCII{} lines. To obtain CO optical depth ($\tau_{\rm CO}$) we use Eq.~\ref{eq:tau} for the CO(2--1) and $^{13}$CO(2--1) line intensities in the each velocity channel and with the same carbon atomic abundance ratio $r=80$ and then determine \tex{} value with the analog of equation~\ref{eq:tex}. Using Eq.~\ref{eq:thinhcop} with \tmb{} of the CO(2--1) line and following constants: $B_0=55101$~MHz, $\mu=0.11$~Debye, $E_{\rm u} / {\rm k}=16.6$~K, we calculate \NCO{} in the optically thin approximation $N_{\rm CO}^{\rm thin}$ for the each channel, and then correct it using the optical depth correction factor introduced by \citet{Frerking1982}, see also \citet{Goldsmith1999}:

\begin{equation}\label{eq:tauco}
N_{\rm CO}= N_{\rm CO}^{\rm thin} \frac{\tau_{\rm CO}}{1-{\rm exp}(-\tau_{\rm CO})} \quad ({\rm cm^{-2}}).
\end{equation}

For the velocity channels where $^{13}$CO(2--1) line intensity is not higher than $3\sigma$ of the noise we use $ \langle T_{\rm ex} \rangle$ averaged over the optically thick channels and Eq.~\ref{eq:thinhcop} to calculate \NCO{}. The uncertainties of $N_{\rm CO}$ were estimated using standard error propagation formulae.

\subsection{Archival continuum data}\label{dustcont}

In order to determine the dust column density from the continuum spectral energy distributions we collected data from two different instruments. First, archival maps of continuum emission from the Bolocam galactic plane survey (BGPS) were downloaded\footnote{https://irsa.ipac.caltech.edu/data/BOLOCAM\_GPS/}. We use the SHARC-II maps at 350~\micron{} \citep{Merello2015} and the BGPS v2.1 data at 1.1~mm \citep{Ginsburg2013} without additional manual reduction. The spatial resolution of the BGPS data is 8.5\arcsec{} and 33\arcsec{} for 350~\micron{} and 1.1~mm, respectively. The SHARC-II camera has 2.59\arcmin$\times$0.97\arcmin{} field of view but is not sensitive to extended emission on scales larger than $\approx 1$\arcmin. For S235~A and S235~C this corresponds to about 0.5~pc.

Second, we use SCUBA-2 data at 450 and 850~\micron{} downloaded from the JCMT Science Archive\footnote{http://www.cadc-ccda.hia-iha.nrc-cnrc.gc.ca/en/jcmt/}, taken as part of the SCUBA-2 Ambitious Sky Survey (SASSy) \citep{Thompson2007,Nettke2017}. The spatial resolution of the data is $\approx 9$ and 14\arcsec, respectively. The SCUBA-2 camera covers a 45\arcmin$^2$ field of view \citep{Holland2013}. We manually reduced the SCUBA-2 data using the 2018A release of Starlink \citep{Currie2014} and versions 1.6.1 of the SMURF \citep{Chapin2013} and 2.5-8 of the KAPPA packages through the ORAC-DR \citep{Cavanagh2008} pipeline.

All continuum data were smoothed to the resolution of 33$''$ provided by the Bolocam data on a common grid. The maps from SCUBA-2 and SHARC-II showed some large-scale negative flux values with a magnitude above the indicated flux errors. These probably stem from the removal of large-scale emission in the chopping observations. We removed those negative values by adding a constant flux to the maps so that no large-scale negative features above the indicated calibration uncertainty remained. This required adding 1.9~mJy/arcsec$^2$ at 350~\micron, 5.0~mJy/arcsec$^2$ at 450~\micron, and 0.1~mJy/arcsec$^2$ at 850~\micron. We added those offsets to the uncertainties in the fitting. The resulting maps then showed typical fluxes around zero outside of S235~AB and S235~C.

The four maps were fitted by a grey-body SED fit using an emissivity law of $\kappa = \kappa_{250} (\lambda/250~\micron{})^{-\beta}$ with $\kappa_{250}=2.16\times 10^{-25}$~cm$^{-2}$ per hydrogen proton and $\beta=1.8$~\citep{Juvela2015}. To convert the hydrogen column density obtained from the dust emission maps or from CO emission maps into an equivalent visual extinction $A_{\rm V}$, measured in magnitudes, we use the standard ratio of $1.87\times 10^{21}$ H-protons per cm$^{-2}/ A_{\rm V}$ from \citet{Bohlin1978}. Variations of $\beta$ have little impact on the derived column density as the main constraints stems from the long-wavelength data. The temperature fit was constrained to values of at least 9~K, a typical value outside of the sources. This excludes column density spikes in regions with low fluxes due to noise in the data. The uncertainties in
the derived equivalent $A_{\rm V}$ from the fit fall well below 1~mag for all observed points in S235~A and grow up to equivalent $A_V\approx 2$~mag for the two most western points in S235~C.

\section{Observational view of the PDR\lowercase{s}}

\subsection{Spatial distribution of the line emission}\label{sec:obsPDRs}

\begin{figure*}
\includegraphics[height=14cm]{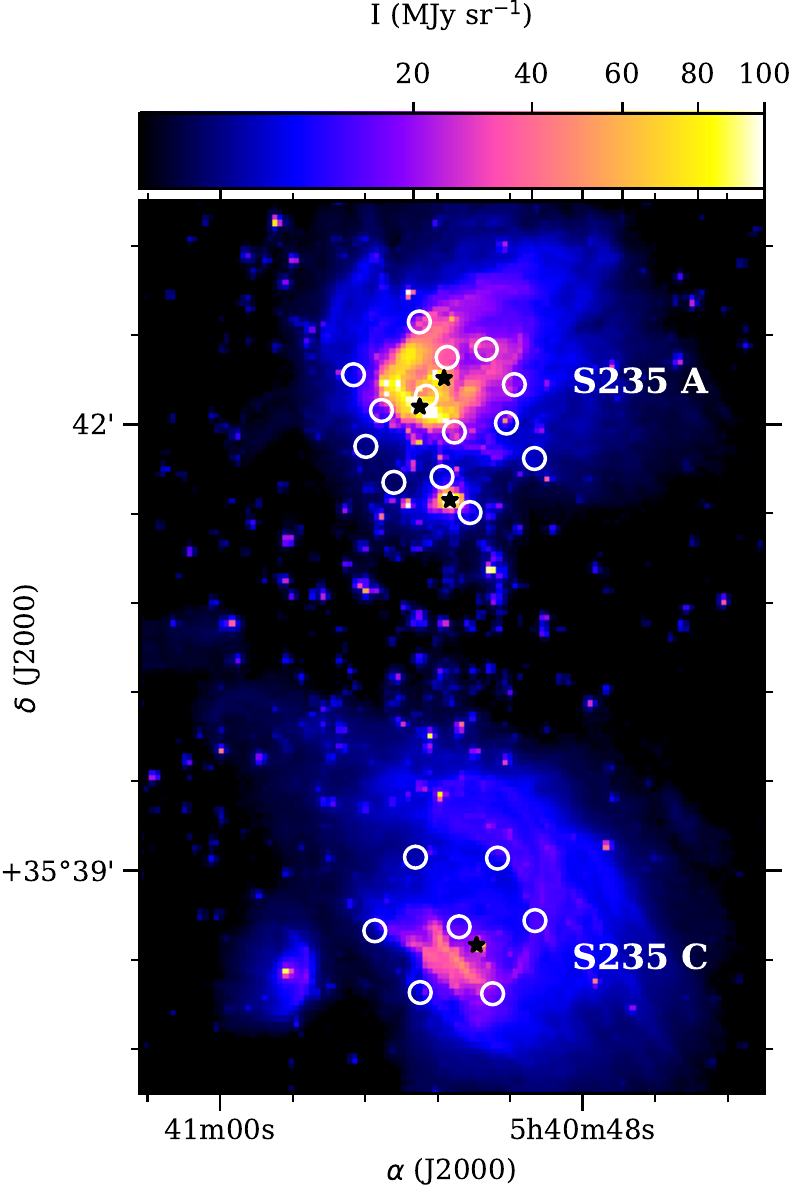}
\includegraphics[height=13.85cm]{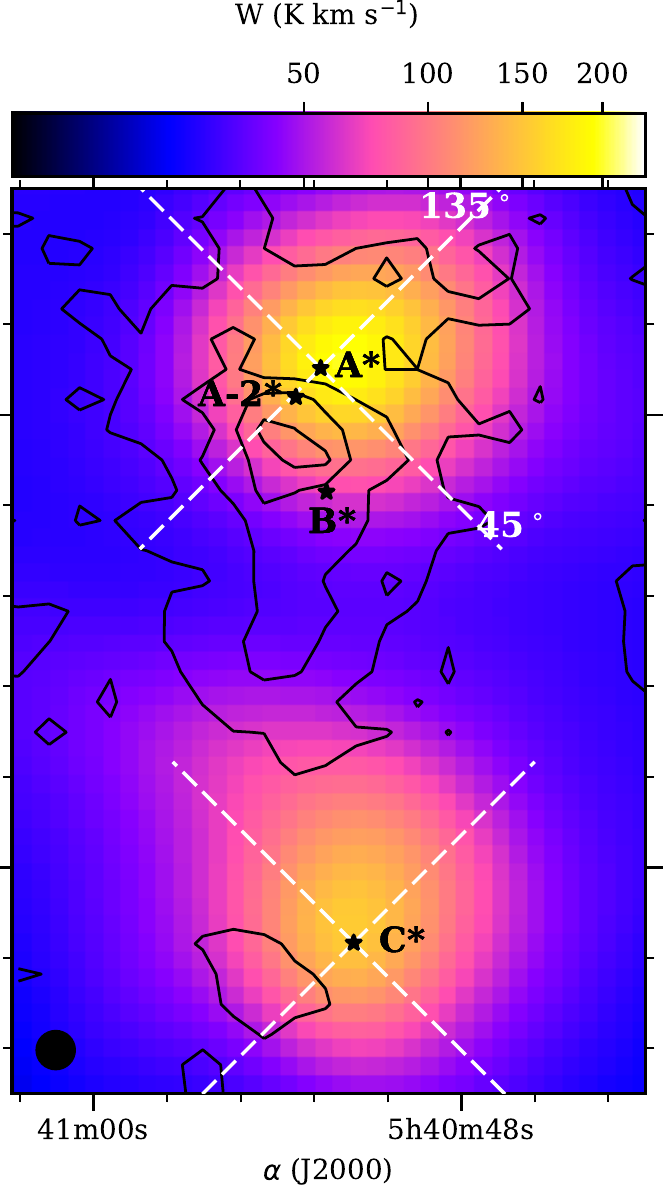}
\caption{Left: The 3.6~\micron{} {\it Spitzer} image of the region \citep[only pixels with S/N ratio higher than $3\sigma$,][]{2004sptz.prop..201F} with the positions where \thirtCII{} line was observed shown by the white empty circles. Right: Integrated intensity of the \CII{} $158~\mu{}$m emission.  White dashed lines show cuts for the pv~diagrams where the numbers give the angles measured west of south, with the ionizing star at the origin. The integrated intensity of the \HCOp{} emission is overlaid in black contours at 2, 6, 10 and 14 K~\kms. Point sources from Table~\ref{tab:ionizstars} are shown by black stars on both panels, labeled by black symbols on the right panel}.
\label{fig:CIImom0}
\end{figure*}

The \CII{} $158~\mu{}$m emission is detected in both \hii{} regions as already reported by \citet{Anderson2019}. The map of the \CII{} $158~\mu{}$m integrated line emission is shown in Fig.~\ref{fig:CIImom0}. Peak of the \CII{} emission is shifted by about 10\arcsec{} to the north-west of the ionizing star in S235~A, corresponding to the physical distance of 0.08~pc, while the emission peak coincides with the ionizing star position in S235~C. Both S235~A and S235~C have elliptical shapes in the \CII{} images, with an axis ratio $\approx 0.8$. S235~C also shows a diffuse tail extending to the north-west. The orientation and size of the \CII{} ellipsoids are similar to the 3.6~\micron{} {\it Spitzer} arc-like images of the objects shown in the Fig.~\ref{fig:CIImom0}. There is no enhancement of the \CII{} emission in the direction of the stellar cluster which is situated between the \hii{} regions.

The spatial distribution of the \HCOp{} emission significantly differs from \CII{}: molecular gas is concentrated in the south and south-eastern part of S235~A and in the south-eastern part of S235~C. There are no peaks of the \HCOp{} emission in the north and north-western parts of the \hii{} regions. The \HCOp{} peak around S235~A is situated between the point sources S235~A-2$^{\star}$ and S235~B$^{\star}$ and partially following the bright shell-like emission at 3.6~\micron{} (Fig.~\ref{fig:CIImom0}a). The estimated angular size of the \HCOp{} cloud around S235~A, calculated through the contour at 2~K~\kms, is 70\arcsec$\times$100\arcsec{} (0.5$\times$0.8~pc) elongated in the north-south direction. 

\begin{figure*}
\includegraphics[width=1.95\columnwidth]{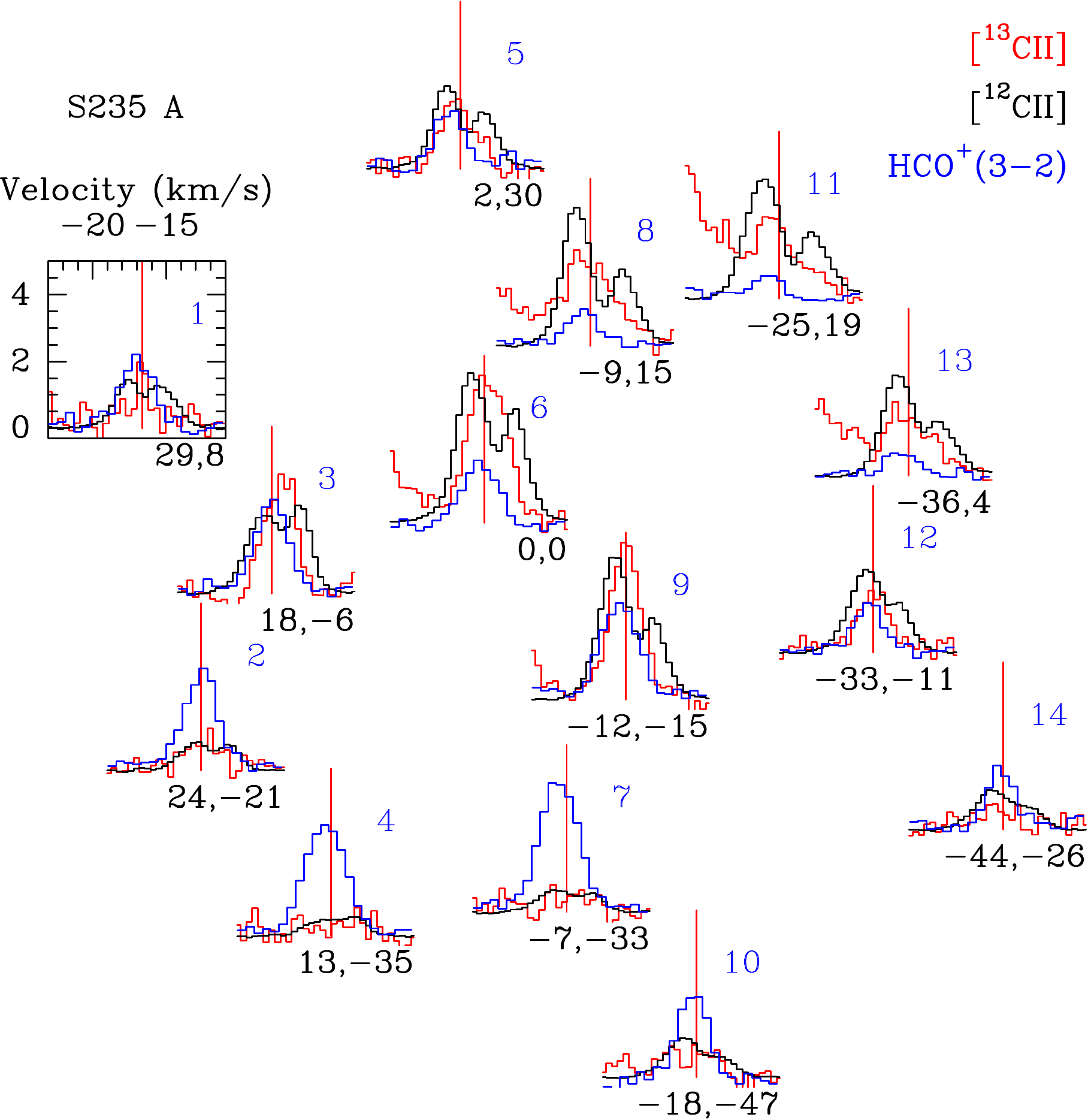}
\caption{\thirtCII{}, \CII{} and \HCOp{} spectra in S235~A shown by red, black and blue lines, respectively. The \thirtCII{} spectra are the result of the combined three hyperfine components. The velocity range is from --23 to --11~\kms{} in the plots. Offsets of the spectra in arcsec (labelled by black) are given relative to $\alpha_1({\rm J2000.0}) = 05^{\rm h}40^{\rm m}53\fs 17^{s}$ and $\delta_1({\rm J2000.0}) = +35^{\circ}42\arcmin11\farcs 03$ (spectrum 6). The \CII{} spectra are divided by a factor of 17. The red vertical line corresponds to a velocity of --16.6~\kms{} to show the velocity shift of the spectra. The excursion at the blue end of some of the \thirtCII{} spectra is due to contamination by the red wing of the nearby \CII{} line.}
\label{fig:13CIIspeA}
\end{figure*}

\begin{figure*}
\includegraphics[width=1.95\columnwidth]{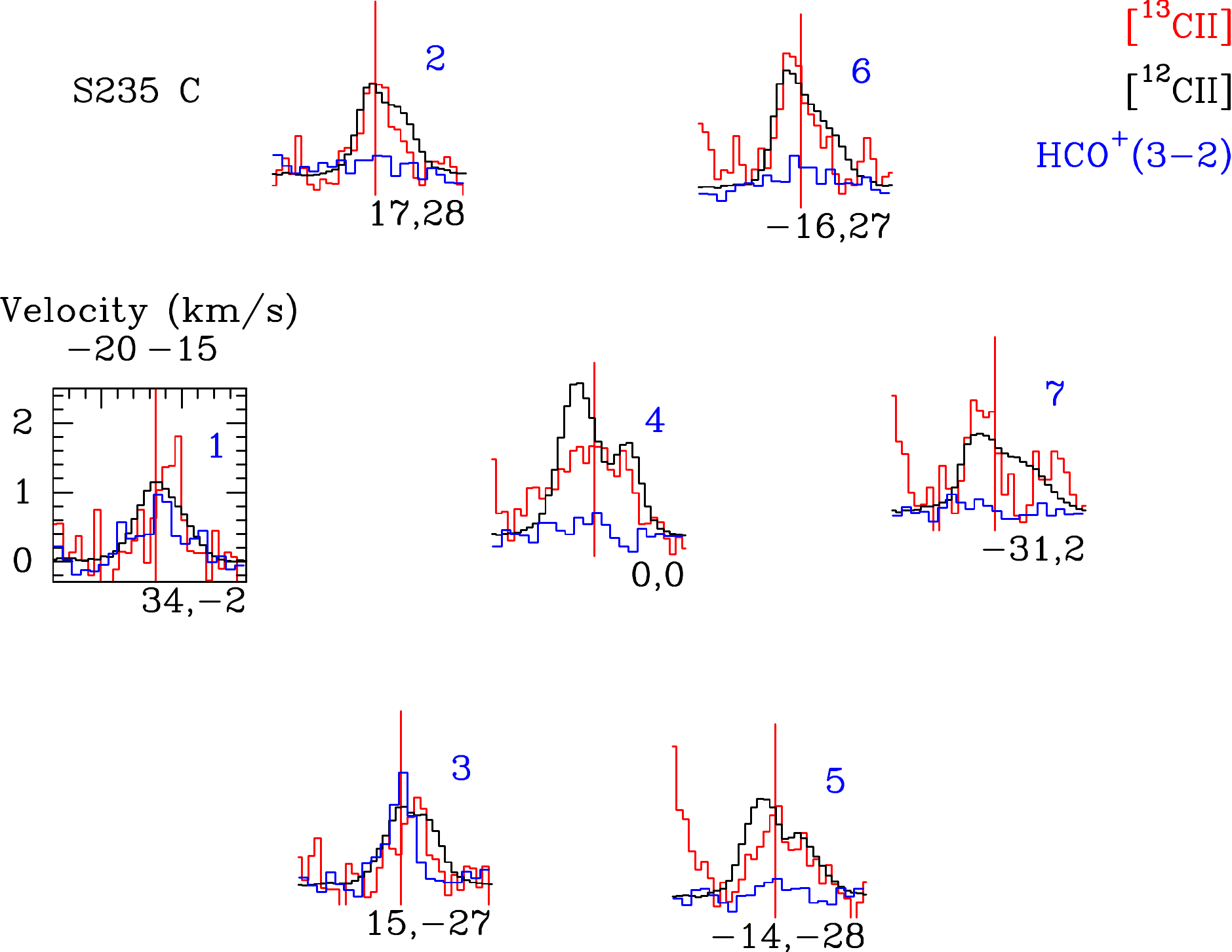}
\caption{\thirtCII{}, \CII{} and \HCOp{} spectra in S235~C (right) shown by red, black and blue lines, respectively. The \thirtCII{} spectra are the result of the combined three hyperfine components. The velocity range is from --23 to --11~\kms{} in the plots. Offsets of the spectra (labelled by black) are given relative to $\alpha({\rm J2000.0}) = 05^{\rm h}40^{\rm m}52\fs 09^{s}$ and $\delta({\rm J2000.0}) = +35^{\circ}38\arcmin37\farcs 03$ (spectrum 4). The \CII{} spectra are divided by a factor of 25. The red vertical line corresponds to a velocity of --16.6~\kms{} to show the velocity shift of the spectra. The excursion at the blue end of some of the \thirtCII{} spectra is due to contamination by the red wing of the nearby \CII{} line.}
\label{fig:13CIIspeC}
\end{figure*}

The \thirtCII{} spectra in 14 positions of S235~A and in 7 positions of S235~C are shown in Fig.~\ref{fig:13CIIspeA} and \ref{fig:13CIIspeC}. The majority of the detected \thirtCII{} lines show a single-peak in contrast to the \CII{} lines which are mostly double-peaked. The brightest \thirtCII{} line has \tmb$=4.7$~K in S235~A at position 9 with \offsets\,=(--12\arcsec,--15\arcsec) while the integrated \thirtCII{} intensities are the same in this position and near S235~A-2$^{\star}$ at position 6 with offset \offsets\,=(0\arcsec,0\arcsec). We have not detected any \thirtCII{} with a signal-to-noise ratio $S/N > 3$ near S235~B$^{\star}$ at position 7 with \offsets\,=(--7\arcsec,--33\arcsec). The brightest \thirtCII{} line in the S235~C PDR has \tmb$=1.7$~K at position 6 with \offsets\,=(--16\arcsec,27\arcsec). The average integrated and line peak intensities are shown in Table~\ref{tab:avervalues}. These values will be used in Sec.~\ref{sec:model} as selection criteria for theoretical models. We note that the scatter in the \CII{} and \thirtCII{} line intensities is significantly higher in S235~A. Parameters of Gaussian fits to the \thirtCII{} line for each position are given in Table~\ref{tab:CIIres}.

\begin{table*}
\caption{Parameters of the \thirtCII{} and \HCOp{} lines fitted with single Gauss function. The parameters are given for those positions where the signal-to-noise ratio $ \geq 3$.}
\begin{tabular}{lrcccccccc}
\hline
N & $(\Delta \alpha$, $\Delta \delta$) & \multicolumn{4}{c}{\thirtCII} & \multicolumn{4}{c}{\HCOp}\\
\cmidrule(lr){3-6}\cmidrule(lr){7-10}
  &                                 &  W        & \vlsr{} & FWHM    & \tmb{}  & W        & \vlsr{} & FWHM    & \tmb{}\\
  & (\arcsec{}      , \arcsec{})      & K~\kms{} & \kms{} & \kms{}   & K & K~\kms{} & \kms{} & \kms{}   & K \\
\hline
\multicolumn{10}{c}{\it S235~A}\\
1* &  (29           ,    8) &  1.7$\pm$0.4 &--16.6$\pm$0.1 & 0.8$\pm$0.3 & 1.9 $\pm$0.2 & 5.1$\pm$0.2& --16.8$\pm$0.1 & 2.7$\pm$0.1 & 1.8$\pm$0.2\\
2 &  (24           , --21) &  2.5$\pm$0.4 &--16.5$\pm$0.2 & 2.8$\pm$0.5 & 0.9 $\pm$0.3  & 8.3$\pm$0.2& --16.6$\pm$0.1 & 2.7$\pm$0.1 & 2.9$\pm$0.2\\
3 &  (18           ,  --6) &  9.7$\pm$0.5 &--16.0$\pm$0.1 & 2.5$\pm$0.1 & 3.6 $\pm$0.5 & 7.9$\pm$0.3 & --16.9$\pm$0.1 & 2.5$\pm$0.1 & 2.9$\pm$0.2\\
4  &  (13           , --35) &  \multicolumn{4}{c}{--} & 11.6$\pm$0.2& --17.0$\pm$0.1 & 3.2$\pm$0.1 & 3.4$\pm$0.1\\
5 &   (2           ,   30) &  5.4$\pm$0.3 &--16.9$\pm$0.1 & 2.6$\pm$0.2 & 1.9 $\pm$0.3 & 3.7$\pm$0.2 & --17.4$\pm0.1$ & 2.2$\pm$0.1 & 1.7$\pm$0.16\\
6 &   (0           ,    0) & 11.6$\pm$0.3 &--16.4$\pm$0.1 & 2.8$\pm$0.1 & 3.9 $\pm$0.3 & 5.1$\pm$0.2 & --17.2$\pm$0.1 & 2.7$\pm$0.1& 1.8$\pm$0.1\\
7 & (--7           , --33) &  \multicolumn{4}{c}{--} & 12.7$\pm$0.3 & --17.2$\pm$0.1 & 3.0$\pm$0.1 & 3.9$\pm$0.2\\
8 & (--9           ,   15) &  6.3$\pm$0.4 &--16.8$\pm$0.1 & 3.2$\pm$0.2 & 1.9 $\pm$0.4 & 2.6$\pm$0.2 & --17.2$\pm$0.1 & 2.4$\pm$0.2 & 1.1$\pm$0.2\\
9 & (--12           , --15) & 11.7$\pm$0.3 &--16.6$\pm$0.1 & 2.4$\pm$0.1 & 4.5 $\pm$0.2 & 8.5$\pm$0.3 & --17.3$\pm$0.1 & 2.9$\pm$0.1 & 2.8$\pm$0.1\\
10  &(--18           , --47) &  \multicolumn{4}{c}{--} & 6.3$\pm$0.2 & --16.8$\pm$0.1 & 2.3$\pm$0.1 & 2.6$\pm$0.2\\
11 &(--25           ,   19) &  9.1$\pm$0.7 &--16.9$\pm$0.1 & 4.0$\pm$0.4 & 2.1 $\pm$0.3 & 1.8$\pm$0.2 & --17.5$\pm$0.1 & 2.2$\pm$0.3& 0.7$\pm$0.2\\
12 &(--33           , --11) &  3.9$\pm$0.3 &--16.6$\pm$0.1 & 2.8$\pm$0.2 & 1.7 $\pm$0.3 & 3.4$\pm$0.2 & --16.9$\pm$0.1 & 2.2$\pm$0.2 & 1.5$\pm$0.1\\
13 &(--36           ,    4) &  2.5$\pm$0.4 &--16.7$\pm$0.2 & 2.4$\pm$0.4 & 1.0 $\pm$0.3 & 1.7$\pm$0.2 & --17.1$\pm$0.1 & 2.5$\pm$0.3 & 0.7$\pm$0.2\\
14  &(--44           , --26) &  \multicolumn{4}{c}{--} & 3.2$\pm$0.2& --16.7$\pm$0.1 & 1.9$\pm$0.2 & 1.6$\pm$0.2\\
\multicolumn{10}{c}{\it S235~C}\\
1 &  (34           ,  --2) &  2.8$\pm$0.5 &--15.7$\pm$0.1 & 1.7$\pm$0.3 & 1.6$\pm$0.4 & 1.9$\pm$0.1 & --16.4$\pm$0.1 & 1.5$\pm$0.1 & 1.3$\pm$0.1\\
2 &  (17           ,   28) &  3.8$\pm$0.4 &--16.5$\pm$0.1 & 2.6$\pm$0.3 & 1.4$\pm$0.3 & \multicolumn{4}{c}{--}\\
3 &  (15           , --27) &  2.5$\pm$0.3 &--15.8$\pm$0.1 & 2.2$\pm$0.4 & 1.1$\pm$0.3 & 2.9$\pm$0.2& --16.7$\pm$0.1 & 1.7$\pm$0.2 & 1.6$\pm$0.2\\
4 &   (0           ,    0) &  6.5$\pm$0.5 &--17.0$\pm$0.2 & 5.2$\pm$0.5 & 1.2$\pm$0.2 & \multicolumn{4}{c}{--}\\
5 & (--14           , --28) &  2.5$\pm$0.5 &--16.4$\pm$0.3 & 2.8$\pm$0.6 & 0.9$\pm$0.2 & \multicolumn{4}{c}{--}\\
6 & (--16           ,   27) &  3.5$\pm$0.4 &--17.1$\pm$0.1 & 2.0$\pm$0.3 & 1.6$\pm$0.3 & \multicolumn{4}{c}{--}\\
7 & (--31           ,    2) &  2.6$\pm$0.5 &--17.5$\pm$0.2 & 1.7$\pm$0.3 & 1.4$\pm$0.5 & \multicolumn{4}{c}{--}\\
\hline
\multicolumn{7}{l}{$^{*}$The \thirtCII{} spectrum was Hanning-smoothed before fitting.}
\end{tabular}
\label{tab:CIIres}
\end{table*}

\begin{table}
    \centering
    \begin{tabular}{c|l|c|c}
    \hline
    line & averaged value & S235~A & S235~C \\
    \hline
    \thirtCII{} & $\langle W \rangle$ (K~\kms)  & $5.0\pm3.8$ & $3.5\pm1.3$\\
    \thirtCII{} & $\langle T_{\rm mb} \rangle$ (K)  & $1.8\pm1.3$ & $1.3\pm0.2$\\
         \CII{} & $\langle W \rangle$ (K~\kms)   & $230\pm80$ & $225\pm42$\\
    \OI{} & $ \langle W \rangle$ (K~\kms)   & $30.8\pm14.4$ & $8.8\pm5.2$\\
     \hline
    \end{tabular}
    \caption{The average observed values over the 14 positions in S235~A and over the 6 positions in the S235~C PDR.}
    \label{tab:avervalues}
\end{table}

\begin{figure*}
\centering
\includegraphics[width=0.95\columnwidth]{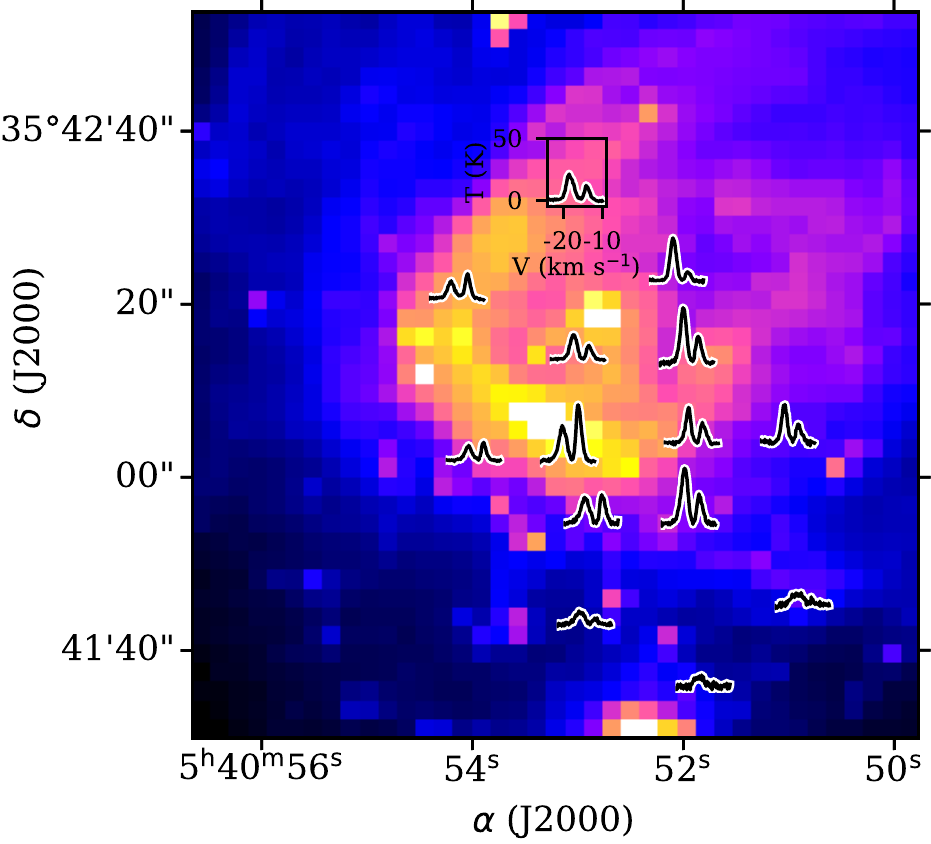}
\includegraphics[width=0.95\columnwidth]{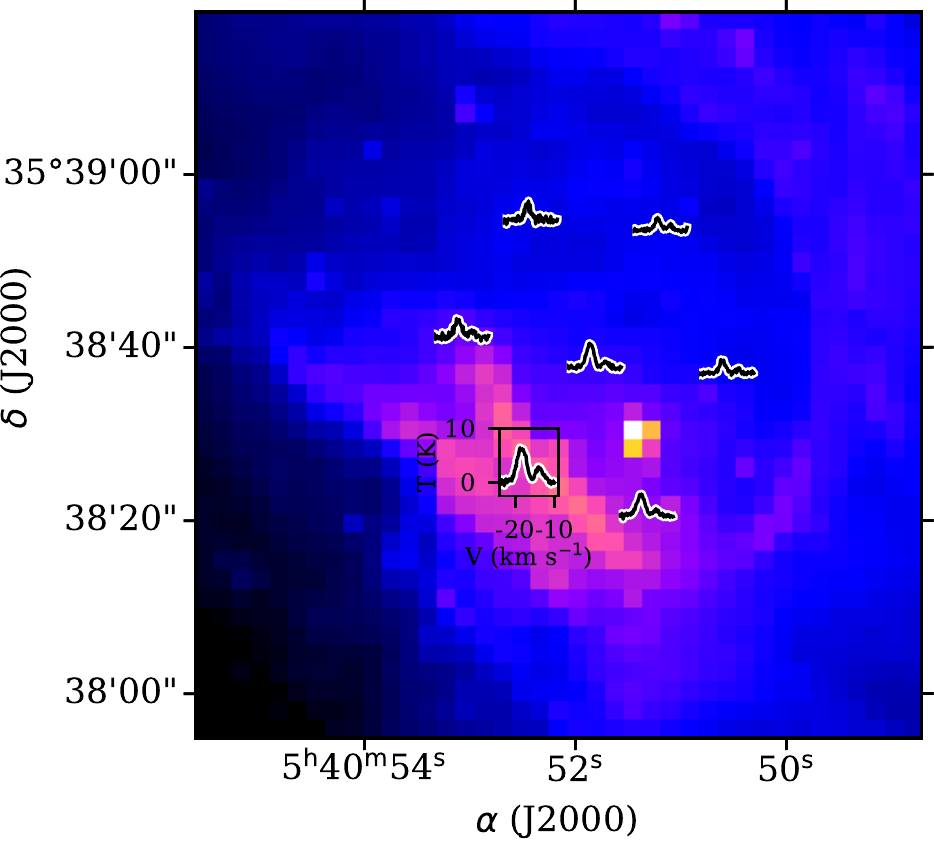}
\caption{The \OI{} spectra in S235~A (left) and S235~C (right) superimposed on the {\it Spitzer} 3.6~\micron{} image. The velocity range is from --24 to --9~\kms{} in the plots.}
\label{fig:OIspe}
\end{figure*}

Due to different array configurations at 63 and $158~\mu{}$m, the offsets of the \OI{} spectra are different from the \thirtCII{} offsets except for the central pixel positions. Fig.~\ref{fig:OIspe} shows the \OI{} spectra, concentrated in a more compact area compared to the \thirtCII{} spectra. The maximum detected intensity of the \OI{} emission has $T_{\rm mb} \approx 45$~K south of S235~A$^{\star}$. The maximum of the \OI{} emission has $T_{\rm mb} \approx 13$~K in S235~C south-east of the ionizing star. The \OI{} lines have double-peaked profiles which seems to be a common feature in massive star-forming regions. They are typically interpreted as being due to self-absorption \citep[see, for example, the recent study of the self-absorbed \OI{} line in S~106 by][]{Schneider2018}. The average values of integrated and line peak intensities are shown in Table~\ref{tab:avervalues}. The averaged values of the \CII{} integrated intensities over the area observed in the \OI{} line give almost the same numbers in both PDRs. The resulting average ratio of the \OI{} to \CII{} integrated intensities is $0.14\pm0.04$ and $0.04\pm0.02$ in S235~A and S235~C, where the difference in the ratio is almost exclusively due to the different \OI{} intensity. Since the \OI{} emission is a good tracer of the gas number density, we conclude that these PDRs have different densities, with the density of neutral gas being higher in S235~A. The \HCOp{} emission, which has a higher critical density compared to the \CII{} line, is strong towards the south-east of both PDRs, with the highest peak around S235~A. In contrast to the \CII{} emission peaks, the peaks on the \HCOp{} emission map appear in the densest part of PDRs, in the same region where we find bright arcs at $3.6~\mu{}$m (Fig.~\ref{fig:CIImom0}). 

\subsection{Kinematics of the PDR\lowercase{s}}

\begin{figure*}
\includegraphics[width=0.99\columnwidth]{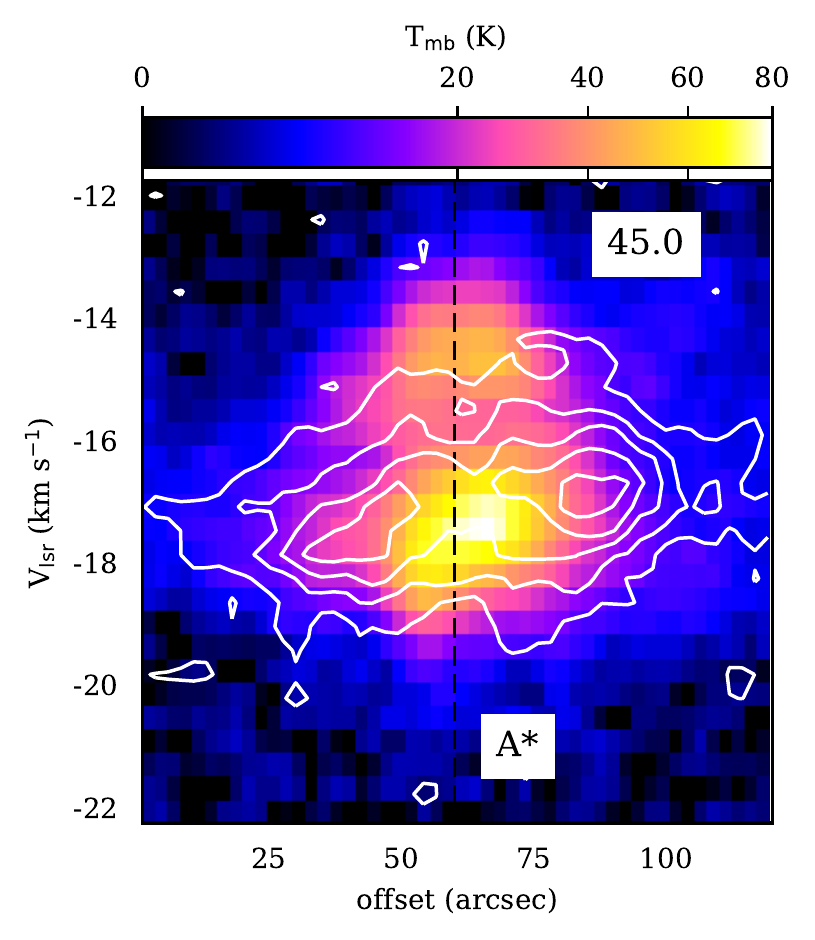}
\includegraphics[width=0.99\columnwidth]{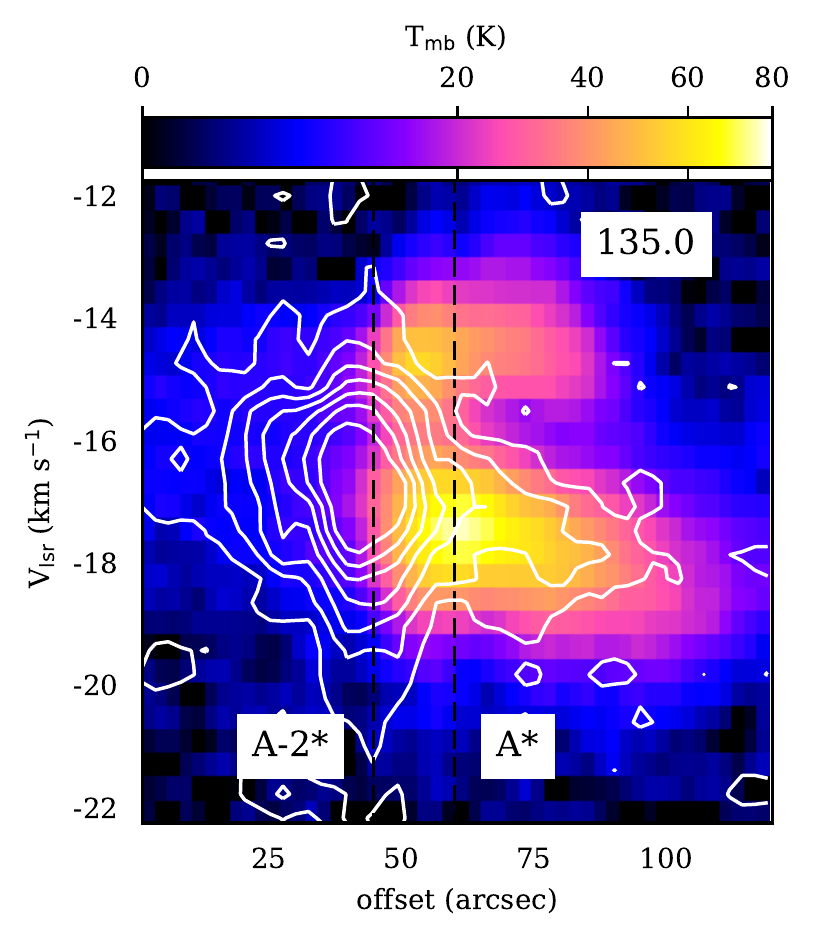}\\
\includegraphics[width=0.99\columnwidth]{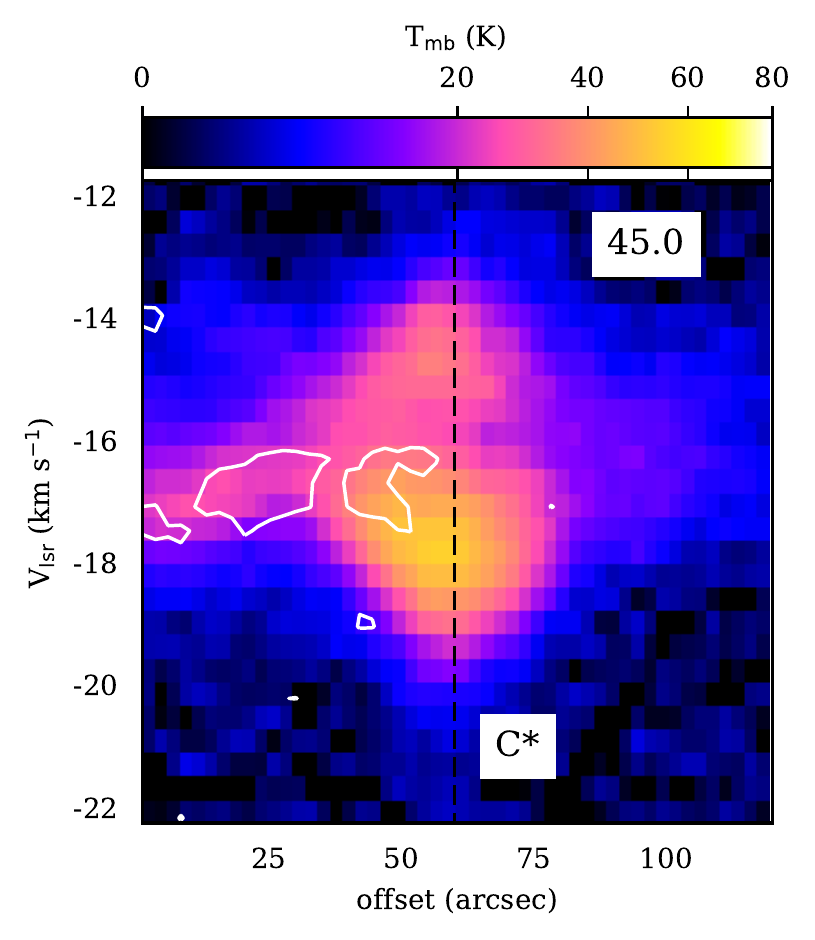}
\includegraphics[width=0.99\columnwidth]{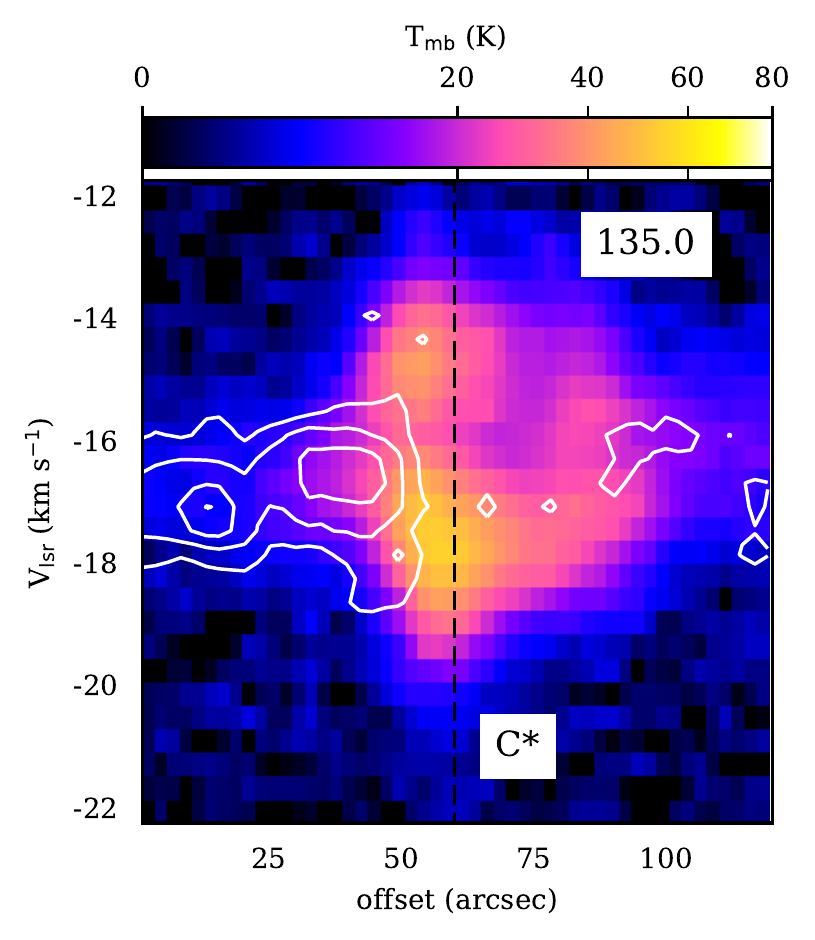}
\caption{pv diagrams of the \CII{} (colour) and HCO$^+$(3-2) (white contours) emission for S235~A (top panels) and S235~C (bottom panels). Dashed vertical lines correspond to positions of the point sources listed in Table~\ref{tab:ionizstars}, from the beginning of the pv~diagrams.
}\label{fig:CIIpv}
\end{figure*}

We use the \CII{} position-velocity (pv) diagrams to study gas kinematics in the PDRs. The pv~diagrams are computed along the major and minor axes of the elongated \CII{} image of S235~A and in the same directions for S235~C as shown in Fig.~\ref{fig:CIImom0} by the dashed lines. The pv diagrams are shown in Fig.~\ref{fig:CIIpv}. The \CII{} lines have double-peaked profiles with the red-shifted component around $-15 < V_{\rm lsr} < -14$\kms{} and the blue-shifted component around $-18 < V_{\rm lsr} < -17$\kms{} in both \hii{} regions. The double-peaked profiles have a dip at $V_{\rm lsr} = -16$~\kms{} corresponding to the velocity peak of hydrogen RRL emission, see \citet{Anderson2019}. On the contrary, the pv diagrams of \HCOp{} emission show single-peaked line profiles with the peaks falling between the \CII{} peaks. A truncated arc-like shape is visible at the high-density regions in the south-east area of the regions. In the 135 degrees pv diagram through S235~A, the position where the \CII{} emission is sharply confined coincides with the direction of the point source S235~A-2$^{\star}$. The source is projected on the bright arc-like structure visible on the {\it Spitzer} images in Fig.~\ref{fig:CIImom0} \citep[see also images in ][]{Dewangan2011}.

The velocity difference between the peaks of the double-peaked \CII{} line and the peak of the \HCOp{} line is $\approx 2$~\kms, as can be seen from the pv~diagrams. A simple combination of the \CII{} and \HCOp{} pv~diagrams seems to indicate an expansion of the \CII-emiting gas of the PDR with $V_{\rm exp} \approx 2$~\kms{} into the quiescent surrounding molecular cloud visible in the \HCOp{} line. However, to study the gas kinematics in the atomic and molecular layers of the PDRs in details, we extracted \HCOp{} spectra at the same positions where we have \thirtCII{}, and show them in Fig.~\ref{fig:13CIIspeA} and ~\ref{fig:13CIIspeC}. The single-peaked \thirtCII{} line profiles shown in these figures bring us to the conclusion that the \CII{} dips are related to self-absorption rather than to the expansion of the front and back neutral walls of the PDRs. 

The brightest \thirtCII{} line in S235~A has a symmetric single-peaked profile at a velocity $V_{\rm lsr}=-16.6$~\kms{}. This value of \vlsr{} is shown by the red vertical line on each spectrum in Fig.~\ref{fig:13CIIspeA} to emphasize the \thirtCII{} velocity gradient across the \hii{} regions. There is an  $\approx 1$~\kms{} global west-east velocity gradient visible in the spectra of the \thirtCII{} lines in S235~C (at positions 1 and 7 the gradient is clearly visible) and a south-west -- north-east gradient of a similar, but more uncertain, magnitude in S235~A (positions 3 and 11). The \thirtCII{} spectra in the other directions of S235~A have a complex shape: flat-top in position 6, red-shifted and red-skewed in position 3, blue-skewed broad profiles to the north-west of the S235~A$^{\star}$ in positions 8, 11 and 13. Properties of the \CII{} line emission in the directions observed in the \thirtCII{} line are given in Table~\ref{tab:CD}. It is seen that the $\tau_{\rm C^+} > 1$ in all directions where we detect \thirtCII{} at a level higher than $3\sigma$. Average \tex{} of the \CII{} and \thirtCII{} line emission is from 40 to about 90~K. We conclude that both \hii{} regions are embedded in dense and warm PDRs and we observed them through the layer optically thick in the \CII{} line.

The detected \OI{} lines have a self-absorption dip, which appears at almost the same velocity (--16~\kms{}) in the \CII{} lines. The self absorption is more significant in the \OI{} spectra than in the \CII{} spectra in both PDRs (see also averaged over the observed positions spectra in Fig.~\ref{fig:averoverpos}). The velocity difference between the two line peaks is the same as in the \CII{} lines: $\approx4$~\kms{}. The double-peaked \OI{} lines have a blue-shifted component brighter than the red-shifted component in almost all positions observed in S235~A and S235~C, except for four directions to the south and east of the point source S235~A-2$^{\star}$. Lines with enhanced blue emission are often attributed to infall in centrally condensed sources \citep{DickelAuer1984}, but can also be due to inhomogeneous density distributions. The region with a brighter red-shifted (blue-shifted) component of the \OI{} lines coincides with the region of the red-shifted (blue-shifted) \thirtCII{} lines in S235~A. This indicates that the effect traces the global velocity gradient in the PDRs instead of local expansion or contraction. The self-absorption dip has the same $V_{\rm lsr}$ for the all spectra in S235~A and S235~C, which implies that a layer of cool gas is situated in front of the warm PDRs, possessing a velocity gradient formed around the \hii{} regions. 

The \HCOp{} lines have a velocity gradient with similar direction and sign as the \thirtCII{} lines in both PDRs. Consequently, this gradient is indeed related to the properties of the parent molecular cloud where the \hii{} regions were formed. We fit the \HCOp{} and \thirtCII{} line shapes with single-peaked Gaussian profiles (see Table~\ref{tab:CIIres}), and find that the \thirtCII{} line peaks are red-shifted relative to \HCOp{} by about 1~\kms{} at positions 3, 6 and 9 south-east of S235~A, and also at positions 1 and 3 east of S235~C. We interpret this velocity difference as relative motion of the \HCOp-emitting dense molecular material, surrounding the \CII-emitting PDR, with $\approx$~1~\kms{} in the south-east parts of the PDRs. We find a similar \HCOp{}-\thirtCII{} velocity shift in the averaged spectra of S235~A in Fig.~\ref{fig:averoverpos}, and an excess of red-shifted \thirtCII{} emission compared to \HCOp{} in S235~C, visible in positions 1 and 3 in Fig.~\ref{fig:13CIIspeC}, as well as in the averaged spectra.

The lack of a velocity difference between \thirtCII{} and \HCOp{} in the western parts of both PDRs may be due to the absence of high-density molecular material. The \CII{}- and \HCOp{}-emitting gas, apparently, flows freely (i.e. with no dense molecular shell formed by shock wave ahead of the ionization front) from the PDRs into the surrounding low-density medium at these locations. The red wings of the \CII{} line visible in positions 8, 11 and 13 in S235~A and in positions 6 and 7 in S235~C indicate that expansion into the low-density medium occurs in the opposite from the observed direction.

The value of $V_{\rm exp} \approx 1$~\kms{} found from the analysis of the \thirtCII{} and \HCOp{} spectra is about two times less than the value which would be estimated from the \CII{} and \HCOp{} pv diagrams. Below, in Sec.~\ref{sec:model}, we compare the observed gas kinematics with predictions of a numerical model of expanding \hii{} regions, and show simulated pv~diagrams.

\subsection{Column densities of C-bearing species}

\begin{figure*}
\includegraphics[height=9.5cm]{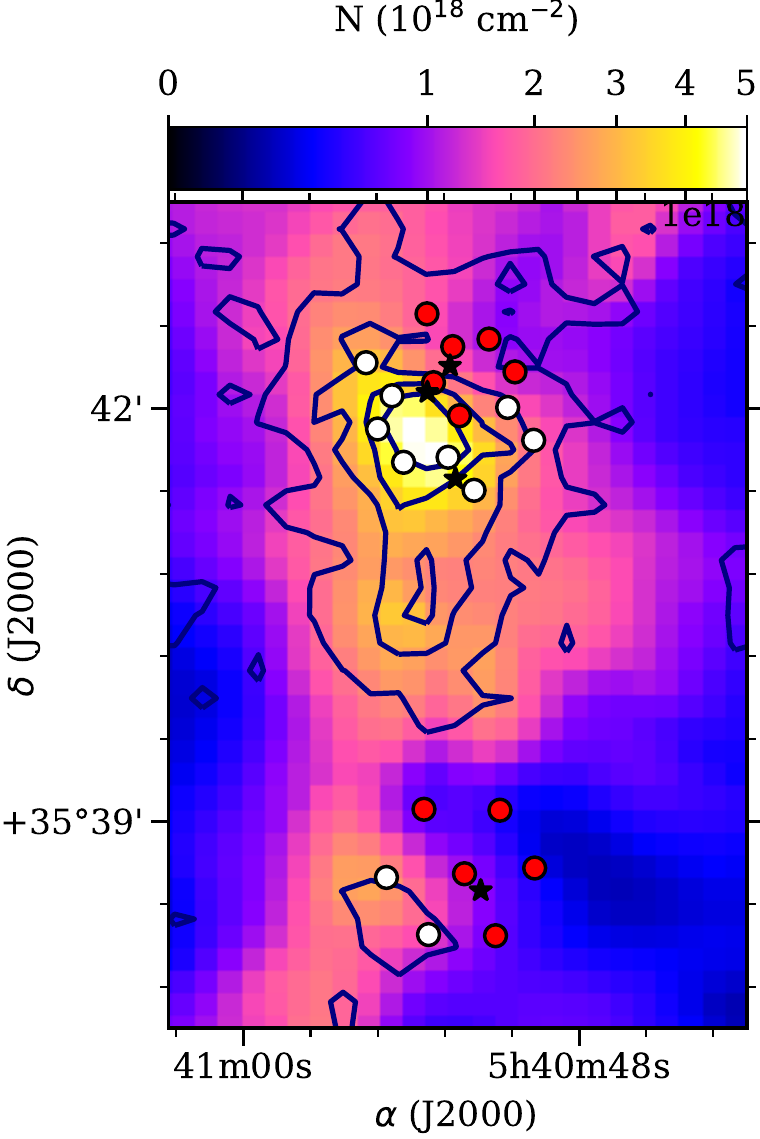}
\includegraphics[height=9.5cm]{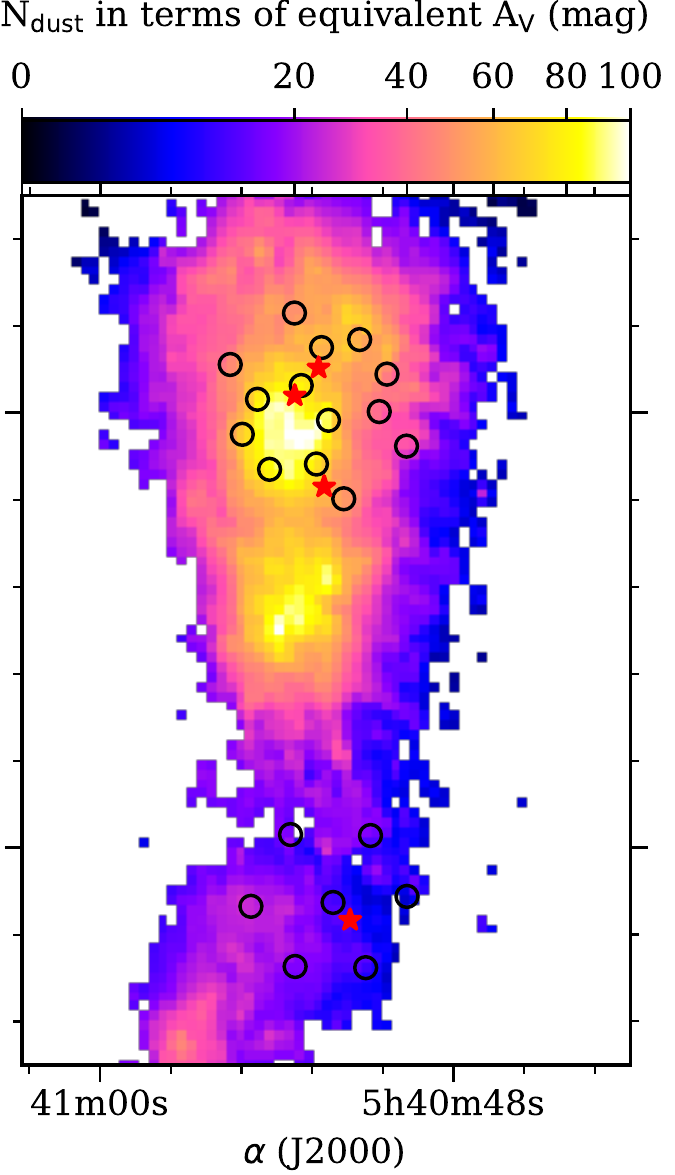}
\includegraphics[height=9.5cm]{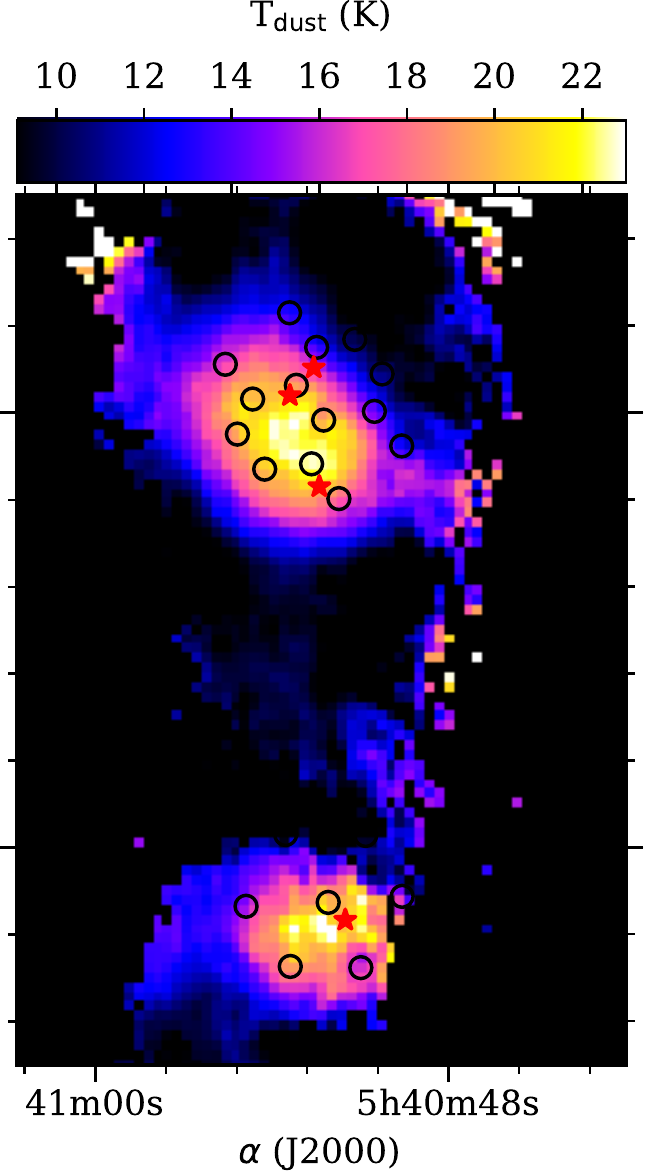}
\caption{Left: Maps of \NCO{} (colour) and \NHCOp{} (contours). The contours correspond to 1.0, 2.5, 4.0 and $5.5\times 10^{12}$~cm$^{-2}$.  Directions of the \thirtCII{} observations are shown by circles. Red circles stand for positions with $N_{\rm C^+} > N_{\rm CO}$, white circles show the opposite relation. Middle: dust column density in equivalent visual extinction $A_{\rm V}$ scale. 
The column density maps are given at the original spatial resolution of the observations (38$''$ for \NCO{} in the left panel, 33$''$ for the middle and right panels). Right: dust temperature. Point sources from Table~\ref{tab:ionizstars} are shown by black stars in the left panel and by red stars in the central and right panel.}
\label{fig:cd}
\end{figure*}

Column density maps of CO and HCO$^+$ molecules are shown in Fig.~\ref{fig:cd}. Both maps reveal a concentration of molecular gas to the south and south-east of S235~A in between the point sources S235~A-2$^{\star}$ and S235B$^{\star}$, and to the south-east of S235~C. While the regions with the highest \NCO{} and \NHCOp{} around S235~A and S235B$^{\star}$ coincide in the plane of the sky, there is a decrease in the intensity of \NCO{} near the secondary peak of \NHCOp{} to the south of S235B$^{\star}$ (around $\alpha \approx 05^{\rm h}40^{\rm m}54^{\rm s}$ and $\delta \approx +35^{\circ}40^{\rm m}30^{\rm s}$).

The dust temperature ($T_{\rm dust}$) and column density ($N_{\rm dust}$) maps are also shown in Fig.~\ref{fig:cd}. The dust column density structure roughly follows the spatial distribution of \NHCOp. The $T_{\rm dust}$ value is enhanced up to about 24~K close to the embedded sources, but the temperature peaks do not agree with the column density peaks but are rather offset. For S235~A we find the temperature peak south of the main ionizing star, probably due to the contribution from the other embedded sources (e.~g. S235~A-2$^{\star}$, S235~B$^{\star}$); in S235~C the temperature peak is offset to the west, probably due to the density gradient allowing for a more efficient heating of the eastern part of the source. The dust temperature agrees with excitation temperature of CO molecules found by \citet{Bieging2016}. These molecules are tracers of the general distribution of gas with relatively low critical density. We will note about higher values of the gas temperatures obtained with ammonia lines in Sec.~\ref{sec:disc}.

\begin{figure}
\includegraphics[width=\columnwidth]{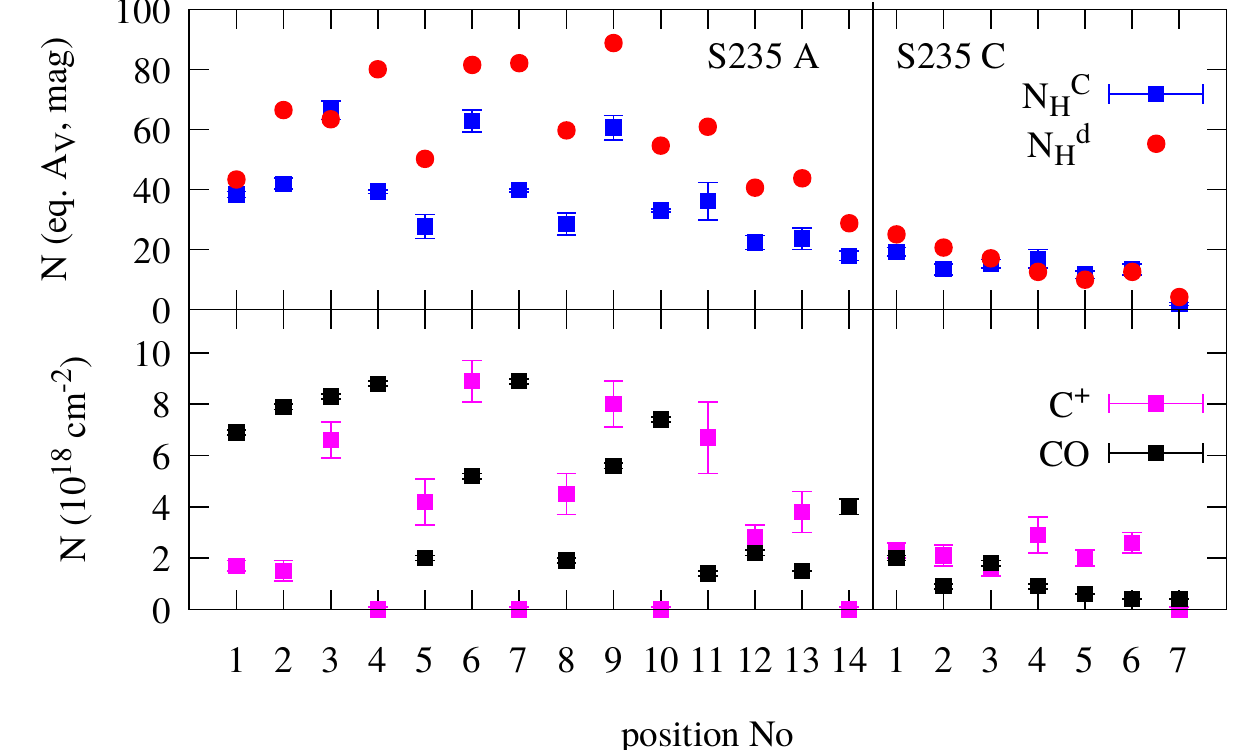}
\caption{Top: carbon and hydrogen column densities in terms of equivalent $A_{\rm V}$. Uncertainty of the hydrogen column density is within 1 and 2~mag in S235~A and S235~C, respectively (see Sec.~\ref{sec:obs}). Bottom: $N({\rm C^+})$ and $N({\rm CO})$.}
\label{fig:cd_table}
\end{figure}

We compare the C$^+$ and CO column densities in the directions where the \thirtCII{} was observed in Table~\ref{tab:CD}. Using the dust continuum emission maps described in Sec.~\ref{dustcont} and assuming that all carbon is in C$^+$ and CO species with the total carbon abundance $1.2\times 10^{-4}$ (see Sec.~\ref{sec:model}), we recalculate hydrogen column density in terms of equivalent~$A_{\rm V}$ value ($N_{\rm H}^{\rm C}$) and compare it with the dust column density ($N_{\rm H}^{\rm d}$) in Table~\ref{tab:CD}, illustrated in Fig.~\ref{fig:cd_table}. While in the majority of the directions both column densities are in agreement, we find several directions where $N_{\rm H}^{\rm d} > (1.5-2) N_{\rm H}^{\rm C}$. These directions surround the dense molecular clump around S235~A from the south, and also to the north-west, as well as in one position to the east of S235~C. 

To explain this disagreement, we checked our assumptions in the derivation of column density. First, using a map of the optical depth in $^{13}$CO(2-1) from \citet{Bieging2016}, we find $\tau_{\rm ^{13}CO(2-1)} \approx 1$ in the molecular cloud around S235~A and C. This approach underestimates the \NCO{} value, which partially explains the missing carbon, at least in positions 4, 7 and 10 in S235~A. Comparison of the panels of Fig~\ref{fig:cd} with the \NCO{} and $N_{\rm H}^{\rm d}$ distributions shows a relative deficiency of CO in the gas phase south of S235~B$^{\star}$, where the peaks of the $N_{\rm H}^{\rm d}$ distribution are not followed by the \NCO{} peaks. Because CO and HCO$^+$ are chemically closely related they should freeze out onto the dust grains simultaneously \citep[see e. g.][]{Bergin1997}. Thus, we associate the decreasing CO column density to the south of S235~B with the optical depth of the $^{13}$CO(2-1) line, rather than with freeze-out onto the dust grains, because a secondary peak of \NHCOp{} is seen at this location.

A second factor contributing to the total carbon column are variations of \tex{} for the \CII{} line along the line of sight. We can estimate them by comparing the maximum of the channel \tex{} with $ \langle T_{\rm ex} \rangle$ in Eq.~\ref{eq:thin}. The maximum channel \tex{} would raise $N_{\rm C^+}$ only within the interval given in Table~\ref{tab:CD} included in the upper error bars in Fig.~\ref{fig:cd_table}. 

Looking at S235~A from the north-west to the south-east we distinguish two different types of environments: the PDR where C$^+$ is the dominant C-bearing species and the molecular cloud where CO is the dominant C-bearing species. Toward S235~A$^{\star}$, S235~A-2$^{\star}$ and in several directions to the west of the source the C$^+$ column density exceeds the CO column density by a significant factor, $N_{\rm C^+} \approx (2-6) \times N_{\rm CO}$. The directions with $N_{\rm C^+} > N_{\rm CO}$ are shown with red circles in the left panel of Fig.~\ref{fig:cd}. The point sources are situated on the border between the PDR and surrounding molecular cloud where the PDR is expanding, see Fig.~\ref{fig:cd}. The column of carbon locked in C$^+$ is higher than in CO in 5 of the 7 observed positions in S235~C. The values of \NCO{} and \NHCOp{} are highest in the directions of the dense molecular clumps where $N_{\rm CO} > N_{\rm C^+}$. 

If we average the ratio of the column density \NCO+\NCp{} to the hydrogen column density computed from the dust emission, we find a gas-phase carbon abundance $7.7\times 10^{-5}$ in S235~A and $1.1\times 10^{-4}$ in S235~C, which is in agreement with the total carbon abundance value used above. Some carbon in the directions with $N_{\rm H}^{\rm d} > N_{\rm H}^{\rm C}$ could be in the gas-phase atomic form, in high-J CO lines, or locked in dust grains. In the directions with $N_{\rm H}^{\rm d} > N_{\rm H}^{\rm C}$ around S235~A we find $10 \leq T_{\rm dust} \leq 20$~K. Taking into account that these positions are situated in the surrounding cold molecular cloud, it is natural to expect that some carbon is depleted from the gas-phase due to freeze-out onto the grain surfaces. A rich young stellar cluster is situated between the \hii{} regions (see Sec.~\ref{sec:disc}), but it does not provide enough heating to make the dust warmer than 10--15~K.

\begin{table*}
\caption{Properties of the \CII{} emission at the positions of the \thirtCII{} measurements: mean values of \tex{} and $\tau_{\rm C^+}$ averaged over the optically thick channels $\pm$ standard deviation over the channels are shown in columns 3 and 4. Value of $N_{\rm C^+}$ calculated for the optically thin and thick channels is shown in column 5. The seventh and eighth columns show gas column density calculated from the carbon-bearing species with the carbon abundance from \citet{Wakelam2008} and dust column density, respectively. The $N_{\rm H}^{\rm C}$ and $N_{\rm H}^{\rm d}$ values are converted into the equivalent $A_{\rm V}$.}
\begin{tabular}{lrcccccc}
\hline
N & ($\Delta \alpha$, $\Delta \delta$) & $ \langle T_{\rm ex} \rangle$ & $ \langle \tau_{\rm C^+} \rangle$ & $N_{\rm C^+}$      & $N_{\rm CO}$  &   $N_{\rm H}^{\rm C}$ & $N_{\rm H}^{\rm d}$  \\
&\arcsec{} ,      \arcsec{}       & K              &                    & $10^{18}$cm$^{-2}$ & $10^{18}$cm$^{-2}$  & eq. $A_{\rm V}$, mag & eq. $A_{\rm V}$, mag\\
\hline
\multicolumn{8}{c}{\it S235~A}\\
1 &  (29           ,    8)  & 47.5$\pm$ 1.3 & 10.3$\pm$2.3 &  1.7$\pm$0.2 & 6.9$\pm$0.1  & 38.3$\pm$1.0 &43.3\\
2 &  (24           , --21)  & 40.7$\pm$ 2.8 &  9.3$\pm$1.7 &  1.5$\pm$0.4 & 7.9$\pm$0.1  & 41.9$\pm$1.8 &66.5\\
3 &  (18           ,  --6)  & 72.6$\pm$ 3.6 &  6.1$\pm$2.3 &  6.6$\pm$0.7 & 8.3$\pm$0.1  & 66.4$\pm$3.1 &63.4\\
4 &  (13           , --35)  & --            & --           &  <0.1        & 8.8$\pm$0.5  & 39.3$\pm$2.3 &80.1\\
5 &   (2           ,   30)  & 63.1$\pm$ 9.2 &  4.3$\pm$1.7 &  4.2$\pm$0.9 & 2.0$\pm$0.2  & 27.6$\pm$4.0 &50.2\\
6 &   (0           ,    0)  & 90.1$\pm$14.0 &  4.8$\pm$2.4 &  8.9$\pm$0.8 & 5.2$\pm$0.1  & 62.8$\pm$3.6 &81.5\\
7 & (--7           , --33)  & -- &  -- &  <0.1  & 8.9$\pm$0.2  & 39.7$\pm$0.5 &82.1\\
8 & (--9           ,   15)  & 75.4$\pm$17.0 &  3.4$\pm$2.0 &  4.5$\pm$0.8 & 1.9$\pm$0.2  & 28.5$\pm$3.6 &59.7\\
9 & (--12          , --15)  & 82.5$\pm$16.7 &  4.9$\pm$2.7 &  8.0$\pm$0.9 & 5.6$\pm$0.1  & 60.6$\pm$4.1 &88.8\\
10&(--18           , --47)  &  --           & --           &  <0.1        & 7.4$\pm$0.1  & 33.3$\pm$0.5 &54.6\\
11&(--25           ,   19)  & 69.3$\pm$16.6 &  4.0$\pm$1.9 &  6.7$\pm$1.4 & 1.4$\pm$0.1  & 36.1$\pm$6.2 &60.9\\
12&(--33           , --11)  & 69.0$\pm$ 5.2 &  3.1$\pm$1.0 &  2.8$\pm$0.5 & 2.2$\pm$0.1  & 22.3$\pm$2.3 &40.6\\
13 &(--36          ,    4)  & 68.4$\pm$13.4 &  3.6$\pm$3.7 & 3.8$\pm$0.8  & 1.5$\pm$0.1  & 23.6$\pm$3.6 &43.7\\
14 &(--44          , --26)  &  --           & --           & <0.1         & 4.0$\pm$0.1  & 17.9$\pm$1.6 &28.7\\
\multicolumn{8}{c}{\it S235~C}\\
1 & ( 34           ,  --2)  & 57.5$\pm$ 3.0 &  5.3$\pm$1.3 & 2.3$\pm$0.3  & 2.0$\pm$0.3  & 19.2$\pm$1.4 &25.0\\
2 & ( 17           ,   28)  & 63.9$\pm$ 2.7 &  3.6$\pm$1.0 & 2.1$\pm$0.4  & 0.9$\pm$0.1  & 13.4$\pm$1.8 &20.6\\
3 & ( 15           , --27)  & 59.2$\pm$ 1.5 &  3.5$\pm$0.6 & 1.6$\pm$0.3  & 1.8$\pm$0.1  & 15.2$\pm$1.4 &17.0\\
4 & (  0           ,    0)  & 67.7$\pm$ 6.7 &  2.1$\pm$0.8 & 2.9$\pm$0.7  & 0.9$\pm$0.2  & 16.9$\pm$3.1 &12.5\\
5 & (--14           , --28)  & 60.7$\pm$ 5.2 &  3.6$\pm$0.9 & 2.0$\pm$0.3  & 0.6$\pm$0.1  & 11.6$\pm$1.4 &9.9\\
6 & (--16           ,   27)  & 72.2$\pm$ 5.0 &  3.0$\pm$0.4 & 2.6$\pm$0.4  & 0.4$\pm$0.1  & 13.4$\pm$1.8 &12.5\\
7 & (--31           ,    2)  & 58.5$\pm$ 5.1 &  4.7$\pm$1.3 & 2.3$\pm$0.4  & 0.4$\pm$0.1  & 1.8$\pm$0.5 &4.1\\
\hline
\end{tabular}
\label{tab:CD}
\end{table*}

\section{Numerical modelling}\label{sec:model}

As the \hii{} regions S235~A and S235~C are deeply embedded in the molecular cloud, we cannot use published predictions from standard plane-parallel PDR models without foreground absorption \citep[e.g.][]{Kaufman1999}. In order to understand how the observational view by SOFIA is matched by a 'classical' model of expanding \hii{} regions, described analytically and numerically by \citet{Spitzer1978, Elmegreen1977, Hosokawa2006, Raga2012, Kirsanova2009, Bisbas2015}, we make simulations with the MARION model \citep{Kirsanova2009, Akimkin2015}. The gas-phase chemical network from \citet{Rollig2007} (mainly based on the UMIST99 ratefile, \citet{LeTeuff2000}) allows us to reproduce the C$^+$, CO and HCO$^+$ abundances in PDRs \citep[see][]{Kirsanova2019}, so we use this network together with the ionization of the atomic species and corresponding recombinations in the \hii{} regions. The cross-sections for most of photoreactions are taken from the Leiden data base of ``Photodissociation and photoionization of astrophysically relevant molecules'' \citep{Heays2017}. We implement the formation of H$_2$ on grain surfaces and accretion and desorption processes of other neutral species, but other chemical reactions on dust surfaces are not considered in the calculations to save computation time. The rates of accretion and desorption processes are based on the work by \citet{Hasegawa1993}, with updated desorption energies from \citet{Garrod2007}. We use the 'high-metallicity' initial elemental abundances based on the 'EA2' set from \citet{Wakelam2008}. The initial conditions are cold, molecular, and solid. This means that we start with all carbon and oxygen in CO and H$_2$O on dust surfaces. 52 chemical species are included: H, H$^+$, H$_2$, H$_2^+$, H$_3^+$, O, O$^+$, O$^{++}$, OH$^+$, OH, O$_2$, O$_2$:d, O$_2^+$, H$_2$O, H$_2$O$^+$, H$_3$O$^+$, C, C$^+$, C$^{++}$, CH, CH$^+$, CH$_2$, CH$_2^+$, CH$_3$, CH$_3^+$, CH$_4$, CH$_4^+$, CH$_5^+$, CO, CO:d, CO$^+$, HCO$^+$, He, He$^+$, S, S$^+$, S$^{++}$, Si, Si$^+$, H:d, H$_2$:d, O:d, OH:d, H$_2$O:d, C:d, CH:d, CH$_2$:d, CH$_3$:d, CH$_4$:d, S:d, Si:d, e$^-$ where the postfix ``:d'' indicates species on dust grain ices. Chemical species containing Si and S are included only to obtain correct gas temperature in ionized region. We recognize that the chemical network is far from complete, but our choise was motivated by limited computation time. The heating and cooling processes included in the model are listed in \citet{Akimkin2015, Kirsanova2019}. 

The dust grains in the model are represented by silicate and carbonaceous dust grains with 24 size bins for each type. The initial dust-to-gas mass ratio is 0.0088 with the size distribution corresponding to model number 16 from \citet{WeingartnerDraine2011}. This model has $R_{\rm V}=5.5$ instead of standard interstellar $R_{\rm V}=3.1$, and we use it because this reddening law implies lower relative amount of small dust grains and, therefore, lower dust extinction and thicker PDR (see below). Values of $R_{\rm V}>5$ were found in PDRs such as the Orion Bar and NGC~7023 \citep[see][respectively]{Marconi1998, Witt2006}. The dust model implies that the carbon abundance locked in dust refractory cores is $3\times 10^{-5}$. Therefore, the total carbon abundance in the model is $1.5\times 10^{-4}$. The Si abundance locked in the refractory cores corresponds to solar value of $3.63\times10^{-5}$. The grain cores are not destroyed in our simulations. The size of the computational domain is 0.5~pc, divided into 550 radial cells. The dust temperature is computed self-consistently from radiative processes and interaction with the gas. The gas temperature is also computed self-consistently, except for the initial moment, when it is set up to the value corresponding to cold molecular clouds.

In the beginning of the simulation it is assumed that a massive star is embedded in a molecular cloud with uniform gas distribution. At $t=0$ the star starts to ionize and heat the surrounding material.  We scanned a range of effective temperatures of the ionizing stars (25000 to 31000~K) and of initial gas number densities ($5\times 10^3$ to $10^5$~cm$^{-3}$) to reproduce the observed properties of the \hii{} regions and surrounding PDRs. In this work we present the simulations in this parameter range which give the best agreement with the observations. Initial conditions of gas and dust in these simulations are summarized in Table~\ref{tab:initial}.

\begin{table}
\caption{Parameters and initial conditions of the MARION model for S235~A and S235~C.}
\begin{tabular}{l|l}
\hline
Parameter & Value \\
\hline
Stellar temperature                 & $25\,000-31\,000$~K \\
$n_{\rm init}$ (H-protons)          & $5\times 10^3 - 10^5$~cm$^{-3}$ \\
Radius of molecular cloud           & 0.5~pc \\
Distance                            & 1.6~kpc \\
$T_{\rm gas}$                       & 10~K \\
Dust-to-gas mass ratio              & 0.0088 \\
Dust model, number of components    & WD16, 48 \\
Cosmic ray rate $\zeta$     & $1.0\times 10^{-16}$~s$^{-1}$\\
\hline
$x(\rm{H_2})$                       & 0.5\\
$x(\rm{He})$                        & $9.0\times 10^{-2}$\\
$x(\rm{CO:d})$                      & $1.20\times 10^{-4}$\\
$x(\rm{H_2O:d})$                    & $1.36\times 10^{-4}$\\
$x(\rm{S})$                         & $1.50\times 10^{-5}$\\
$x(\rm{Si})$                        & $1.70\times 10^{-6}$\\
\hline
\end{tabular}\\
\label{tab:initial}
\end{table}

In order to simulate the profiles of molecular and atomic line emission for the model physical parameters and chemical abundances calculated with MARION, we apply the spherically symmetric line radiative transfer code SimLine by \citet{Ossenkopf2001}. The main partner for the collisional excitation are electrons in the \hii{} region and hydrogen atoms and molecules in the PDR. We treat this in the radiative transfer computation by switching from electrons to hydrogen when the gas temperature drops below $T_{\rm gas} \leq 500$~K. For HCO$^+$ the electron excitation in the \hii{} region is ignored as there is not significant molecular abundance there. Coefficients for collisional excitation of C$^+$~\citep[][Table~8]{Sternberg1995}, atomic O~\citep{Jaquet1992} and HCO$^+$~\citep{Flower1999} line emission with molecular hydrogen were used. The code uses two parameters for the turbulence description: the width of the turbulent velocity dispersion and the turbulence correlation length. They were set to 2.4~\kms{} (based on typical \CII{} and \HCOp{} line widths; see below) and 0.1~pc \citep{Xie1995, Ossenkopf2001}, respectively.

\section{Model predictions for S235~A and S235~C}\label{sec:modelpred}

As MARION is a chemo-dynamic model, we can trace the evolution of the physical parameters of the \hii{} regions and select the moment of the best agreement between the simulated and observed values of the parameters. The following properties need to be matched by the model:

\begin{itemize}
\item Radii of the \hii{} regions are from 0.1 to 0.3~pc,
\item \nel=1000 and 500~cm$^{-3}$ within a factor of 2 for S235~A and S235~C, respectively,
\item The profiles of the \thirtCII{} and HCO$^+$(3--2) lines are single-peaked,
\item The average W(\thirtCII{}), W(\OI{}) and the \thirtCII{} line peak emission is close to the observed values (see Tab.~\ref{tab:avervalues}) within a factor of 2.
 \item Expansion velocity is $\approx 1$~\kms{},
    \item $3 \leq \tau_{\rm C^+} \leq 10$ and $3\leq \tau_{\rm C^+} \leq 5$ for S235~A and S235~C, respectively
\end{itemize}

We used the output from MARION every five or ten thousand years of model time and calculated the resulting pv~diagrams. We compared the model peak line intensity, integrated intensity and the line profiles with the observed values. 
We find that models with \teff=27000~K and initial $n_{\rm gas}=5\times 10^4$ and $5\times 10^3$~cm$^{-3}$ provide a reasonable agreement for S235~A and S235~C, respectively. The model age of S235~A is $6\times10^4$ and of S235~C is $3\times10^4$ years. Below we compare the observed view of the \hii{} regions and their PDRs and the simulated models. 

\begin{figure}
\includegraphics[width=0.99\columnwidth]{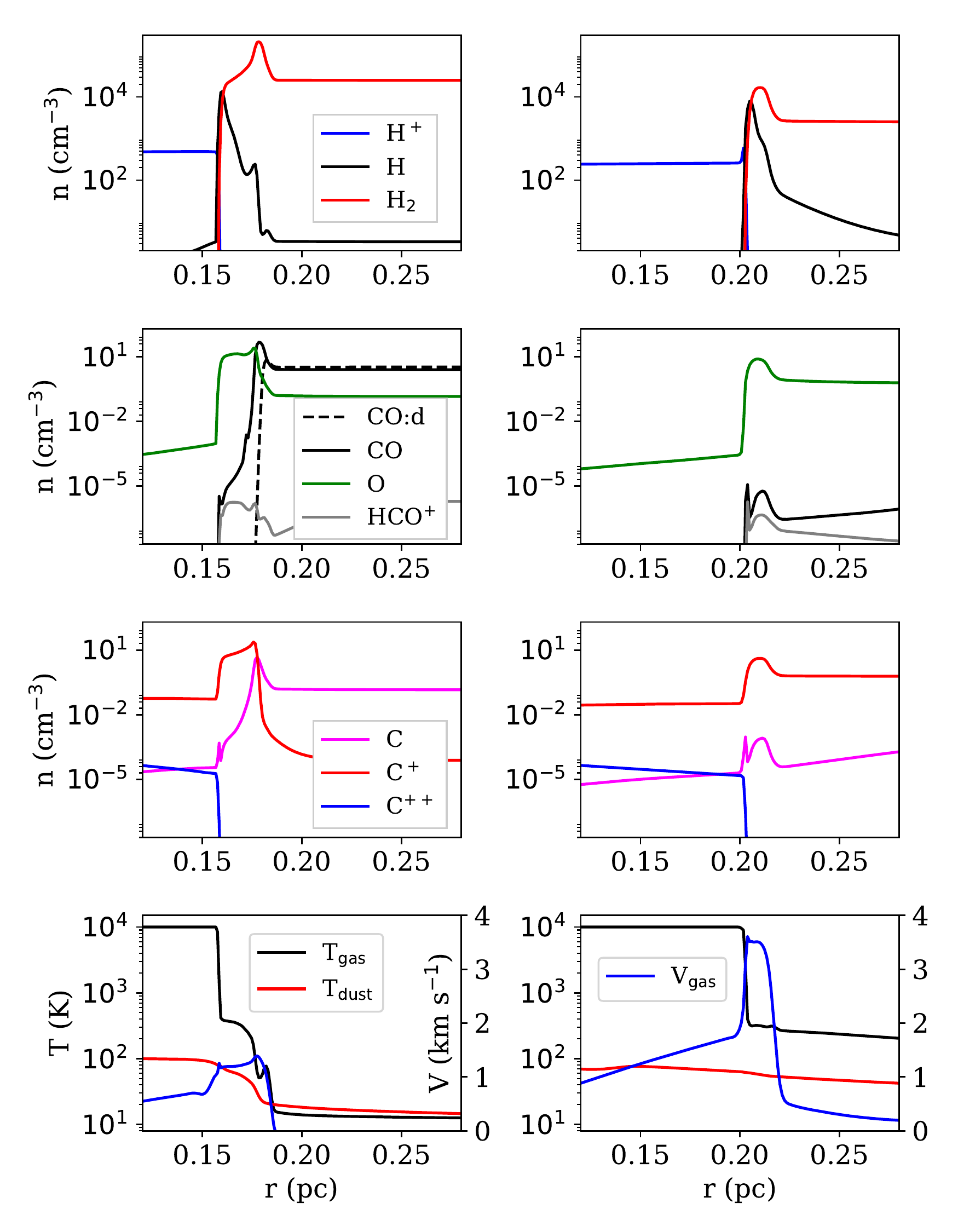}
\caption{Physical conditions in the models of S235~A (left column) and S235~C (right column): number densities of species (three top rows of plots), gas and dust temperature, and also gas velocity (bottom row).}
\label{fig:modelphys}
\end{figure}

\paragraph*{Physical properties of \hii{} regions.} The spatial distributions of the physical conditions and chemical abundances for both \hii{} regions are shown in Fig.~\ref{fig:modelphys}. The values of \nel~$\approx 500$ and 260~cm$^{-3}$ for S235~A and S235~C, respectively. Each \hii{} region is surrounded by a dense molecular shell, with maximum number densities of $n_{\rm H_2}=2\times 10^5$ and $2\times 10^4$~cm$^{-3}$ for S235~A and S235~C, respectively. The shells are created by the shock waves which spread ahead of the ionization fronts on the border of the \hii{} regions. The width of the compressed shells is less than 0.1~pc in both models. The expansion velocities are about 1 and 3~\kms{} for S235~A and S235~C, respectively. 

\paragraph*{Chemical composition.} The temperature contrast between the PDR and the surrounding molecular region is higher in the S235~A model compared to the model for S235~C due to relatively low gas density in the latter model. The central source of UV radiation heats the gas and dust over a larger distance in the S235~C model than in the case of S235~A. Carbon is in the form of C$^+$ inside of the \hii{} regions, and also in the neutral dense shells, but the outer border of the shell consist of molecular CO in the model for S235~A. The surrounding gas contains CO in the form of grain ice in the model of S235~A, but mostly consist of C$^{+}$ in the model for S235~C. The abundance of atomic carbon is smaller than the abundances of CO or C$^+$ by up to several orders of magnitude in both models, except for a very thin layer in the model for S235~A, where the abundance of atomic C is only 5 times lower than CO and 3 times lower than C$^+$. Oxygen is in atomic form in the dense shell in both PDRs and also in the undisturbed gas in S235~C. The value of \tgas{} changes from 20 to 300~K in the PDR of the S235~A model ($20 < T_{\rm dust} < 100$~K there), but falls below 50~K only at a distance of 0.45~pc in the model for S235~C (corresponding $T_{\rm dust} < 30$~K.)

\begin{figure}
\includegraphics[width=0.49\columnwidth]{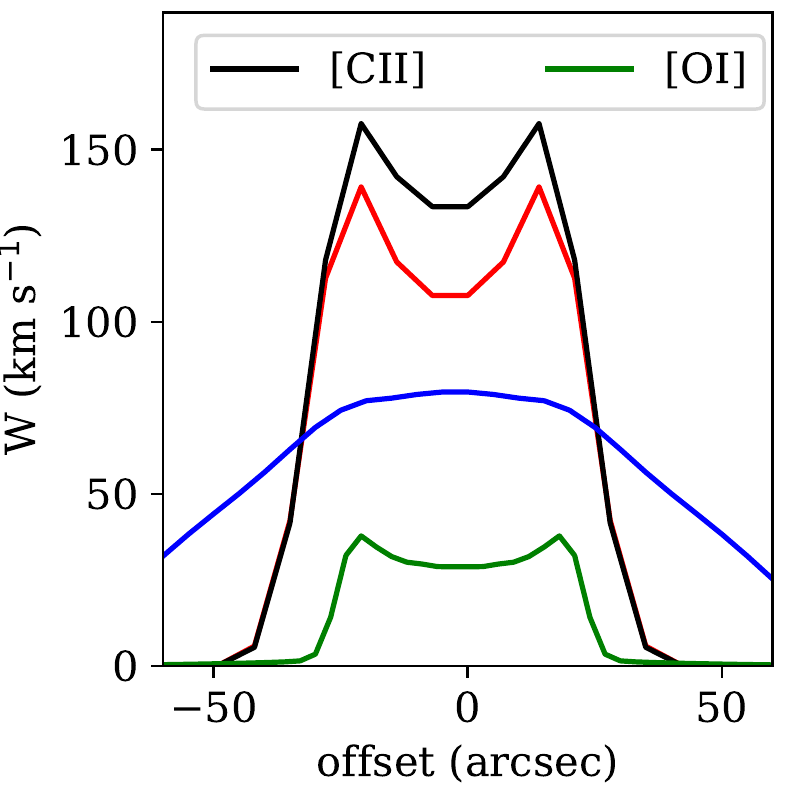}
\includegraphics[width=0.49\columnwidth]{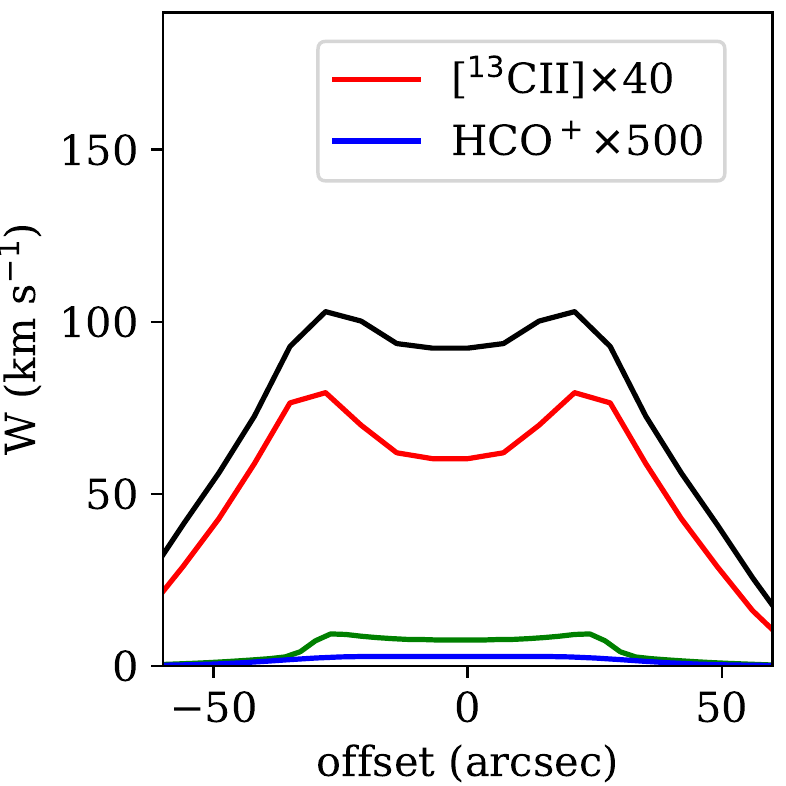}\\
\vspace{1mm}
\includegraphics[width=0.49\columnwidth]{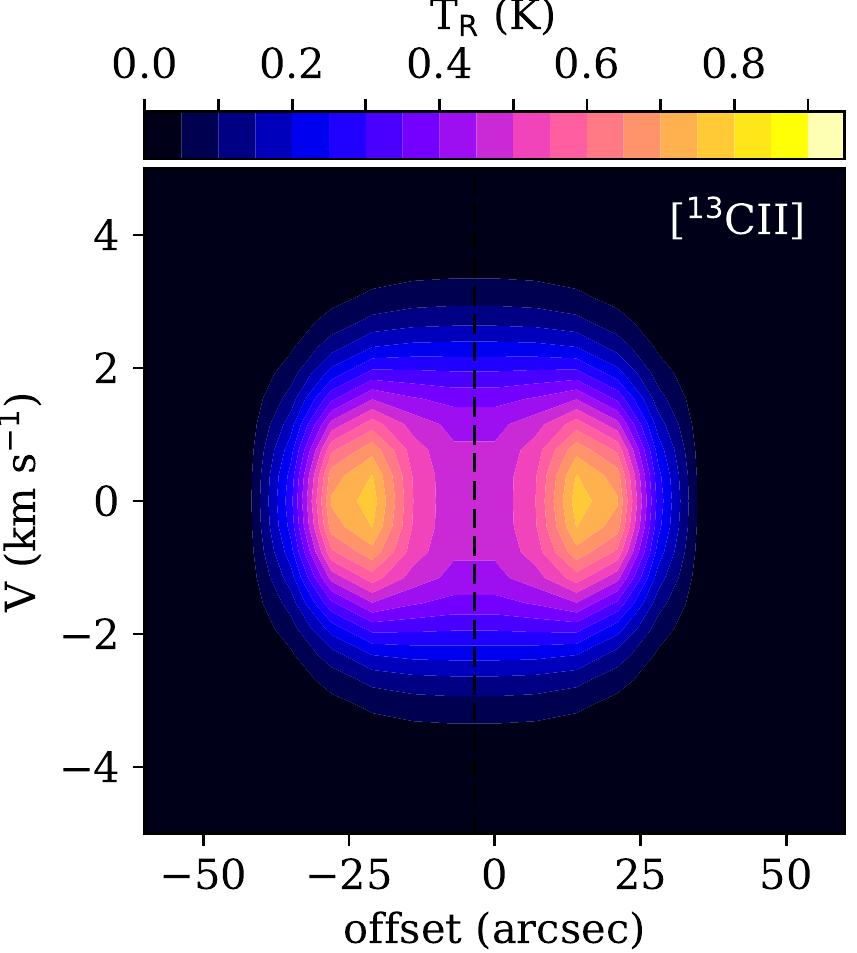}
\includegraphics[width=0.49\columnwidth]{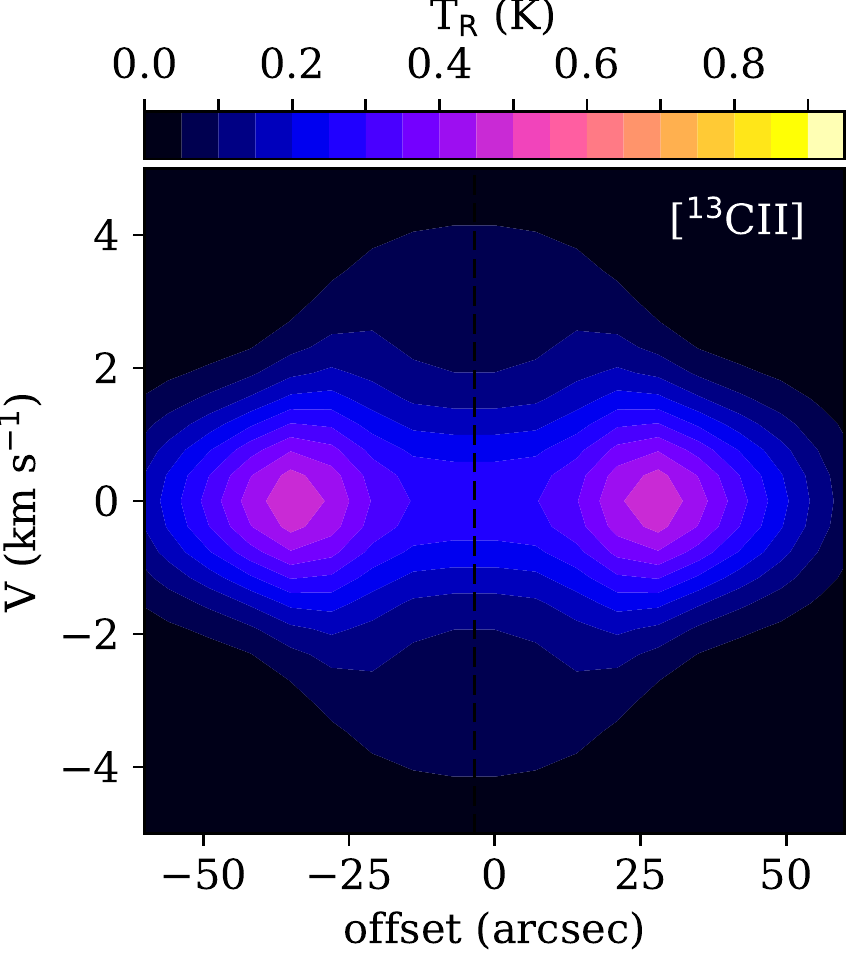}\\
\vspace{1mm}
\includegraphics[width=0.49\columnwidth]{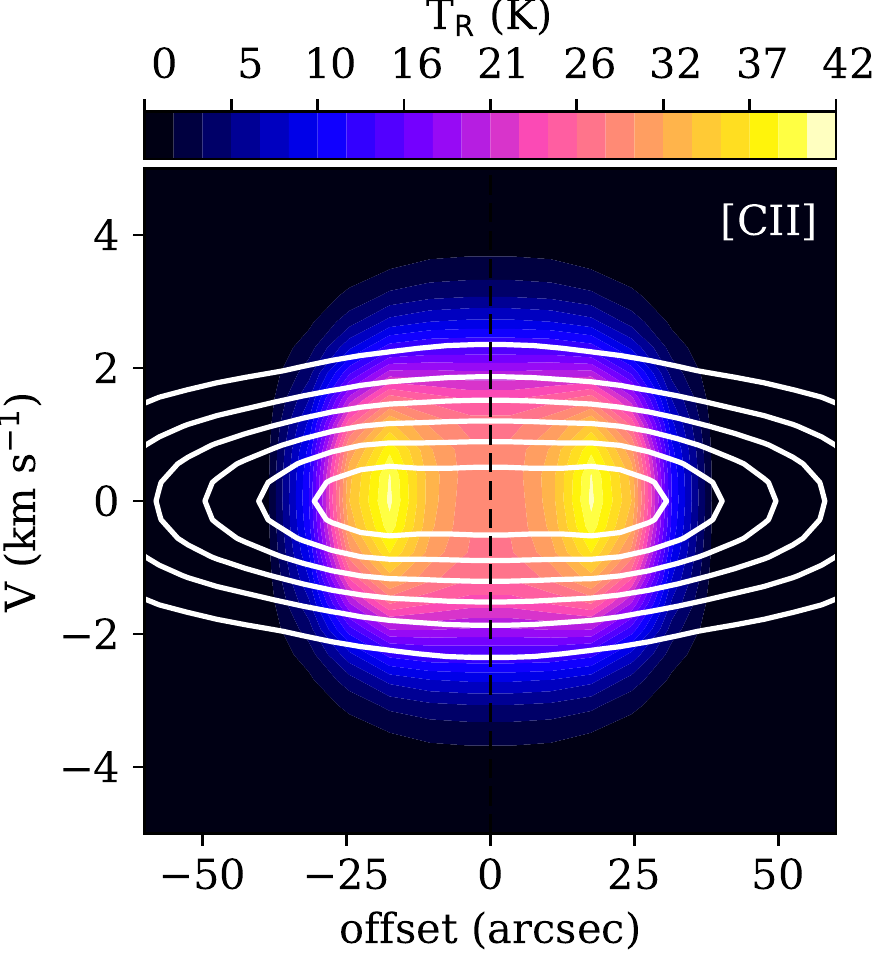}
\includegraphics[width=0.49\columnwidth]{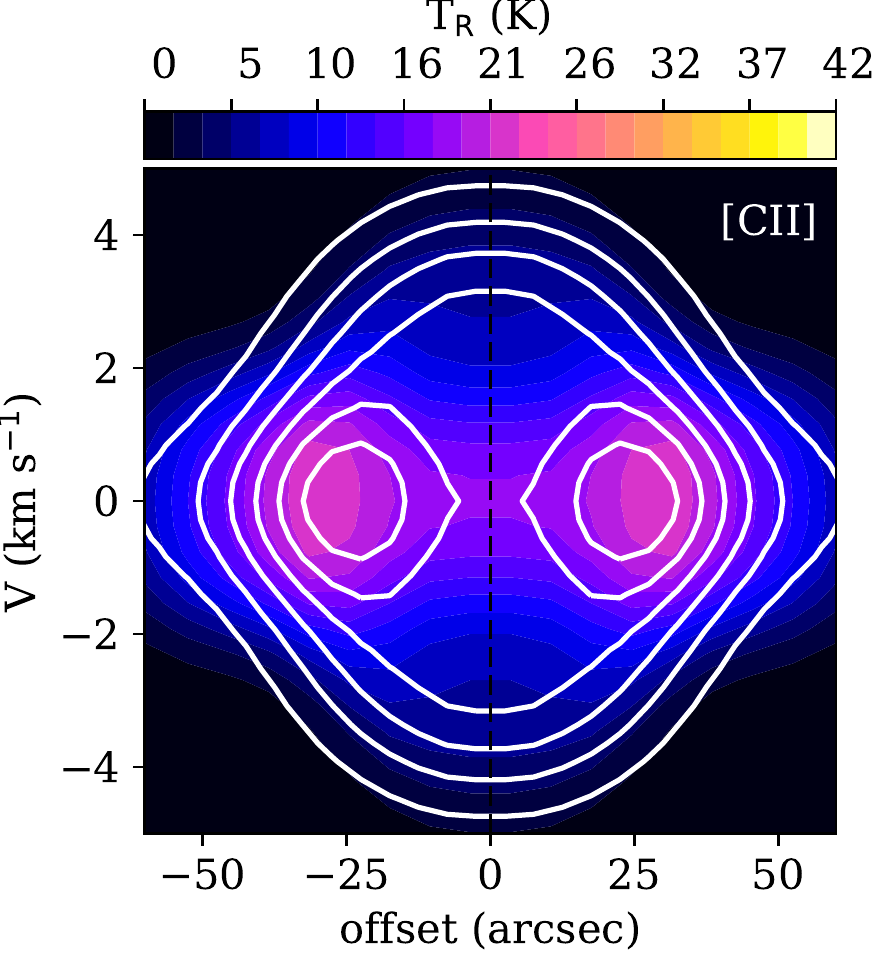}\\
\vspace{1mm}
\includegraphics[width=0.49\columnwidth]{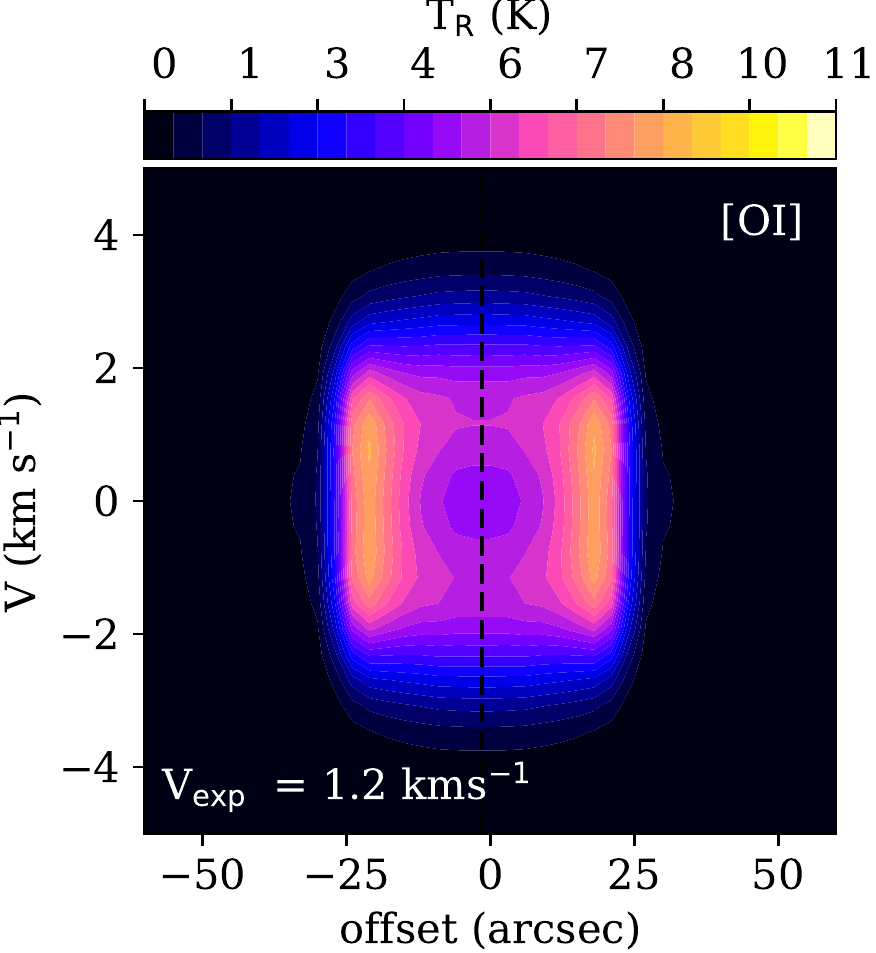}
\includegraphics[width=0.49\columnwidth]{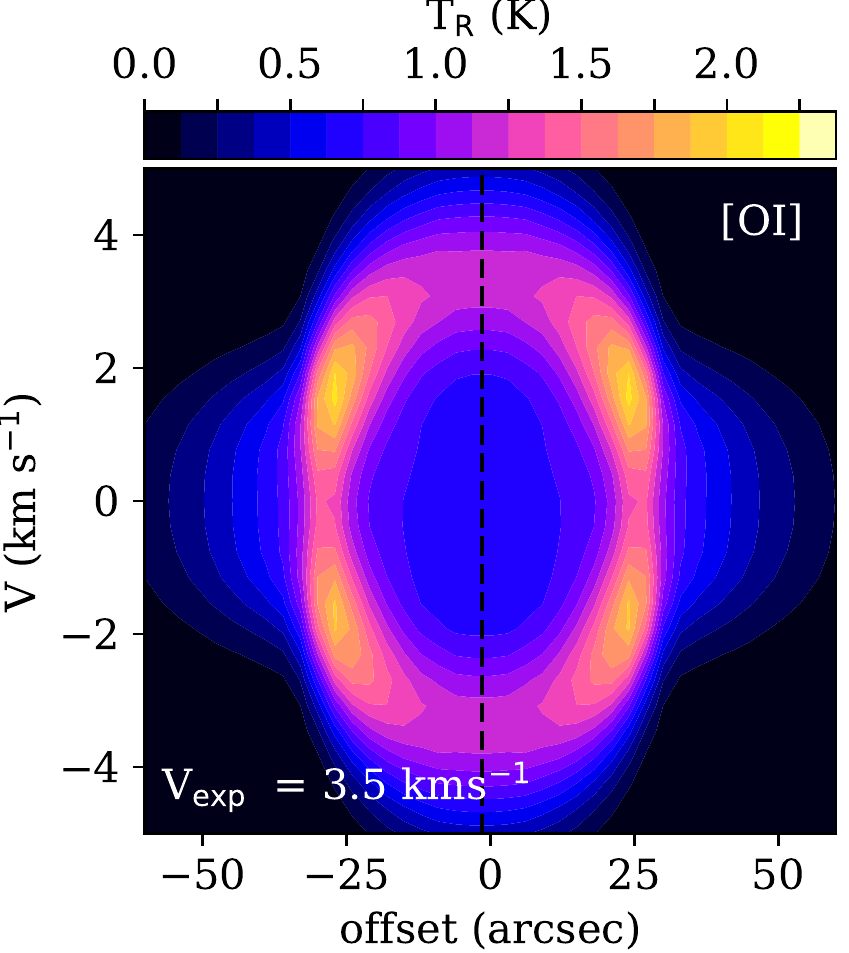}
\caption{Simulated radial integrated intensity distributions (top row) and pv diagrams in S235~A (left) and S235~C (right). The \HCOp{} pv~diagrams are shown by white contours on the top of the \CII{} pv~diagrams. The contours are given every 14\% from the normalized \HCOp{} peak.}
\label{fig:modelpv}
\end{figure}

\paragraph*{Averaged intensities.} The radial distributions of the integrated line intensities are shown in the top panel of Fig.~\ref{fig:modelpv}.  The selected models predict that the \thirtCII{} line intensities are brighter on the periphery of the PDRs than in the direction on the ionizing star in agreement with the observations in Fig.~\ref{fig:13CIIspeA} and \ref{fig:13CIIspeC}. The simulated peak values of W(\CII{}) are 75 and 50\% from the observed in S235~A and S235~C, respectively. Our simulated radiation temperatures of the \thirtCII{} lines about 0.5-1 and 0.3-0.7~K for both PDRs respectively, which is in agreement within factor of two with the average values of \tmb{} from Table~\ref{tab:avervalues}. Integrated intensities of the simulated \OI{} lines are in good agreement with the average values from Table~\ref{tab:avervalues}. The integrated intensities of the \HCOp{} lines in both models are lower than those observed by about two and three orders of magnitude in S235A and S235C respectively.

In order to check whether collisions with electrons can solve the discrepancy by enhancing the \HCOp{} line emission, we performed test described in Appendix~\ref{app:electrons}. We found that the collisional excitation by electrons can increase the peak \HCOp{} intensity by a factor of 2 and 5 times in S235~A and S235~C, respectively. However, this does not solve the problem of the low intensity of the \HCOp{} line in the models. A match of the observed intensity requires a significantly enhanced column of HCO+. We have tested this in two ways. We re-simulated the \HCOp{} lines with an enhanced value of molecular hydrogen number density, but an unchanged value of HCO$^+$ relative abundance compared to Fig.~\ref{fig:modelphys} (i.~e. the column density was enhanced) and found a simulated radiation temperature of the \HCOp{} line of several Kelvins. Namely, the peak of the \HCOp{} line was 2 and 6~K for a column density enhancement by a factor of 10 and 100, respectively, in the model for S235~A. These values are in agreement with the observations, as the maximum peak of \HCOp{} intensity is 4K in S235~A (Fig.~\ref{fig:13CIIspeA}). If we re-scale the relative abundance of HCO$^+$ accordingly, while holding \NHCOp{} constant, we obtain a peak intensity of 0.4 and 1.6~K for a number  density enhancement by a factor of 10 and 100, respectively. The latter value is also within a factor of 2.5 of the peak observational value. The first appraoch combined a column density enhancement with a density enhancement similar to the effect from the additional collisions with electrons. The relatively small difference between the two cases confirms the conclusion that the molecular excitation has a small effect only. It cannot explain the discrepancy between the observations and the model predictions.

\paragraph*{\CII{} and \OI{} optical depths.} The simulated profiles of the \CII{} emission in both selected models are single-peaked. The value of \tauCII{} does not exceed 0.4 and 2.0 in the models for S235~A and C, respectively. In spite of the significant optical depth in the S235~C model, the amount of cold gas ($T_{\rm gas} \approx 30$~K) with abundant C$^+$ is too small to create double-peaked line profiles. In the model of S235~A, we find only slightly red-skewed \CII{} line profiles, instead of the observed double-peaked profiles. Apparently, the low optical depth of the \CII{} emission in the S235~A model is due to the absence of C$^+$ beyond the dense shell in the undisturbed molecular gas, where all the carbon is in the form of solid-phase CO, because the value of $T_{\rm dust}$ in the model is too low to evaporate molecules into the gas phase. 

\begin{figure}
\includegraphics[width=0.49\columnwidth]{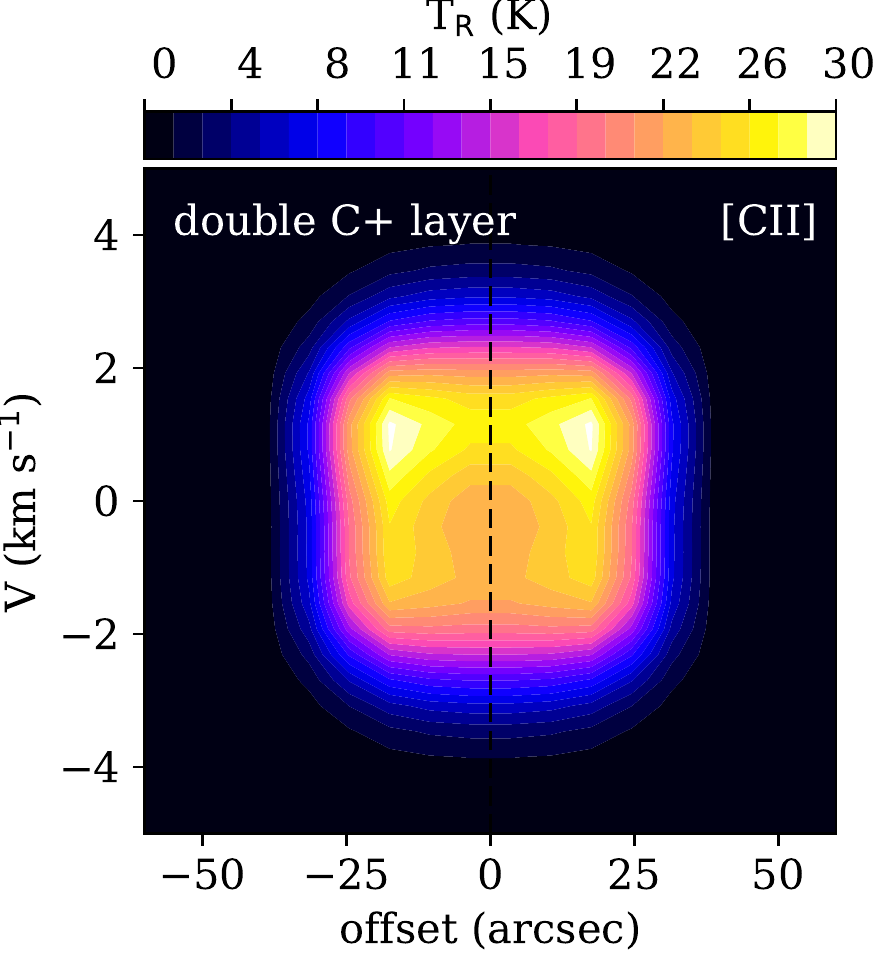}
\includegraphics[width=0.49\columnwidth]{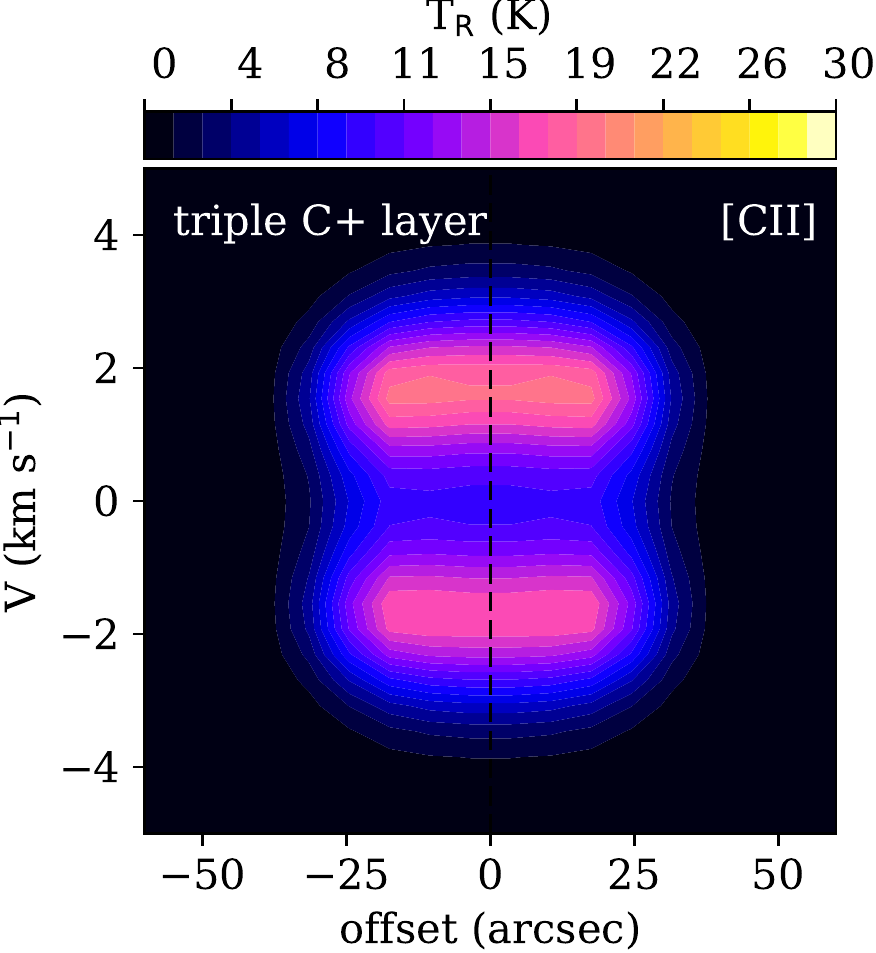}\\
\caption{Simulated pv diagrams for the S235~A model with enhanced width of the C$^+$-layer up to 0.2~pc (original width multiplied by 2, left) and 0.25~pc (original width multiplied by 3, right).}
\label{fig:coldPV}
\end{figure}

To simulate how the \tauCII{} grows with the width of C$^+$-layer in PDR, we increased relative abundance of C$^+$ in front of the expanding envelope up to maximum possible value and preserved all other physical values unchanged. Two test models were considered, there the C$^+$-layer is twice and three times as thick as in original model shown in Fig.~\ref{fig:modelphys}. Calculated pv~diagrams for considered test models are shown in Fig.~\ref{fig:coldPV}. The \CII{} pv~digrams become significantly skewed in the twice thick C$^+$-layer and double-peaked in the three times thick C$^+$-layer, respectively. The peak-to-peak velocity difference $\approx 4$~\kms{} between the blue and red components is defined not by thermal but by turbulent velocity dispersion in the model. The maximum channel value of \tauCII{} becomes 1.6 and 4.5 for the twice and three times thick models, respectively, in agreement with the observations, presented above. We note that in these test calculations the \thirtCII{} line also becomes double-peaked due to significant optical depth which is not in agreement with observations. Therefore, the self-absorption features observed in S235~A and S235~C might not be formed in the PDRs, where relative abundance of C$^+$ is high, but outside of them where the C$^+$ abundance is lower but the absorbing layer is thicker.

The simulated \OI{} lines have double-peaked profiles due to the significant optical depth (about 3-4 in the velocity channels corresponding to the self-absorption feature). 

\paragraph*{Expansion.} In the simulated pv diagrams we find single-peaked \thirtCII{} line profiles in both of the selected models, despite the expansion of the \hii{} regions. For the model for S235~A, the expansion velocity is lower than the width of the turbulent velocity dispersion assumed in the radiation transfer calculations with SimLine, {therefore we do not even see the line wings on the pv~diagram in the direction to the ionizing star}. For the S235~C model, we have significant amount of stationary, undisturbed gas where C$^+$ is the dominant C-bearing species. Broad \thirtCII{} lines {with extended wings} in the direction of the centre of the the pv~diagram indicate the expansion in this model. The \HCOp{} pv~diagrams demonstrate the same behavior in the selected models (but see Fig.~\ref{fig:modelpvappendix} and Fig.~\ref{fig:modelpvappendix2} to find more about the \thirtCII{} and \HCOp{} pv~diagrams.)

The peak-to-peak velocity difference between the near and far neutral walls on the simulated \OI{} line profiles (2 and 6~\kms{} in the models for S235~A and S235~C) corresponds to the expansion with \Vexp{}=1 and 3~\kms{} respectively. The simulated \Vexp{} value is in agreement with the observations of S235~A, but exceeds the values observed in S235~C for a factor of three.

As has been shown in various studies \citep[e.~g.][]{2004R&QE...47...77T, Pavyar_2008}, moderately optically thick lines can be used to discriminate between expansion (double-peaked profiles with a brighter red-shifted component), collapse (double-peaked profiles with a brighter blue-shifted component), and other effects (e.g. rotation). The \OI{} lines are the best tracers of the PDR expansion in the model due to their double-peaked profiles, with a bright red-shifted component, caused by a significant optical depth. We note that the simulated \OI{} profiles in the expanding PDRs are double-peaked, although the other PDR tracers, namely, the \thirtCII{} and \CII{} lines, can be single-peaked. 

In the observations we found the same peak-to-peak velocity difference on the \CII{} and \OI{} line profiles up to 4~\kms. As shown above, the \CII{} self-absorption feature is explained by the layer of cool absorbing C$^+$ on the front of the PDRs. We suggest that the broad and deep \OI{} self-absorption feature is related to the same cold absorbing layer and, unfortunately, hide the internal gas kinematics in the PDRs. As recently shown by \citet{Guevara_2020}, the dense and cold foreground C$^+$ gas leads to the self-absorption effects in several objects. While they note that the origin of the cold gas is unknown, this effect may prevent the detection of expansion and/or contraction in the PDRs. Therefore, further theoretical studies of reliable PDR expansion tracers among other lines, which do not suffer from self-absorption in the cold foreground, are required.

While we reproduce the average integrated intensities of the observed \thirtCII, \CII{} and \OI{} lines, we do not reproduce the detailed line shapes of \CII{} and \OI{}. The model predicts \OI{} (double-peaked) and \CII{} (single-peaked) line profiles with brighter red-shifted components, however the observed profiles are blue-shifted in the majority of the positions. From the comparison of the optically-thin \thirtCII{} line profiles with the double-peaked self-absorbed \CII{} and \OI{} profiles, we know about the east-west velocity gradient, and also about the gas density gradient from the maps of the CO and \HCOp{} emission. Therefore, the agreement between the observed and simulated line profiles is not perfect. Change from the red-shifted to the blue-shifted asymmetry on the observed line profiles might be related with the gas velocity gradient in the PDRs. Foreground absorbing cold material forms the self-absorption feature at $\approx -16$~\kms{} on the observed profiles.

\paragraph*{Summary on the simulated line profiles.} The \OI{} emission is not only a reliable tracer of the gas kinematics, but also of the gas density in PDRs. The \OI{} line at 63\micron{} was the most useful in our model fitting procedure. Both the observed and simulated \OI{} line profiles have the largest peak-to-peak velocity difference (see also averaged spectra in Fig.~\ref{fig:averoverpos}). While the HCO$^+$ lines are more sensitive to the gas number density, we can not obtain the brightness temperatures of several K which are observed in S235~A and S235~C due to insufficient number density of molecular hydrogen. The HCO$^+$(3--2) line becomes a reliable kinematic tracer if $V_{\rm exp}$ is comparable, or exceeds, the turbulent velocity dispersion (models with $n_{\rm init} \leq 10^4$~cm$^{-3}$). The \CII{} and \thirtCII{} lines show no self-absorption in the considered models due to insufficient column density of cold C$^+$ around the PDRs.

While we can reproduce the observed physical conditions and the integrated line intensities with these models (as these properties are determined by UV photons input from the ionizing stars and by the initial number density of the surrounding medium), the gas kinematics and the detailed line profiles strongly depend on the geometry of the regions.

\section{Discussion}\label{sec:disc}

In the previous sections, we considered S235~A and S235~C just as \hii{} regions surrounded by PDRs and molecular clouds with no star formation activity, which can affect the properties of the gas and dust. However, the vicinity of these \hii{} regions is characterized by active star formation. \citet{Felli1997} found a highly obscured young stellar cluster between S235~A and B, as well as a highly variable H$_2$O maser. \citet{Felli2004} observed various molecular transitions as well as mm and sub-mm continuum emission, and found dense molecular cores with new YSOs responsible for driving the maser activity. Aperture photometry of IRAC point sources done by \citet{Felli2006} showed that more-evolved Class~I and II YSOs surround the PDR of S325~A, and there is a younger intermediate-mass protostar embedded in the dense molecular gas and located further from the PDR than the YSOs. With spatially resolved VLA observations, they found a burst of active star formation in the region of S235~A and B, where different evolutionary phases co-exist: from proto-stellar cores to \hii{} regions. \citet{Felli2006} compared the {\it Spitzer}~IRAC images of S235~A and B with the spatial distribution of the HCO$^+$(1-0) emission and proposed that the \hii{} region S235~A and the surrounding molecular cloud might be interacting. They also proposed that dynamic action of S235~A may have triggered a second generation of star formation in the surrounding molecular gas. \citet{Kirsanova2008} found that the region of S235~A, B and S235~C contains the densest and richest cluster of young stars among the whole giant molecular cloud G174+2.5 \citep[see also ][ who identified 55 Class~I, 134 Class~II and 51 Class~III objects around S235~A and B using {\it Spitzer} photometry]{Dewangan2011}. We suggest that ionization of CO-bearing molecular gas around the \hii{} regions by nearby young stellar sources might provide cool C$^+$, which is required to form the self-absorbed \CII{} line profiles and also provide additional HCO$^+$. One of the most promising examples of this is the poorly studied Class~I YSO S235~A-2$^{\star}$ identified by \citet{Dewangan2011} (their source number 5 in Table~C5). Its contribution to the heating of the gas is unknown, although its proximity to the peaks in density and dust temperature suggest that it might provide a significant contribution.

If we remove the assumption of spherical symmetry, the velocity difference between the optically-thin \thirtCII{} and \HCOp{} emission in both PDRs can be interpreted in terms of expansion of the PDRs into the surrounding molecular gas. As \thirtCII{} is red-shifted relative to the \HCOp{} line, and the total column in the front wall is smaller than in the rear wall (compare $A_{\rm V}=10$~mag for the foreground gas from \cite{Thompson1983} and the equivalent $A_{\rm V}$ values from Fig.~\ref{fig:cd_table}), the PDRs expand into the front molecular HCO$^+$-emitting walls of the \hii{} regions. The absence of the signature of gas motion from the back walls is related to the high column density of the backround gas or to the geometry of the PDRs. In S235~A and S235~C the transitions between diffuse and dense gas occurs in an obviously highly inhomogeneous medium. Moreover, we observe a transition between the CO-rich and CO-frozen molecular gas in S235~A. This compact \hii{} region appears as a good example of a PDR with a snow line, proposed by \citet{Goicoechea2016}, while the \hii{} region is embedded and not visible face on (e.g. like the Orion~Bar PDR).

The model of S235~A does not reproduce the observed high C$^+$ column density, while the relatively low gas density in the model of S235~C allows us to obtain the moderately optically thick C$^+$ environment of the \hii{} region, due to large width of C$^+$-emitting layer. We propose that there is a highly inhomogeneous and clumpy medium around the \hii{} regions which is unresolved in the \CII{} observations. The clumpy structure may be composed, for example, of evaporating globules. These globules could allow the photons from the ionizing star to penetrate deeper into the surrounding cloud and increase the C$^+$ column. Signatures of the photon penetration through the clumpy medium were found in, for example, the extended \hii{} region RCW~120 by \citet{Anderson2010, Anderson2015} and \citet{KirsanovaPavyar2019}, instead of uniform expansion; and also at the very early stage ultracompact \hii{} region in Mon~R2 by \citet{Trevino-Morales2016}. A uniform spherically symmetric \hii{} region, considered in the classic theoretical studies (see Sec.~\ref{sec:intro}), has not been found so far. 

Interestingly, not only S235~A and S235~C have higher columns in the back wall, but also other \hii{} regions situated in the giant molecular cloud G174+2.5. \citet{Anderson2019, 2019arXiv191104551K} found the same for the extended \hii{} region S235. \citet{Ladeyschikov2015} determined that the \hii{} region Sh2-233 is also situated on the near side of the molecular cloud. Visual inspection of the CO emission maps from \citet{Bieging2016} shows that two other \hii{} regions from G174+2.5, Sh2-231 and Sh2-232, apparently already dispersed their front and back neutral walls. Thus, all four \hii{} regions in G174+2.5 which are (partly) embedded into molecular cloud have less material on the front than on the rear side.

Besides geometry, our chemical model contains several assumptions which might affect the \thirtCII{} and \CII{} pv~diagrams. First, the 'high-metallicity' initial elemental abundances from \citet{Wakelam2008}, used in this work for the total gas-phase and grain mantle carbon budget, might overestimate the amount of carbon locked in refractory grain cores and, consequently, underestimate the amount of carbon available for chemical reactions in the model. The carbon abundance, used as a reference in this elemental abundance set, is based on observations of the \CII{} line at $\lambda=2325$~\AA{} towards $\zeta$~Oph: $x({\rm C})=1.32\pm0.32\times10^{-4}$, while similar observations of $\chi$~Per give $x({\rm C})=2.45\times10^{-4}$~\citep{Cardelli1993, Savage1996}. \citet{Sofia2004} estimated $x({\rm C})=1.62\pm0.39\times10^{-4}$ for $\zeta$~Oph from \CII{} absorption measurements, and concluded that the diffuse line-of-sight gas-phase value of $x({\rm C})$ might be as high as $2\times10^{-4}$. Both stars are situated within 250~pc of the Sun, while the closest \hii{} region, the Orion Nebula, is at about 400~pc~\citep{Menten2007, Grossschedl2019, Kounkel2019, Kuhn2019}. The value of $x({\rm C})$ measured towards $\zeta$~Oph is successfully used in astrochemical models of the Orion region \citep[e.~g. recent work by][]{Cuadrado2017}. However, \citet{Rubin1991} and \citet{Osterbrock1992} found higher values of $x({\rm C})$ up to $x({\rm C})=2.7\times10^{-4}$ in the Orion Nebula. For the S106 region, which is about 1.3~kpc from the Sun, \citet{Schneider2018} used $x({\rm C})=2.34\times10^{-4}$ based also on analysis of the Orion region by \citet{SimonDiaz2011}. A gas-phase elemental abundance $x({\rm C})=8\times10^{-5}$, significantly lower than the values mentioned above, was found by \citet{Fuente2019} for the nearby dark cloud TMC~1. We see that variations of the gas-phase $x({\rm C})$ in the simulations can be justified observationally in broad limits, and can be tuned to improve the agreement between the observations and simulations. Using higher elemental carbon abundance in our calculations with the 'ISM' set from CLOUDY~\citep{Ferland2013} allows us to reproduce the \CII{} double-peaked line profile in the model for S235~C without addition of the external layer of cold C$^+$, see Fig.~\ref{fig:altcarbonabund}.

Another uncertainty in the initial conditions of the model lies in choosing low gas and dust temperature, and also fixing all carbon in the form of solid CO. Some carbon in the grey positions in Fig.~\ref{fig:cd} might indeed be hidden from the gas phase on the grain surfaces due to the low dust temperature. Therefore, our choice of solid CO as initial conditions for the model is, at least partially, supported by the observations. \citet{Kirsanova2014} determined the value of \tgas{} in the direction of S235~A and the young stellar cluster south of it. They found values up to 60~K in S235~A and $20 \leq T_{\rm gas} \leq 32$~K towards the cluster. Similar values were recently found by \citet{Burns2019}. \citet{Kirsanova2014} also found a gas density higher than 10$^4$~cm$^{-3}$ in the molecular cloud around S235~A and in the direction of the embedded stellar cluster. The value of \tgas{} is higher in the direction of the young stellar cluster than \tdust{} by up to 10-20~K, and the difference is even higher in S235~A, despite the high density.  A similar situation was found, for example, in the direction of S140 PDR by \citet{Koumpia2015}. Our model of the expanding structure ``\hii{} region + PDR + molecular cloud'' for S235~A can explain such a discrepancy between the \tgas{} and \tdust{} values for the gas with density $\geq 10^4$~cm$^{-3}$, see Fig.~\ref{fig:modelphys}. We note that various methods for the temperature determination have their own specific features \citep[e.~g.][]{Schneider2016}.

As we have shown above, an additional external layer of cold C$^+$ would produce the self-absorption gap in the model, but we do not try to reproduce this layer using, for example, external UV radiation in the model, because it requires additional parameters. The \CII{} emission in our models for S235~A and S235~C appears from the PDRs \citep[in agreement with previous theoretical studies by][]{Kaufman2006}, but the self absorption can be created in the undisturbed gas with lower $T_{\rm ex}$, the density and temperature of which represent the cold neutral medium \citep[CNM, e.g.][]{Wolfire2003}. The CNM in the external envelope of the S235~C model may consist of C$^+$ and H$_2$ instead of atomic hydrogen, as suggested in recent theoretical calculations of the thermal and chemical steady state by \citet{Bialy2019}. We do not start from steady state abundances, although we emphasize that different initial conditions might change the pv~diagrams considered in the present study, i.e. increase or decrease the \CII{} self-absorption. 

\section{Conclusions}

In this work, we presented observations of the PDRs around the compact \hii{} regions S235~A and S235~C in the \CII, \thirtCII{} lines at 158~\micron, \OI{} at 63~\micron{} and \HCOp{} at 267 GHz, tracing different layers around the volume of the ionized hydrogen. We summarize our main results below.

\begin{itemize}

\item The double-peaked \CII{} and \OI{} line profiles are due to high optical depth (\tauCII{} is up to 10 in several directions in S235~A), not directly tracing the expansion of the front and back walls of the PDRs. The foreground absorbing material has $V_{\rm lsr} \approx -16$~\kms{}, which remains almost unchanged over the observed area. We find an expanding motion of the \CII-emitting layer into the molecular layer with a velocity up to 1~\kms{} in both PDRs comparing velocities of the \thirtCII{} and \HCOp{} lines.

\item Gas components neither visible in the \CII{} lines, nor in low-J CO lines, may contribute to the total column in several points across S235~A. Due to the low dust temperature in those directions, we expect CO depletion due to the CO freeze-out onto dust grains. Apparently, S235~A represents a good example of an \hii{} region surrounded by a PDR, molecular cloud and a snow line, where molecules are absorbed from the gas phase into the dust grains. The \hii{} regions are situated on the border between the dense molecular and more diffuse atomic gas.

\item We are able to reproduce the physical parameters of the \hii{} regions and integrated intensities of the \thirtCII, \CII{} and \OI{} lines from the PDRs with a spherically symmetric chemo-dynamical model as these properties are defined by spectral type of the ionizing stars and initial gas number density. However, the model does not reproduce the double-peaked \CII{} line profiles related with the self-absorption  and high optical depth of the line. To be in agreement with the observations, we need to enhance C$^+$ column density outside of the PDRs. We propose that the PDRs are surrounded by cold neutral medium composed of evaporating globules. While the \CII{} line emission originates in the model PDRs, the cold neutral medium is responsible for the self-absorption features in the line profiles.

\item The \OI{} line profiles are the best tracers of the expanding motion in the considered PDR models in comparison with the \CII, \thirtCII{} and \HCOp{} lines. However, the absorbing cold neutral medium might distort the \OI{} line profiles originating from the PDRs and contaminate the observed profiles.

\end{itemize}

\section*{Acknowledgements} 

We thank the anonymous referee for critique and suggestions which led to the improvement of this manuscript.

This work is based on observations made with the NASA/DLR Stratospheric Observatory for Infrared Astronomy (SOFIA). SOFIA is jointly operated by the Universities Space Research Association, Inc. (USRA), under NASA contract NAS2-97001, and the Deutsches SOFIA Institut (DSI) under DLR contract 50 OK 0901 to the University of Stuttgart.
 
M.~S.~Kirsanova and P.~A.~Boley were supported by Russian Science Foundation, research project 18-72-10132, during their work with the observational data. Theoretical simulations by M.~S. Kirsanova and Ya.~N. Pavlyuchenkov were supported by the Russian Foundation for Basic Research, research project 18-32-20049. A.M.S. was supported by the Ministry of Science and Education, FEUZ-2020-0030.

V.O. was supported by the Collaborative Research Centre 956, sub-project C1, funded by the Deutsche Forschungsgemeinschaft (DFG), project ID 184018867.
N.S. acknowledges support by the Agence National de Recherche (ANR/France) and the Deutsche Forschungsgemeinschaft (DFG/Germany) through the project ``GENESIS'' (ANR-16-CE92-0035- 01/DFG1591/2-1) and from the BMBF, Projekt Number 50OR1714 (MOBS - MOdellierung von Beobachtungsdaten SOFIA).

The James Clerk Maxwell Telescope is operated by the East Asian Observatory on behalf of The National Astronomical Observatory of Japan; Academia Sinica Institute of Astronomy and Astrophysics; the Korea Astronomy and Space Science Institute; Center for Astronomical Mega-Science (as well as the National Key R\&D Program of China with No. 2017YFA0402700). Additional funding support is provided by the Science and Technology Facilities Council of the United Kingdom and participating universities in the United Kingdom and Canada. We used the data from SASSY project (Program ID MJLSY02)

This research has made use of NASA's Astrophysics Data System Bibliographic Services; SIMBAD database, operated at CDS, Strasbourg, France~\citep{Wenger_2000};  Aladin web page~\citep{2000A&AS..143...33B}; Astropy, a community-developed core Python package for Astronomy~\citep{Astropy_2013}; APLpy, an open-source plotting package for Python \citep[(http://aplpy.github.com)][]{APLpy_2012}. 

\bibliographystyle{mnras}
\bibliography{mybi_v3_clear} 

\appendix

\section{Averaged spectra}\label{app:averagedspectra}

\begin{figure}
\includegraphics[width=0.9\columnwidth]{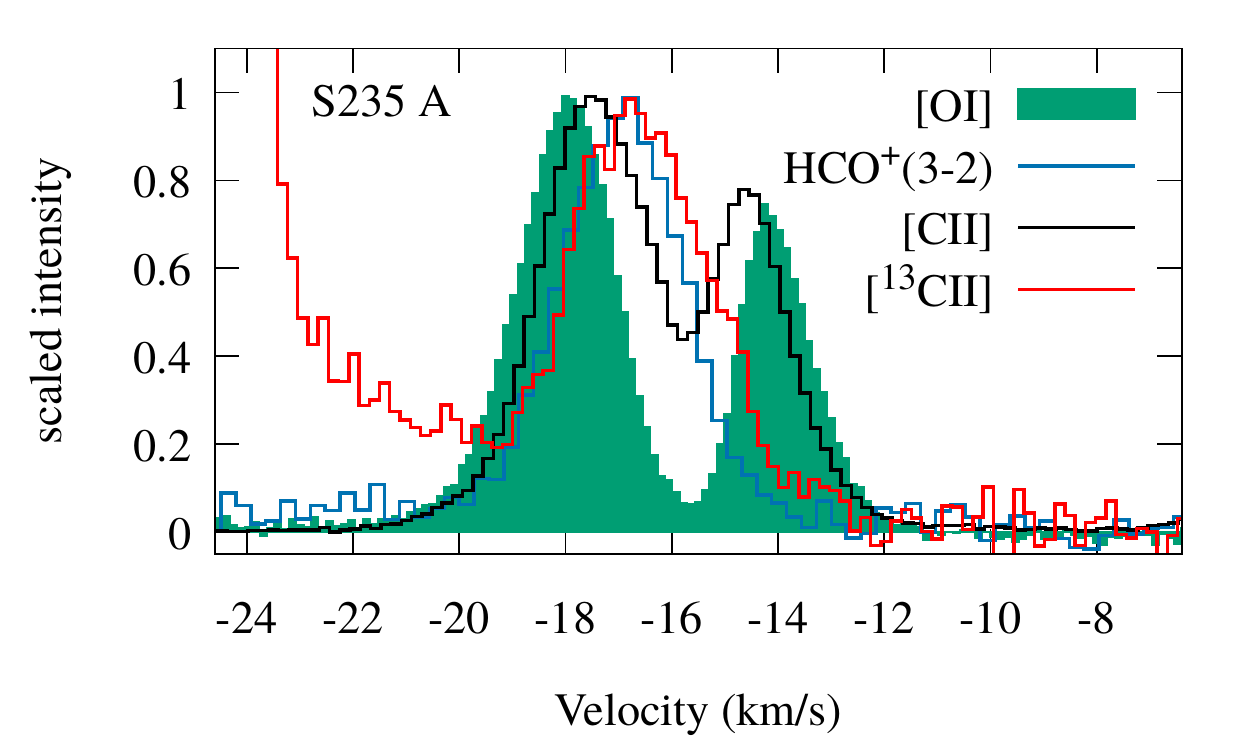}\\
\includegraphics[width=0.9\columnwidth]{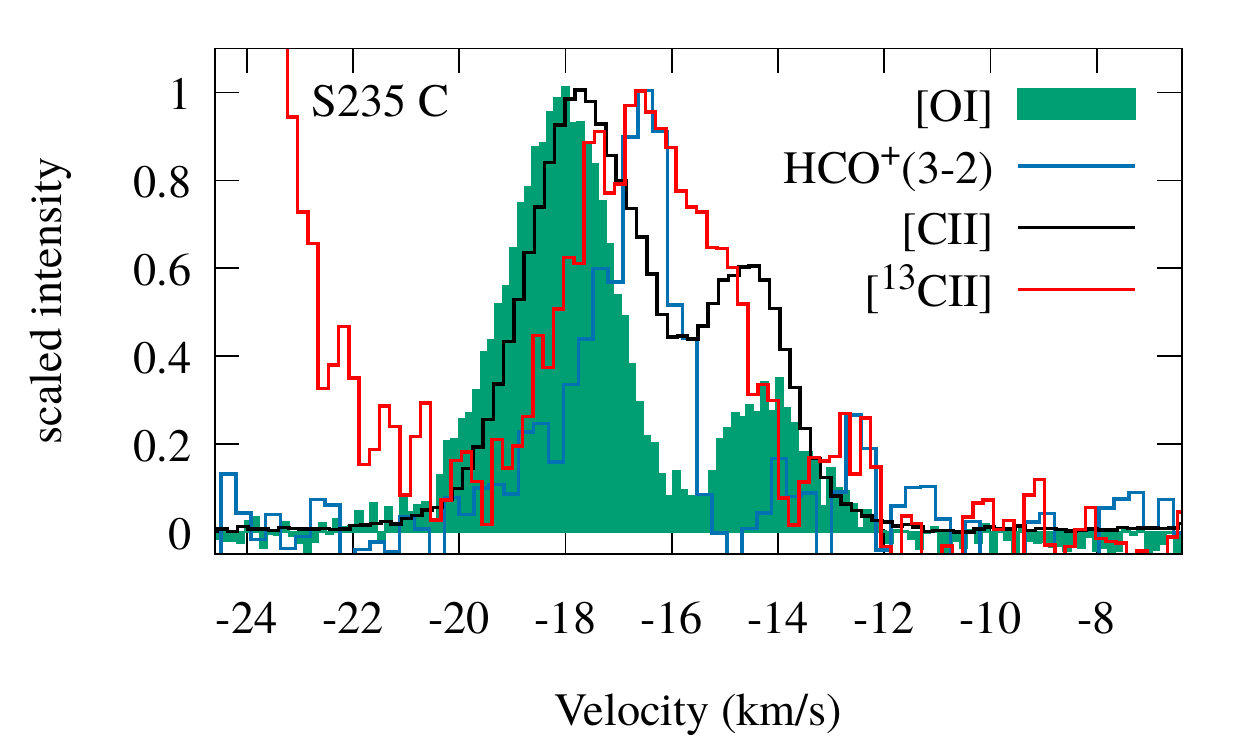}
\caption{The averaged scaled \CII, \thirtCII, \OI{} and \HCOp{} spectra. }
\label{fig:averoverpos}
\end{figure}

We show in Fig.~\ref{fig:averoverpos} scaled spectra of the \thirtCII, \CII, \HCOp{} and \OI{} lines, averaged over positions 14 and 7, which were observed in the \thirtCII{} line in S235~A and S235~C respectively. All spectra are given with their original spectral resolution and are scaled in order to show the difference in the line profiles more clearly. The red wing of the \CII{} line is visible in the \thirtCII{} spectra at $V_\mathrm{LSR} \la -20$~\kms{}.

\section{Additional pv diagrams}\label{app:additionalpv}

\begin{figure*}
\includegraphics[width=0.48\columnwidth]{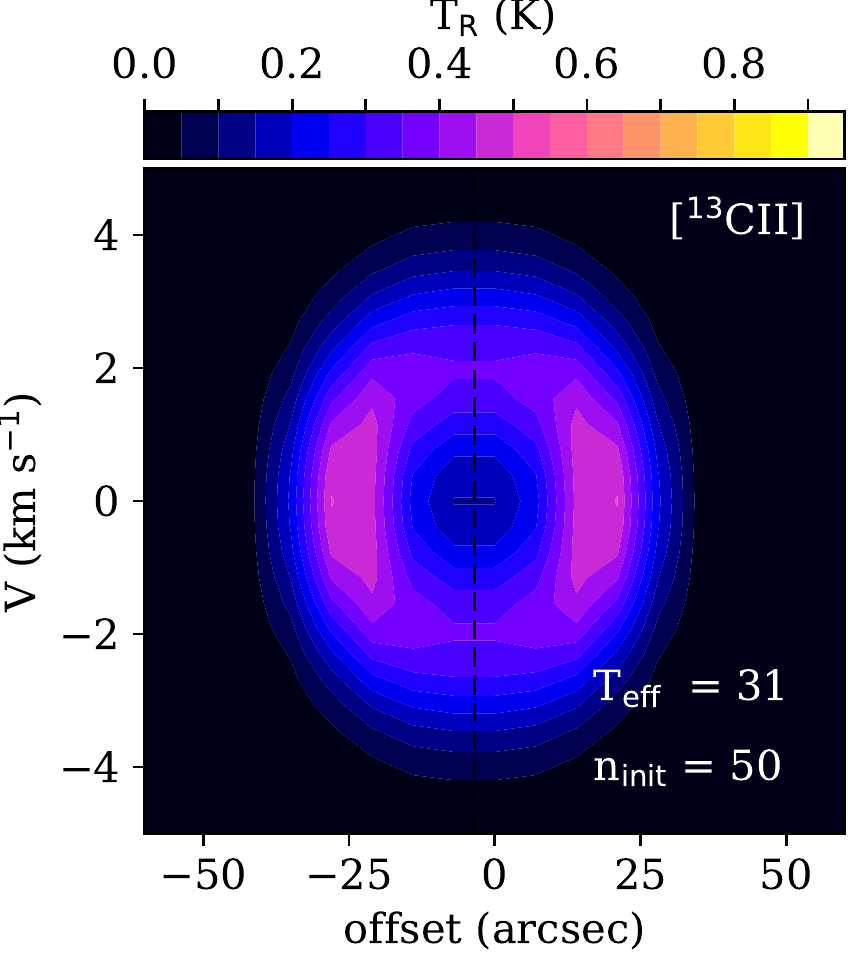}
\includegraphics[width=0.48\columnwidth]{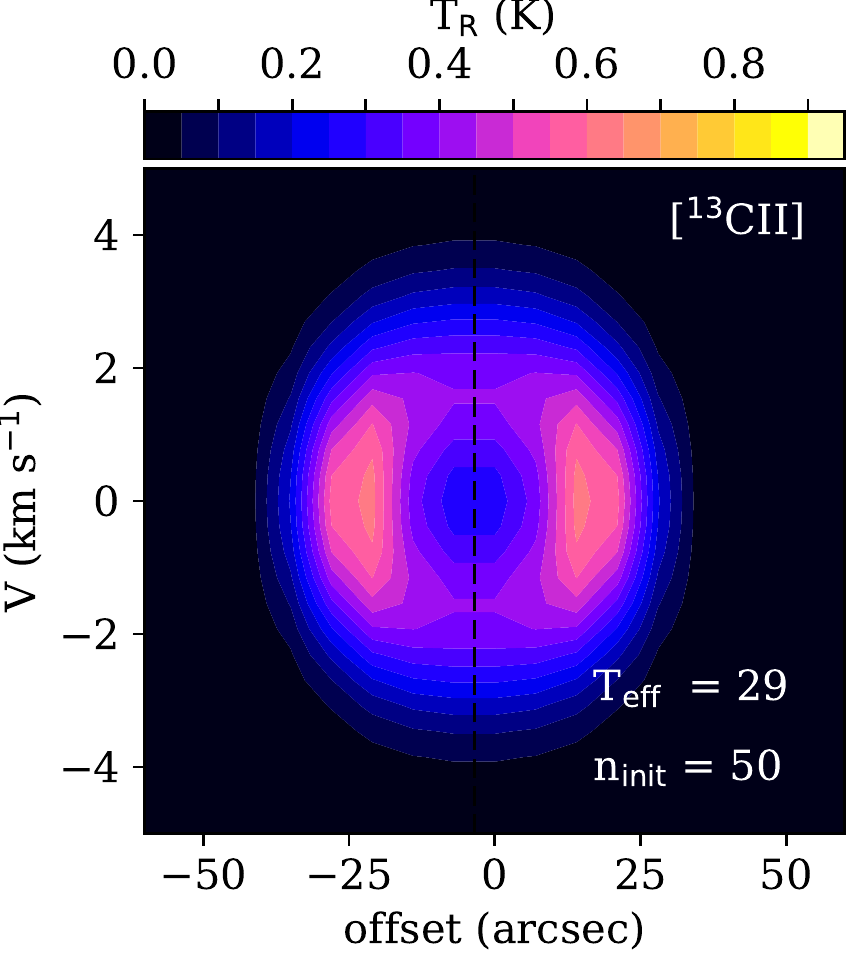}
\includegraphics[width=0.48\columnwidth]{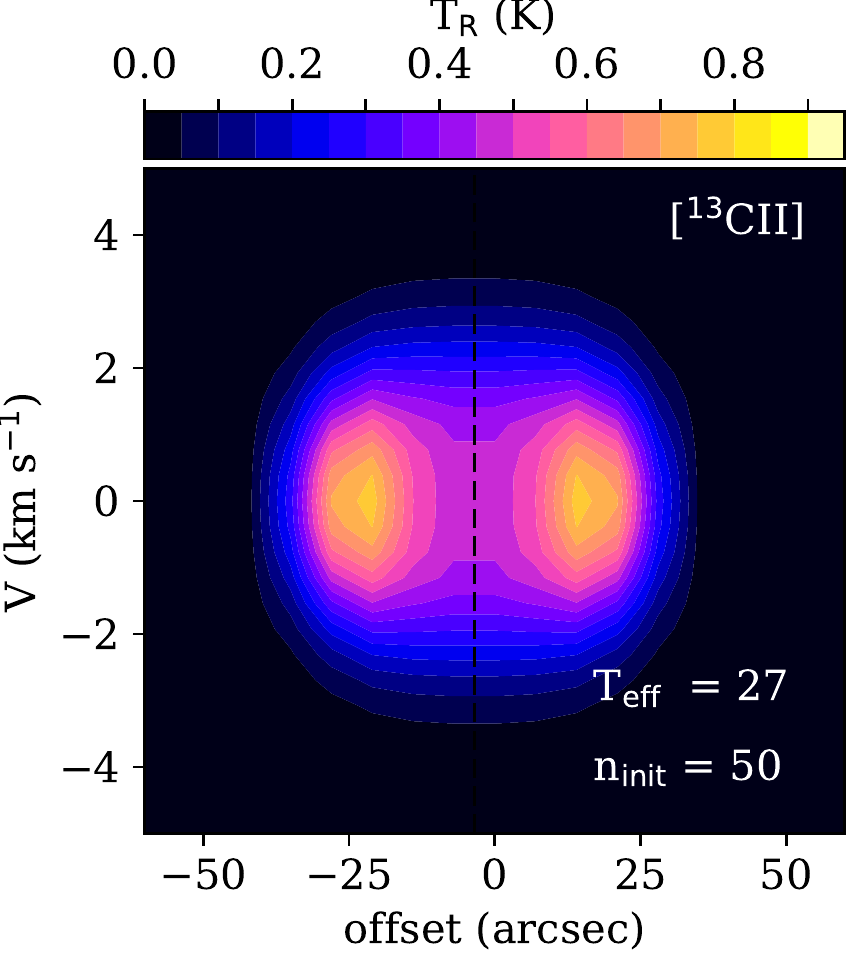}
\includegraphics[width=0.48\columnwidth]{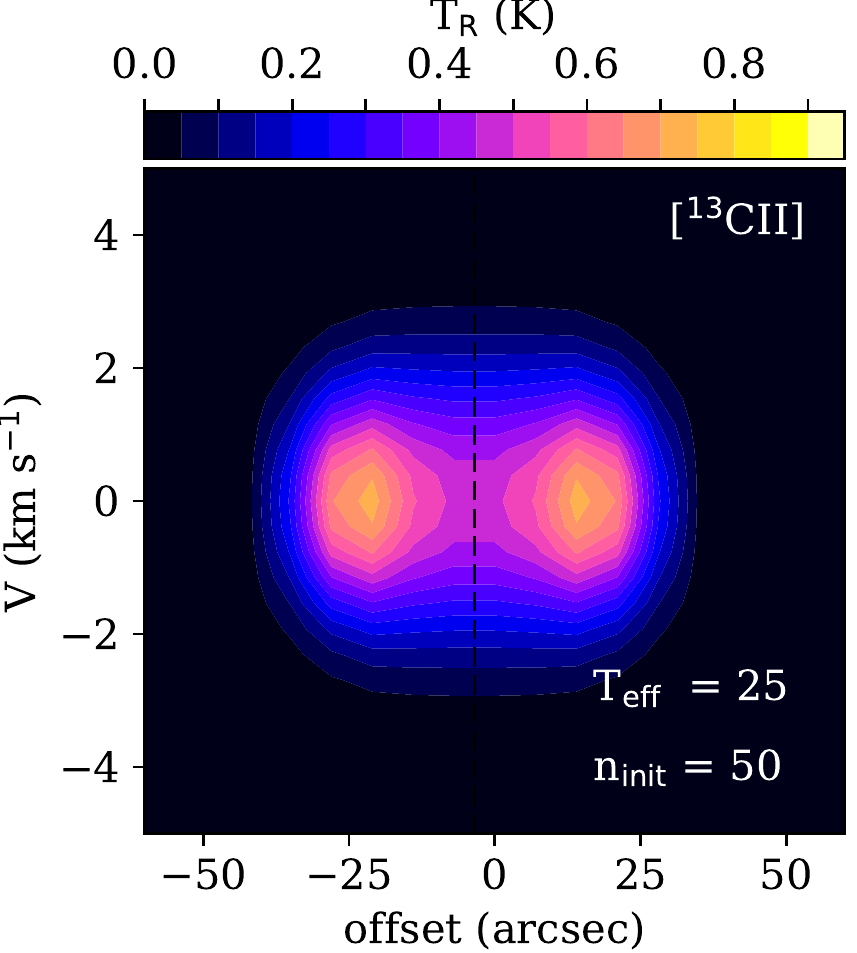}\\
\vspace{2mm}
\includegraphics[width=0.48\columnwidth]{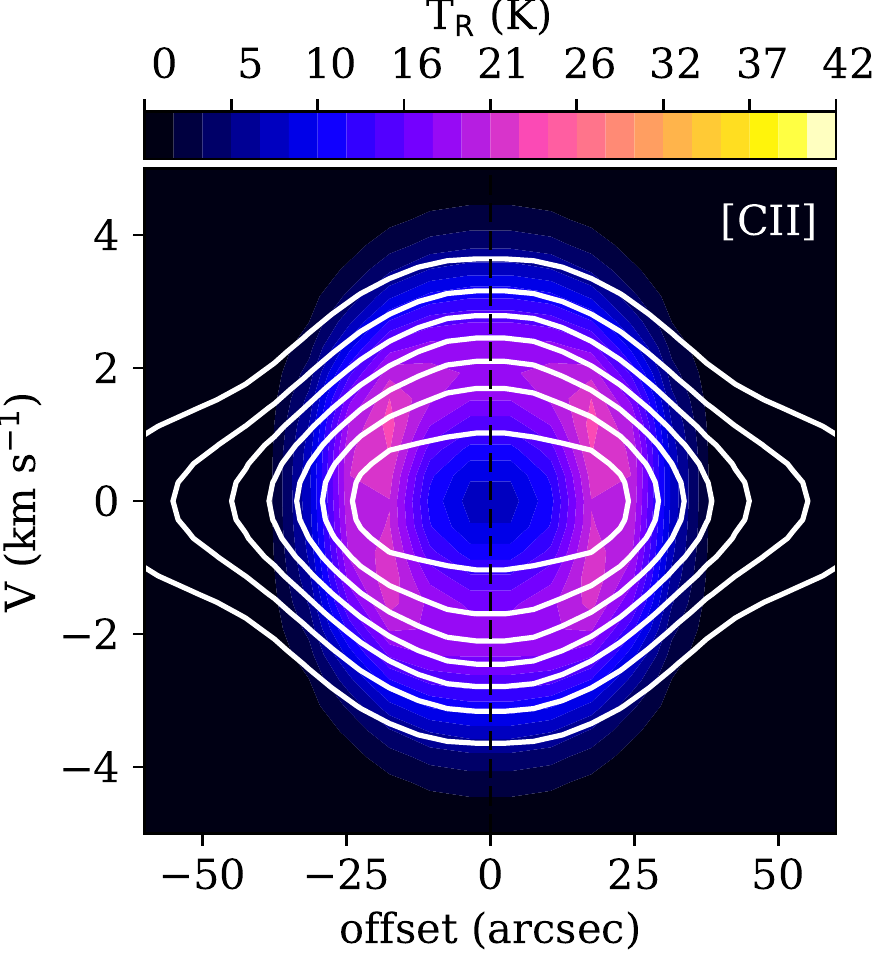}
\includegraphics[width=0.48\columnwidth]{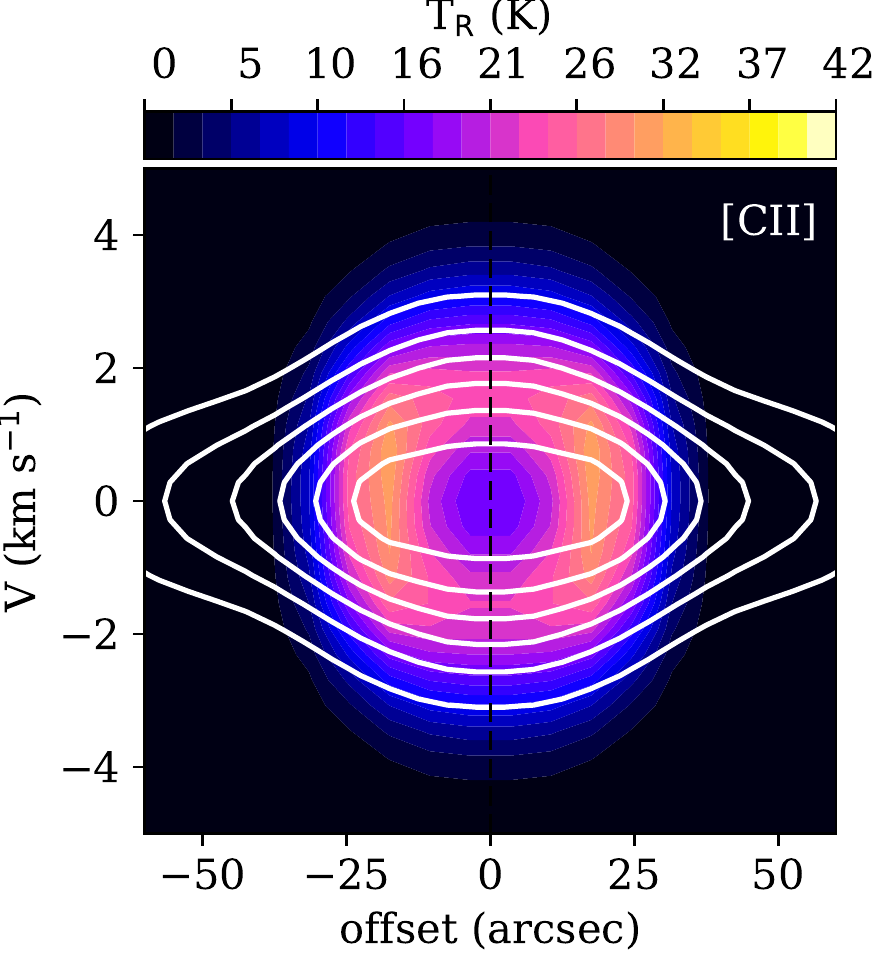}
\includegraphics[width=0.48\columnwidth]{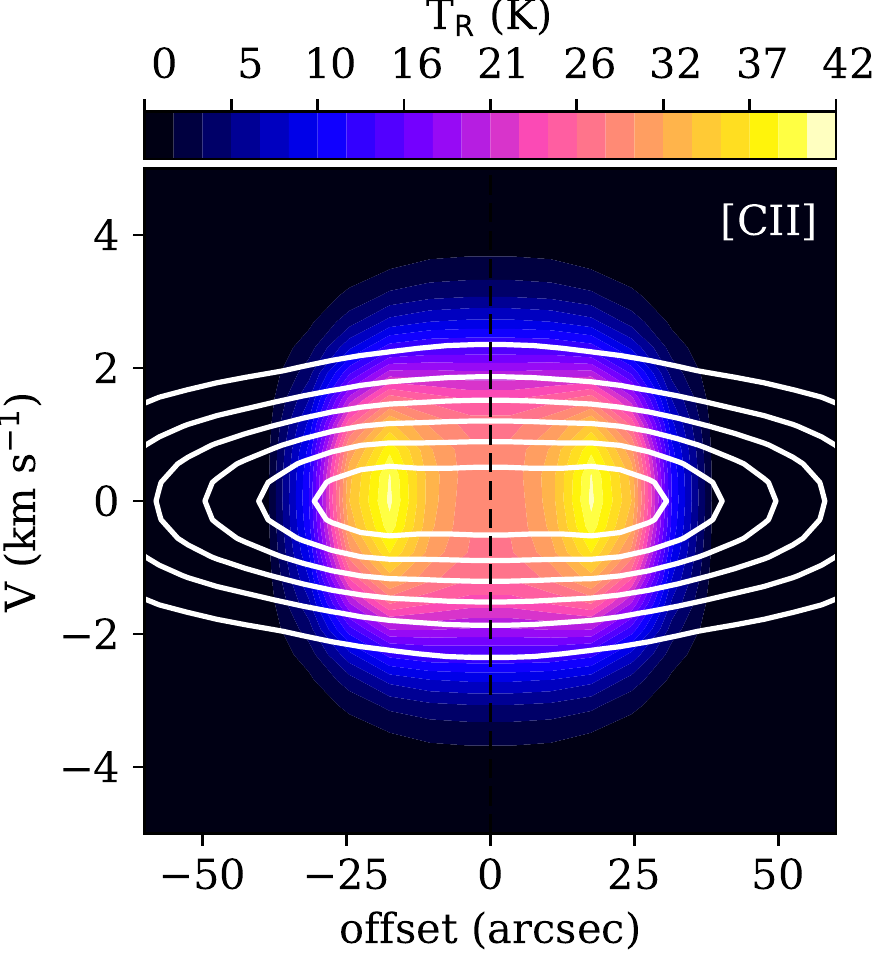}
\includegraphics[width=0.48\columnwidth]{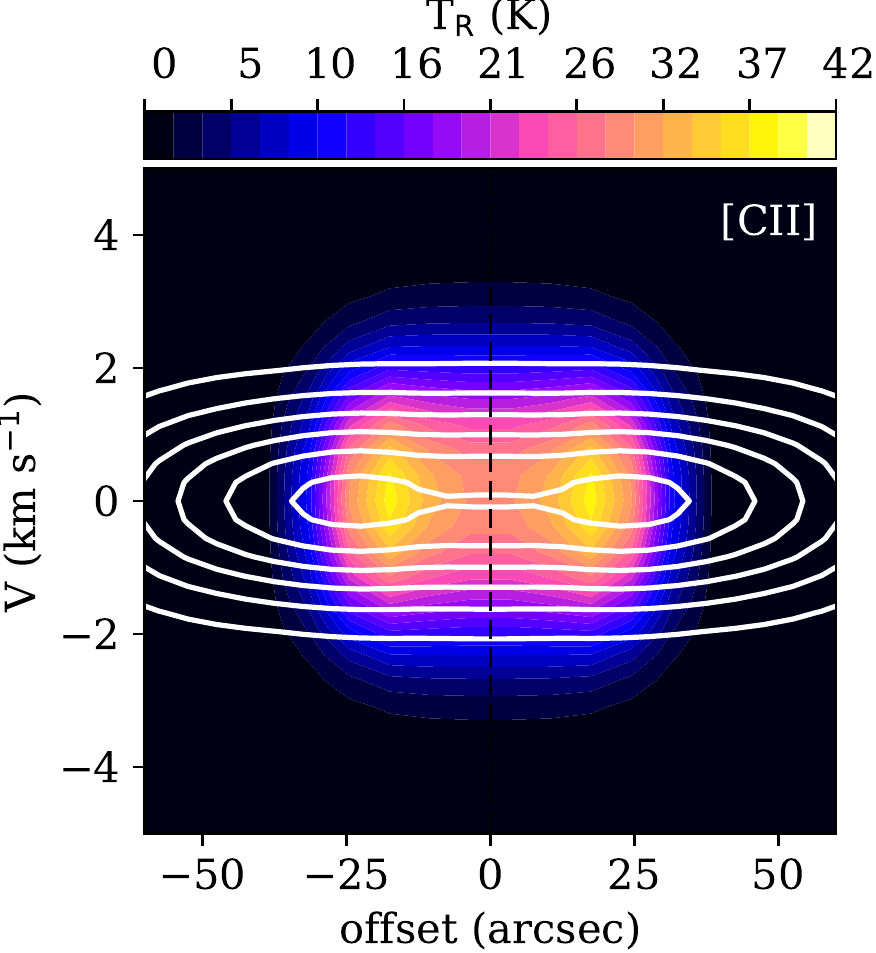}\\
\vspace{2mm}
\includegraphics[width=0.48\columnwidth]{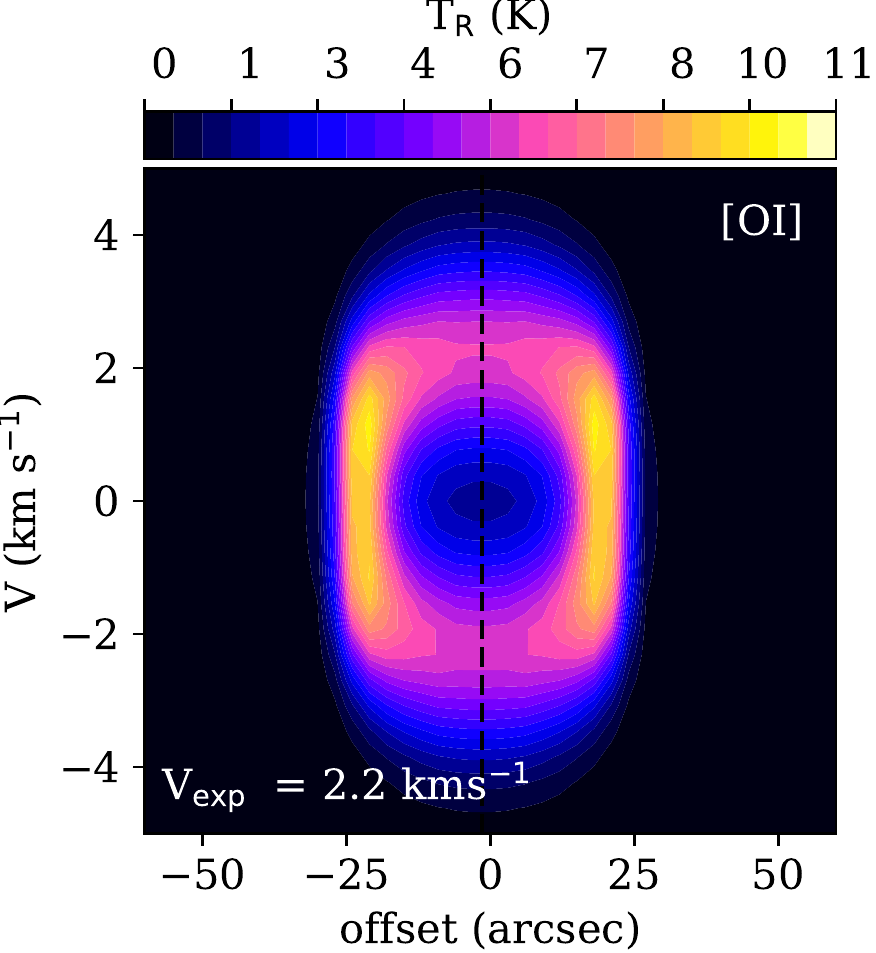}
\includegraphics[width=0.48\columnwidth]{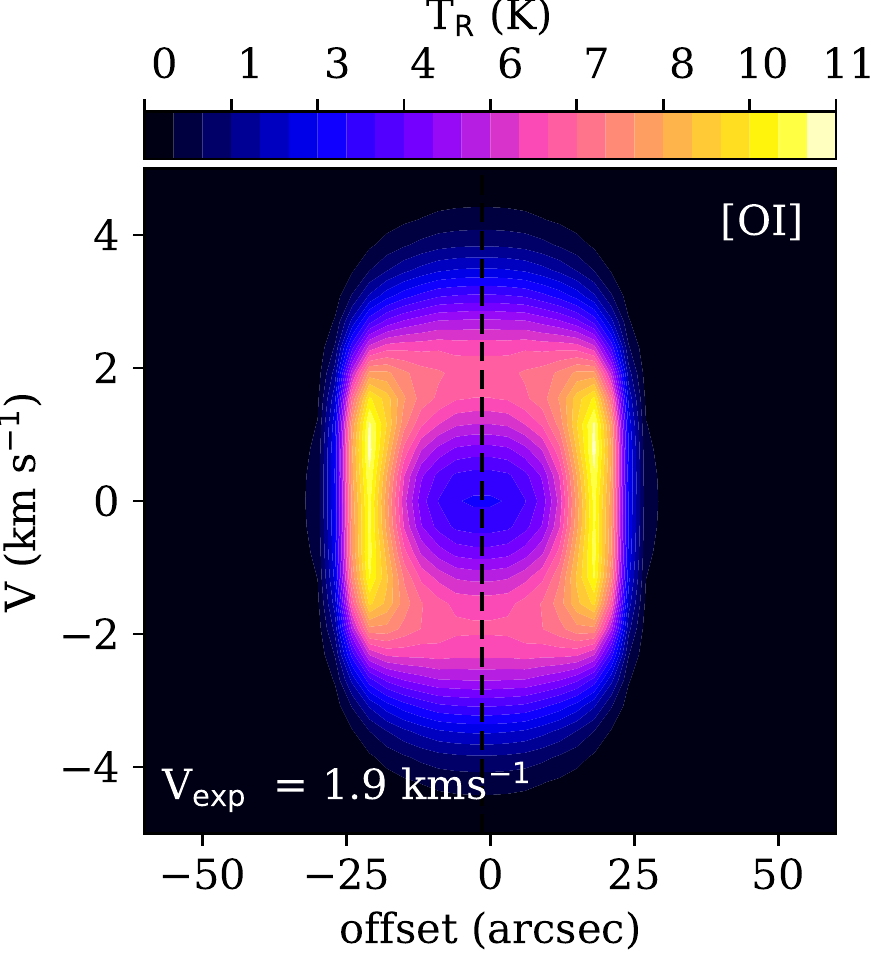}
\includegraphics[width=0.48\columnwidth]{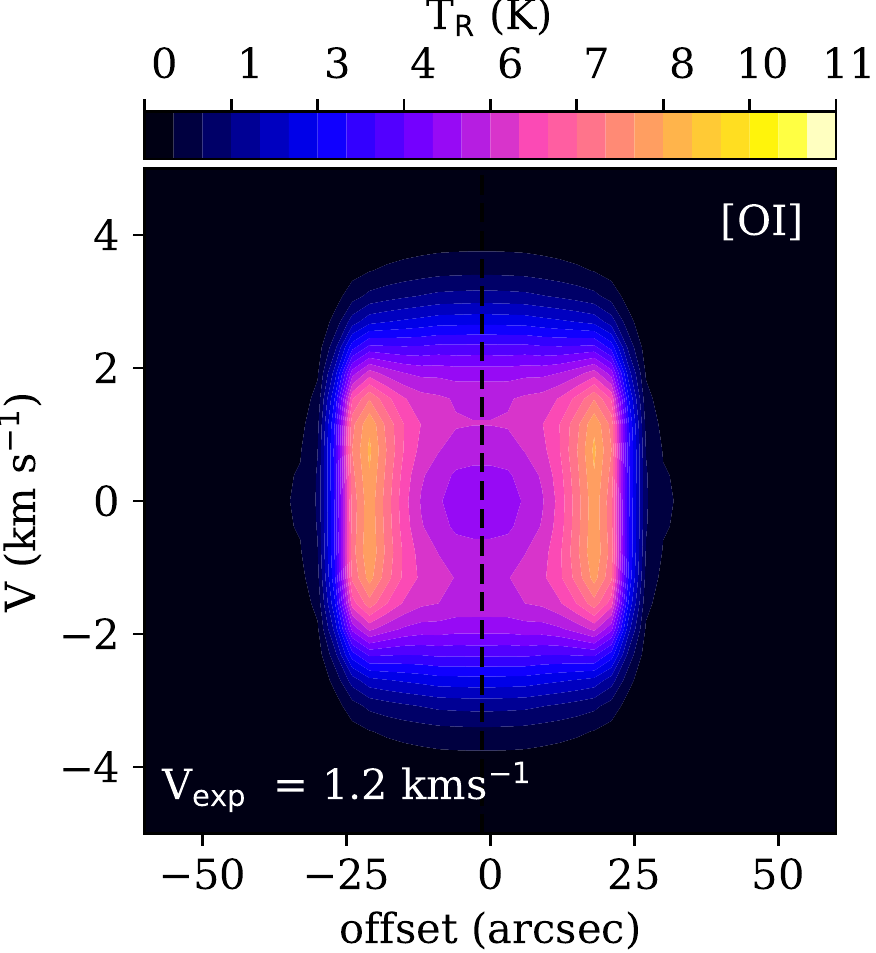}
\includegraphics[width=0.48\columnwidth]{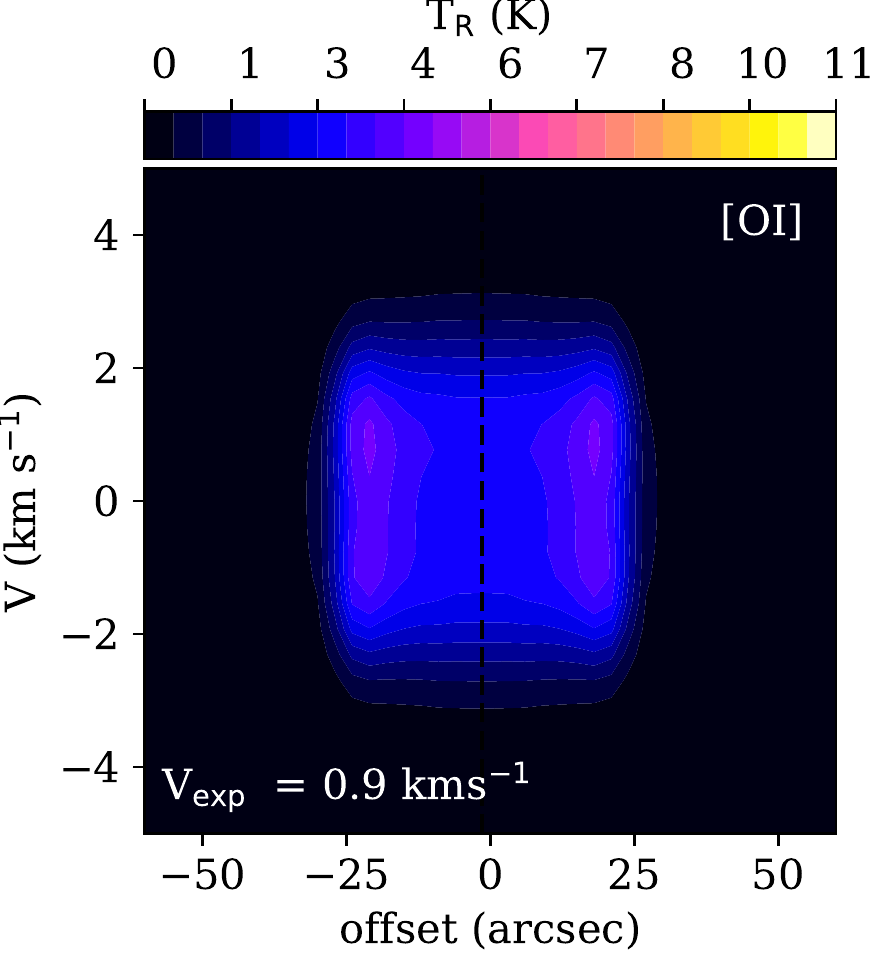}\\
\caption{Simulated pv diagrams in the model with initial gas number density $n_{\rm init}=5\times10^4$~cm$^{-3}$ and $T_{\rm eff} = 31 000$, 29000, 27000 and 25000~K. The model age is 36000, 41000, 60000 and 82000 years, respectively. The \HCOp{} pv~diagrams are shown by white contours on the top of the \CII{} pv~diagrams. The contours are given every 14\% from the normalized \HCOp{} peak.}
\label{fig:modelpvappendix}
\end{figure*}

In Fig.~\ref{fig:modelpvappendix}, we present pv diagrams from alternative models with different stellar temperatures and densities to demonstrate the dependence of the line emission profiles on these parameters. We show the \thirtCII{} and \CII{} lines at 158~\micron, the \OI{} line at 63~\micron{} and the \HCOp{} line.  All lines were simulated with spatial and spectral resolutions corresponding to the observations in Sec.~\ref{sec:obs}. The line profiles were calculated for the moment when the radius of the simulated \hii{} region reaches 0.15~pc, as in Fig.~\ref{fig:modelpv} for S235~A. The values of $T_{\rm eff}$ (in thousands of K) and $n_{\rm init}$ (in 10$^3$~cm$^{-3}$) for the models considered are shown in the top panels with the \thirtCII{} pv~diagrams. The maximum value of the radial velocity is shown in the panels with the \OI{} pv~diagrams, to make comparison of the gas kinematics with the simulations easier.

Fig.~\ref{fig:modelpvappendix} shows that the \thirtCII{} and the \CII{} line profiles become double-peaked in models with $T_{\rm eff}=31000$ and 29000~K. These lines are optically thin, and the peak-to-peak velocity difference corresponds to twice the value of $V_{\rm exp}$. The value of $V_{\rm exp}$ in these two models is comparable with the turbulent velocity dispersion, therefore these lines trace the gas kinematics. The \thirtCII{} and the \CII{} lines are optically thin in all four models from Fig.~\ref{fig:modelpvappendix}. The \OI{} line, as shown above, is a reliable kinematic tracer, as the peak-to-peak velocity difference corresponds to twice the value of $V_{\rm exp}$ in the models with $T_{\rm eff}$ from 25000 to 31000~K. The HCO$^+$ lines are single-peaked in the four models, but their full widths at half maximum corresponds to the twice the value of $V_{\rm exp}$.

\begin{figure*}
\includegraphics[width=0.48\columnwidth]{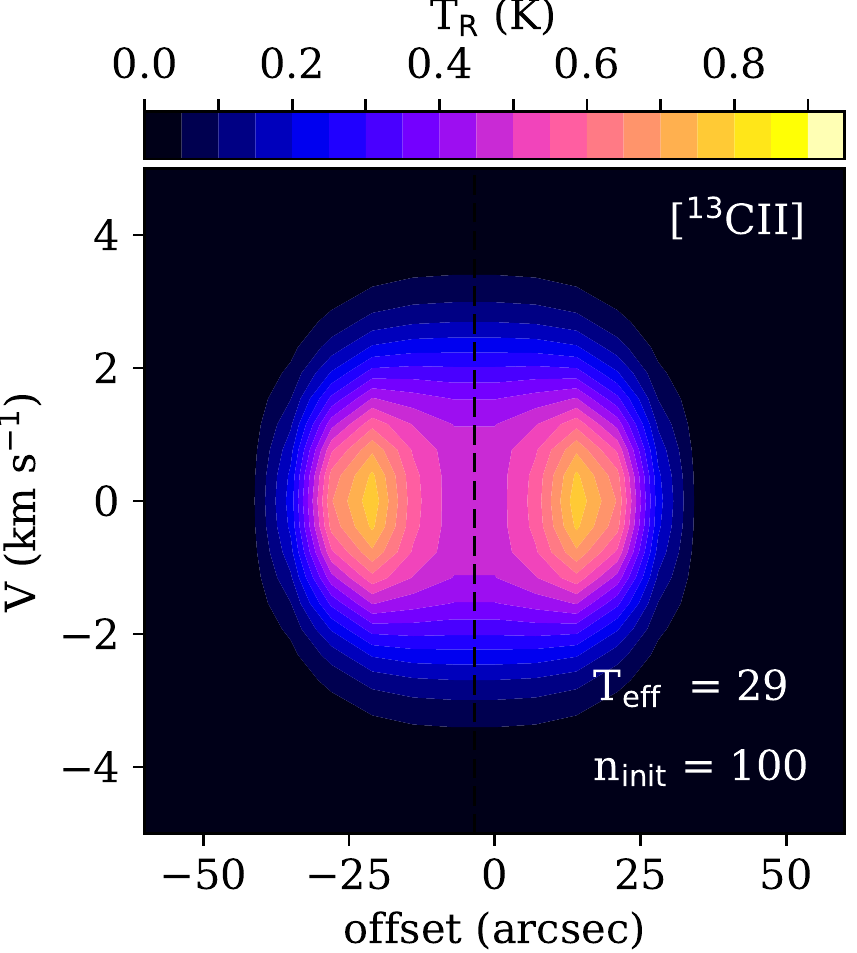}
\includegraphics[width=0.48\columnwidth]{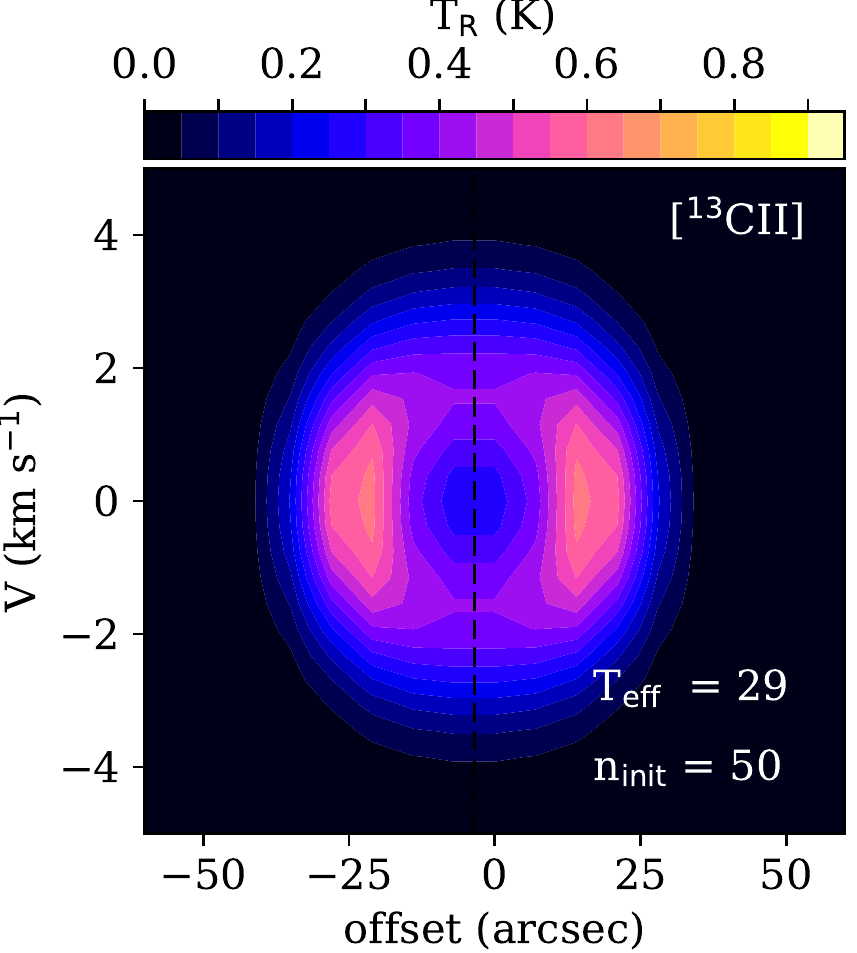}
\includegraphics[width=0.48\columnwidth]{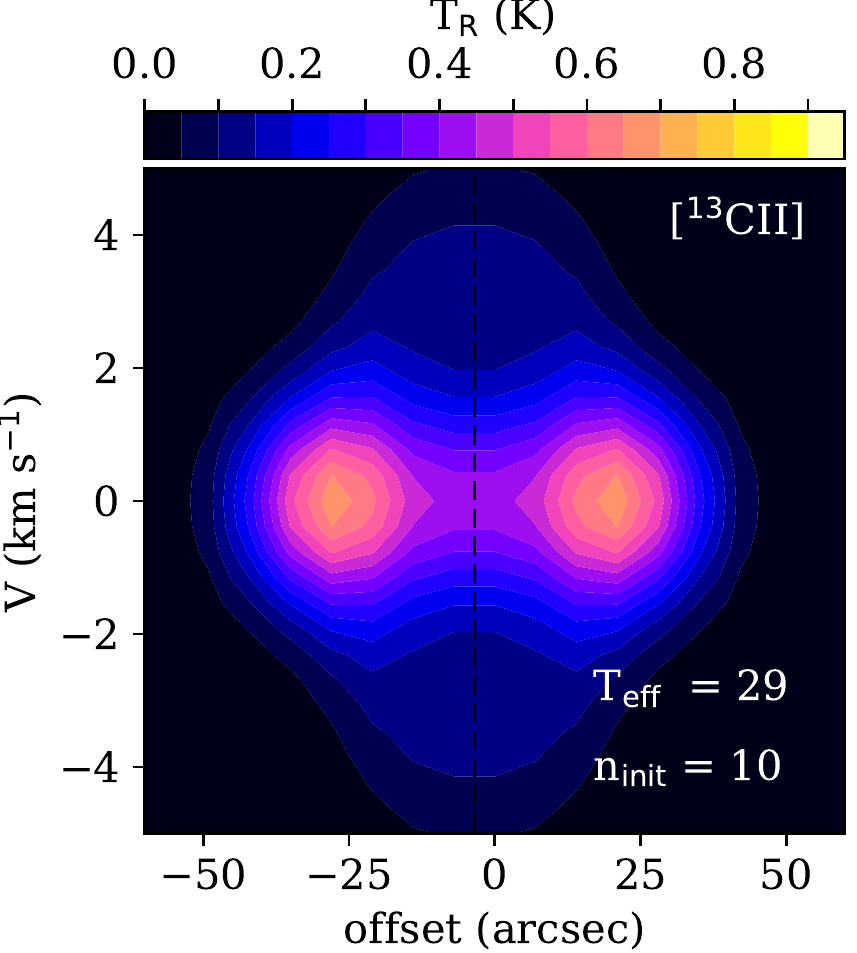}
\includegraphics[width=0.48\columnwidth]{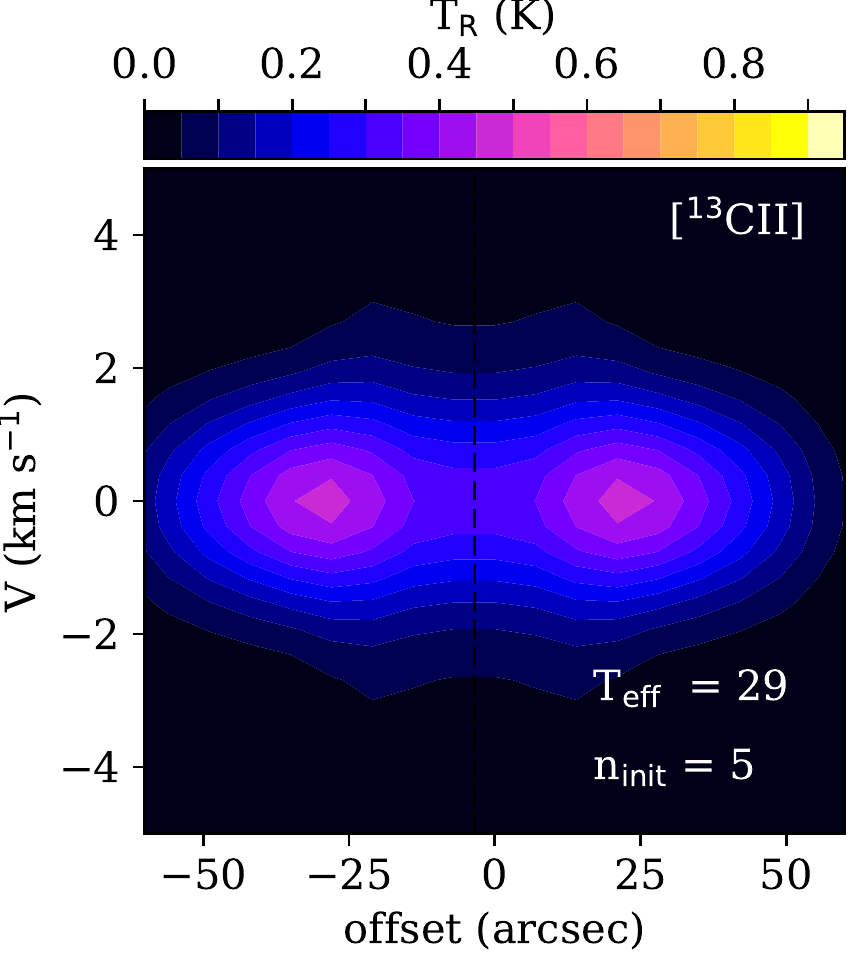}\\
\vspace{2mm}
\includegraphics[width=0.48\columnwidth]{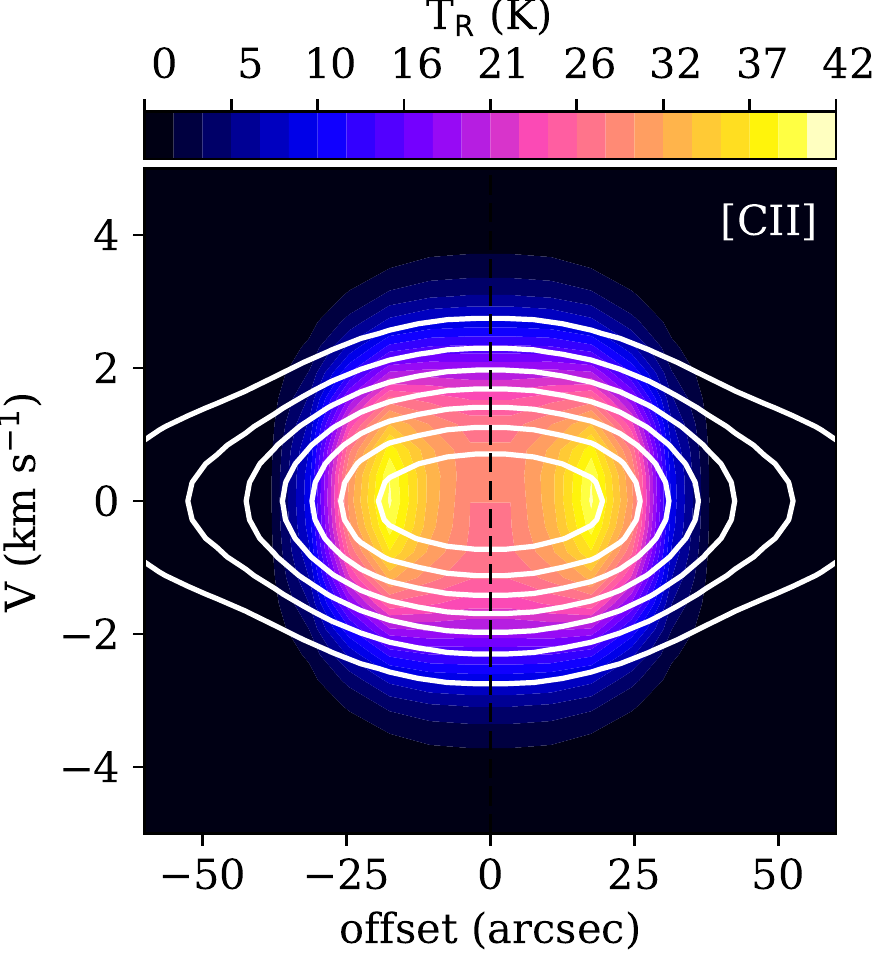}
\includegraphics[width=0.48\columnwidth]{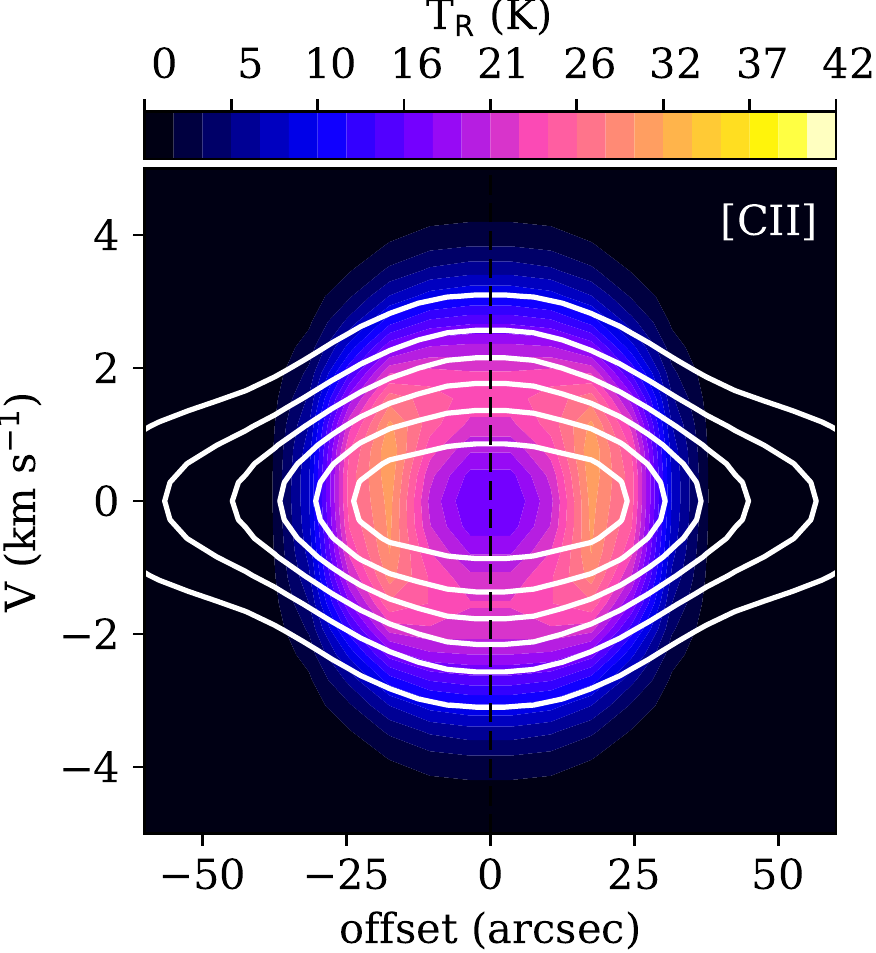}
\includegraphics[width=0.48\columnwidth]{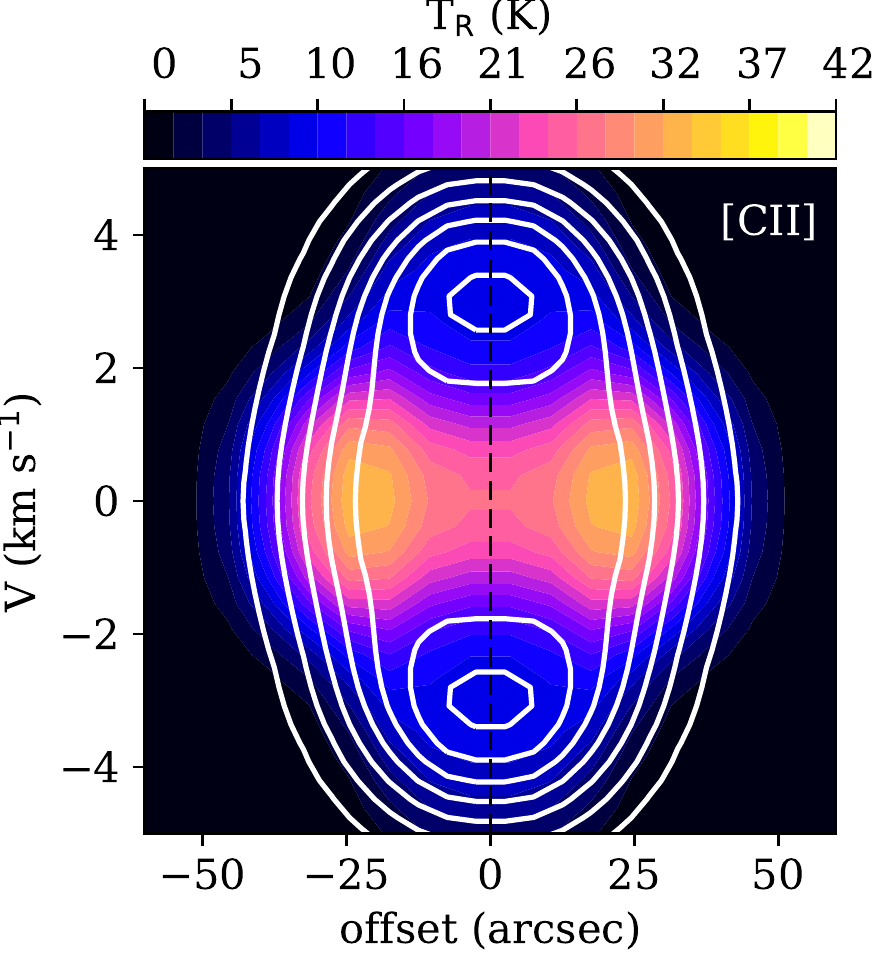}
\includegraphics[width=0.48\columnwidth]{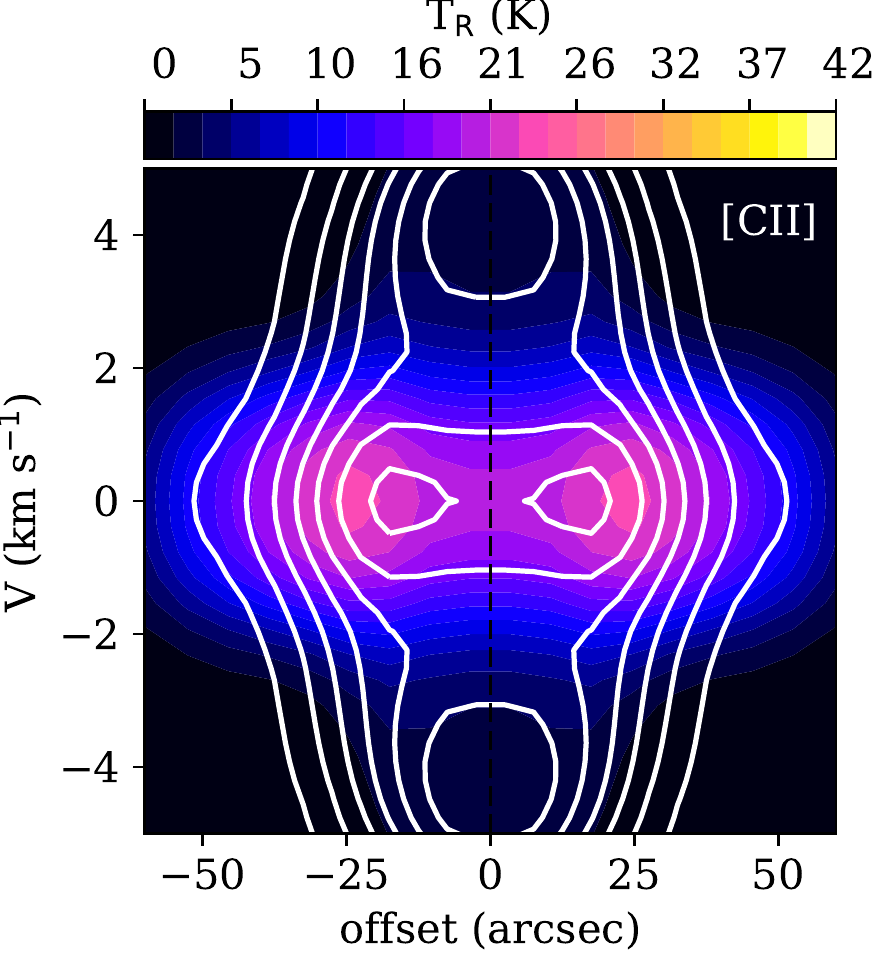}\\
\vspace{2mm}
\includegraphics[width=0.48\columnwidth]{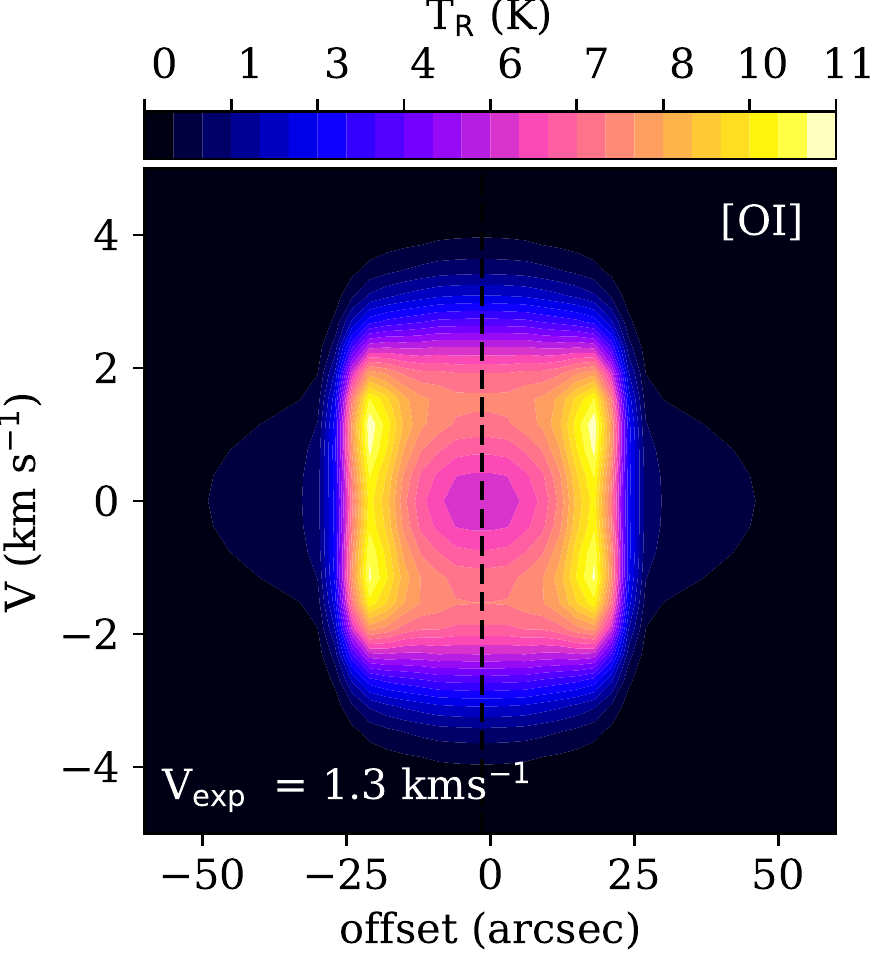}
\includegraphics[width=0.48\columnwidth]{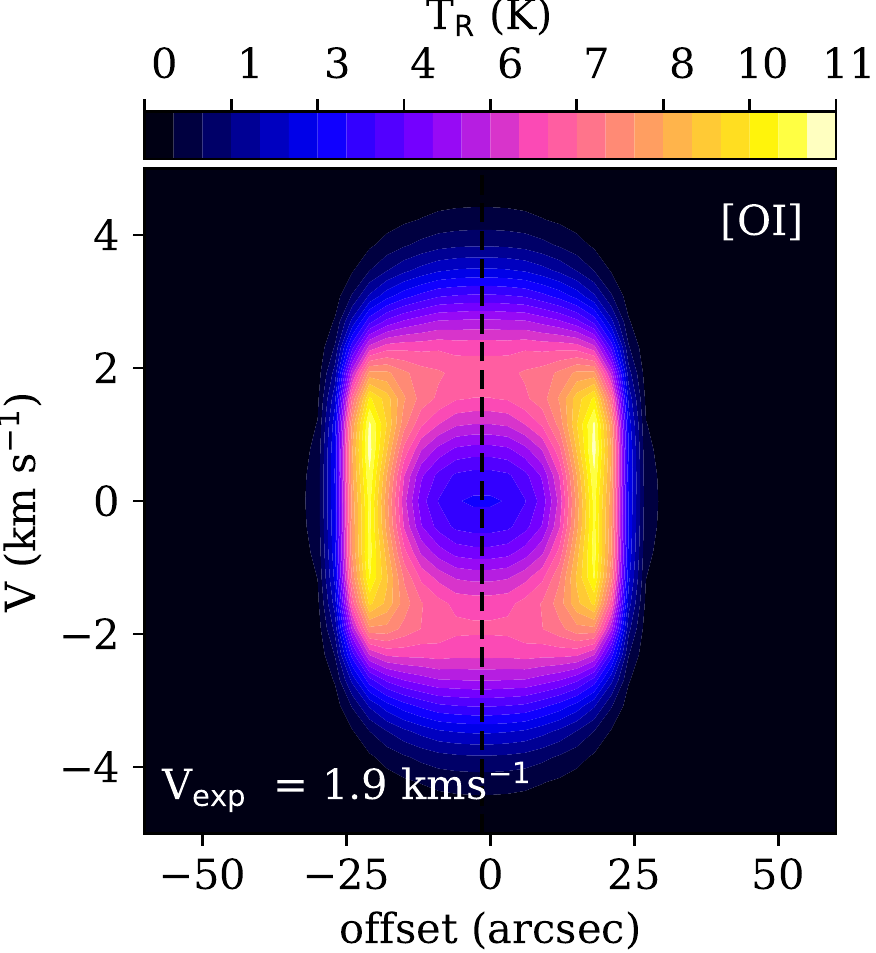}
\includegraphics[width=0.48\columnwidth]{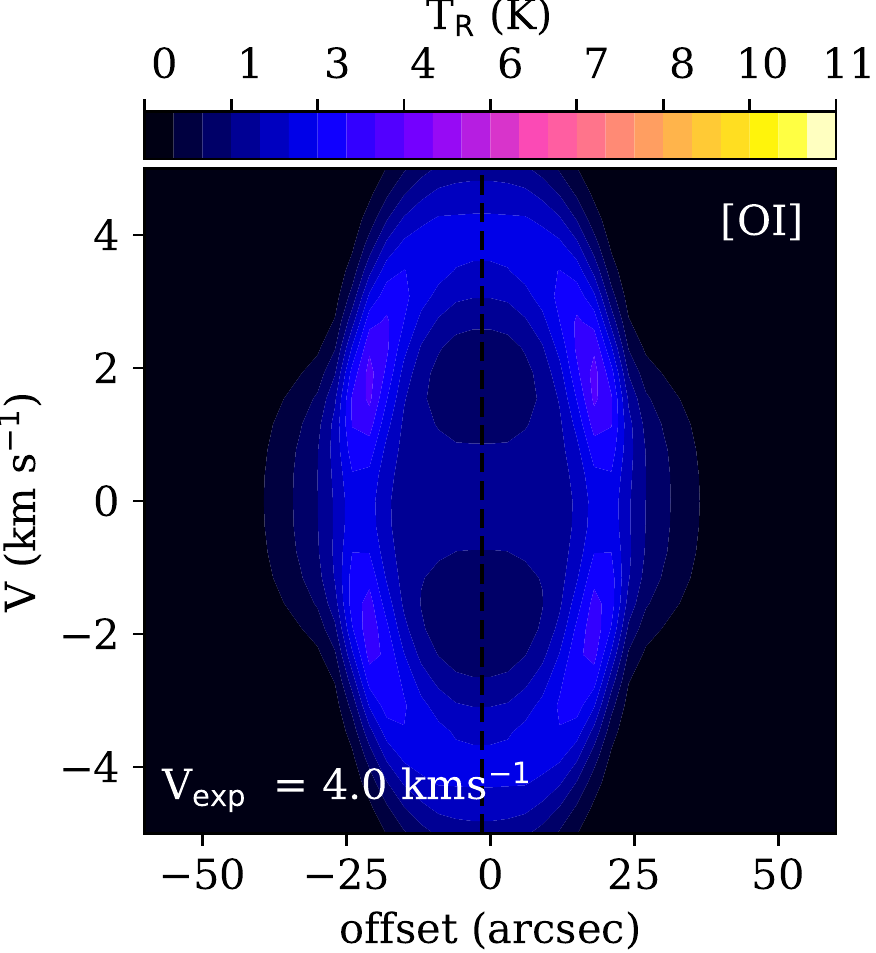}
\includegraphics[width=0.48\columnwidth]{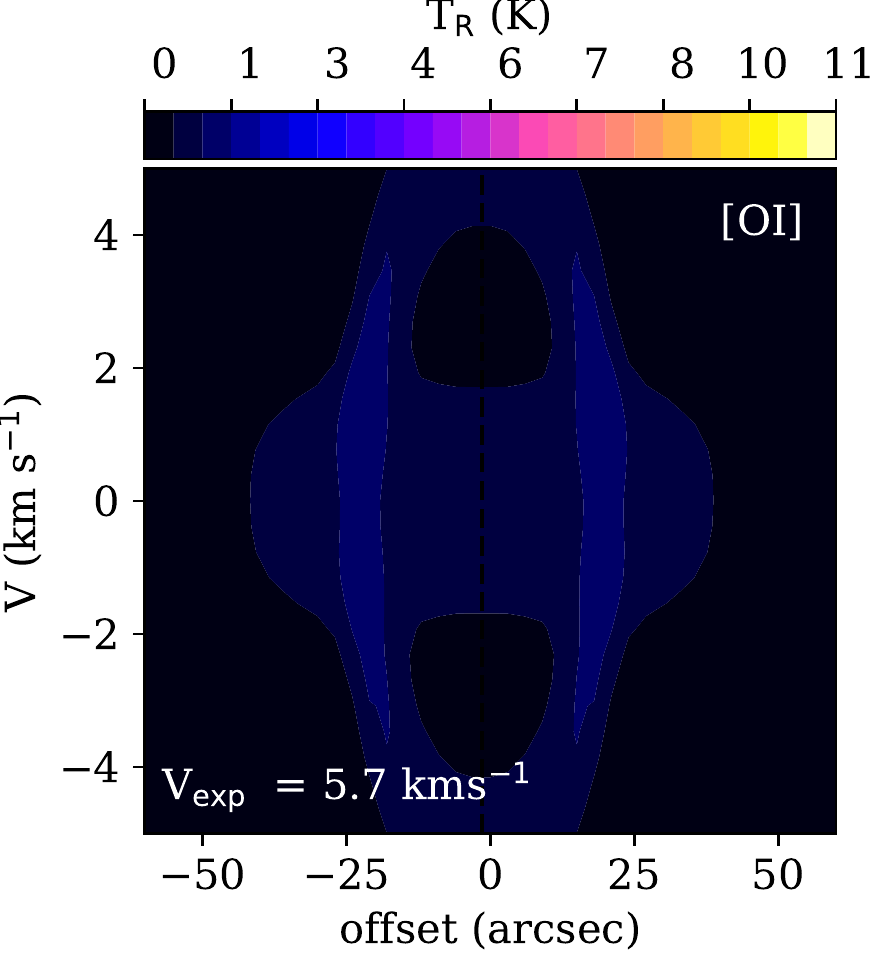}\\
\caption{Simulated pv diagrams in the model with $T_{\rm eff} = 29 000$ and initial gas number density $n_{\rm init}=10^5$, $5\times10^4$, $10^4$, $5\times10^3$~cm$^{-3}$. The model age is 59000, 41000, 19000 and 13000 years, respectively. The \HCOp{} pv~diagrams are shown by white contours on the top of the \CII{} pv~diagrams. The contours are given every 14\% from the normalized \HCOp{} peak.}
\label{fig:modelpvappendix2}
\end{figure*}

The systematic effects related to the variations of $n_{\rm init}$ are shown in Fig.~\ref{fig:modelpvappendix2}. The \thirtCII{} and the \CII{} lines show double-peak structure or have extended wings, which arise due to expansion, if the gas density is not too high and $V_{\rm exp}$ is comparable to or exceeds the turbulent velocity dispersion (models with $n_{\rm init} \leq 5\times10^4$~cm$^{-3}$). We note that while the value of $n_{\rm init}$ is varied by a factor of 20 in Fig.~\ref{fig:modelpvappendix2}, the peak brightness temperature of the \thirtCII{} and the \CII{} lines changes only by a factor of about 2, although the decrease is an order of magnitude for the \OI{} line.

\section{Collisions with electrons}\label{app:electrons}

For HCO$^+$ the radiative transfer computations considered only 
collisions with molecular hydrogen, based on the collision rates 
given in the LAMDA data base \citep{2005A&A...432..369S}. However, 
\citet{Goldsmith2017} showed that the excitation by collision
with electrons can be important. Unfortunately, collision rates
are only available for transitions with $J\le 2$ \citep{Faure2001}
but a solution of the excitation problem for the $J=3-2$ transition
needs to include collisional excitations for the levels well above
$J=3$. For a rough estimate of the effect of collisions with electrons
on the excitation of the HCO$^+$ (3–2) emission, we can however exploit
that the rates for transitions among neighbouring levels have only a 
relatively weak dependence on $J$ and the kinertic gas temperature.
\citet{Faure2001} and \citet{Goldsmith2017} note that
the collision cross sections scale with the effective dipole moment of 
the transition, so that we can extrapolate the given numbers to $J=3-2$
and obtain coefficients of 16.4, 11.6, 8.4 , 6.1, and 5.4 in units
of $10^{−6}$ cm$^{3}$s$^{-1}$ at 10, 20, 40, 80 and 100 K, respectively.
Therefore, electrons are about 20000 times as efficient in the collisional
excitation as H$_2$ (25000 for 20 K, 15000 for 100 K). The
number density of HCO$^+$ has two peaks in the model (see Fig.~\ref{fig:modelphys}).
Collisions with H$_2$ always dominate in the position of the second peak,
while the first peak appears in the thin layer near the ionization
front, where electrons are less abundant than H$_2$ only by a factor 
3000, so that their collisions will dominate. To simulate their contribution,
we added an electron density increased by the factor 20000 to
the H$_2$ density, and recomputed the SimLine models with the
H$_2$ collision rates as before.

\begin{figure}
    \centering
    \includegraphics[width=0.99\columnwidth]{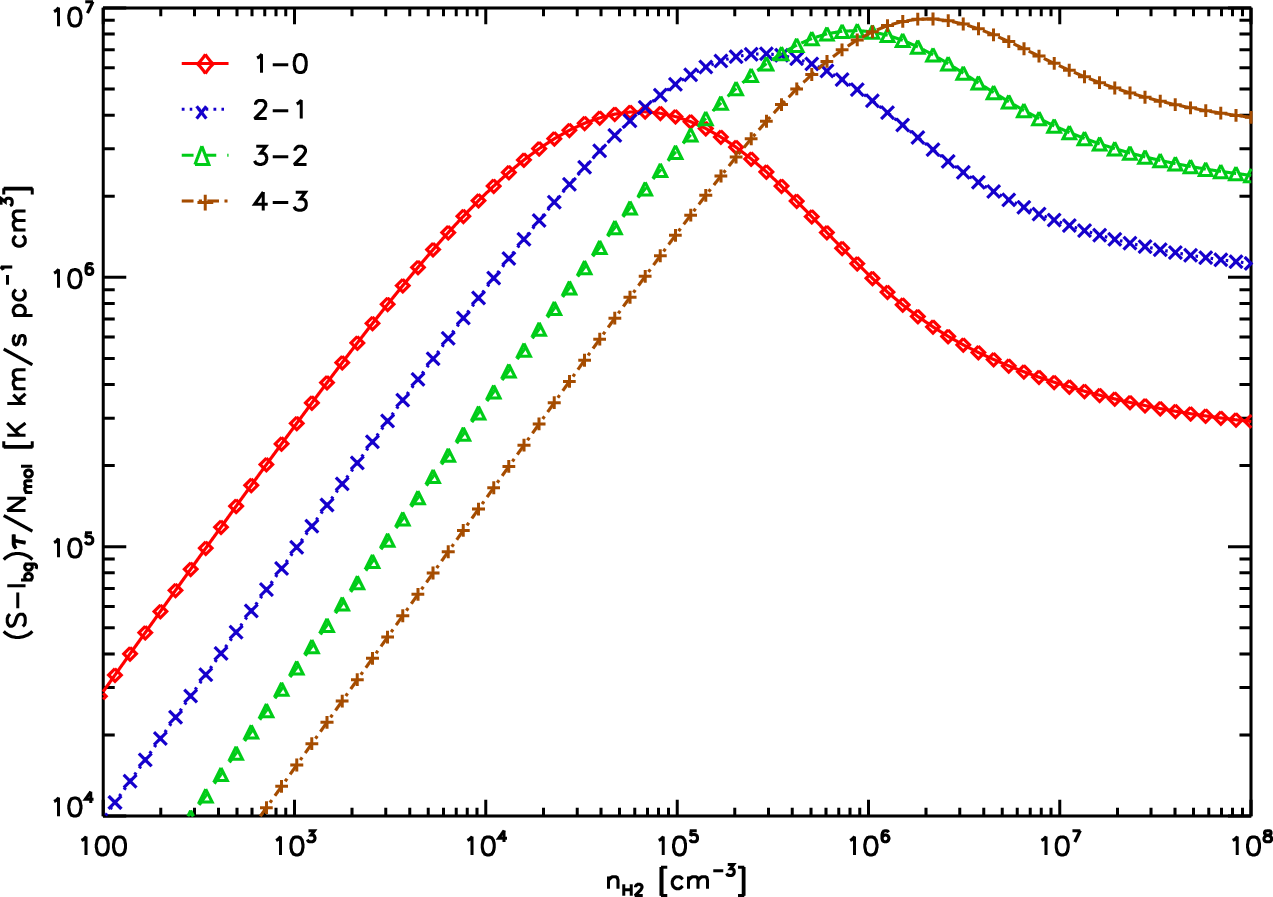}\\
    \includegraphics[width=0.99\columnwidth]{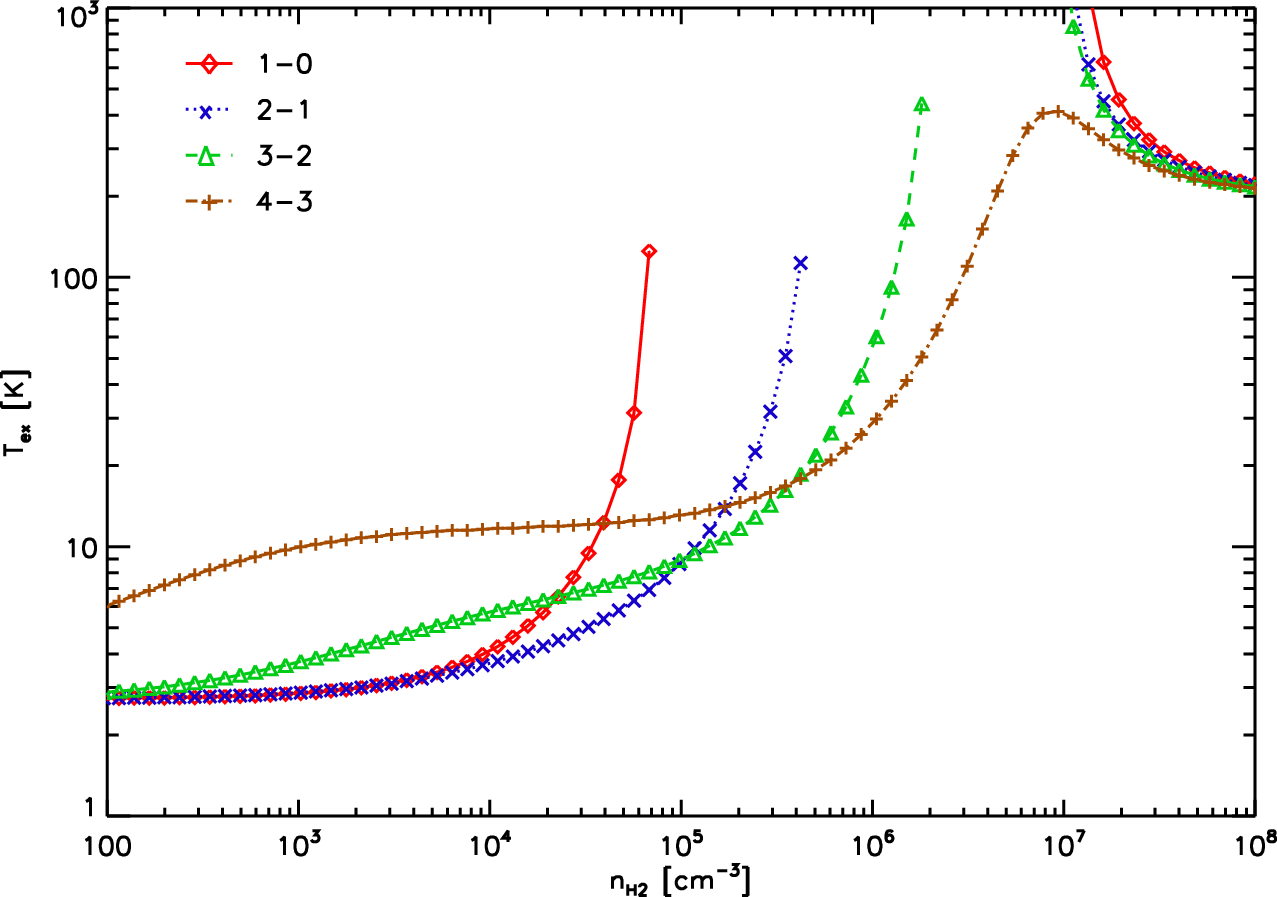}
    \caption{Top: specific emissivity per column of HCO$^+$ molecules for transitions with $J \leq 4$. Bottom: $T_{\rm ex}$ value for the same transitions.}
    \label{fig:general_hcop}
\end{figure}

To understand the impact of the additional collisions on the 
excitation of the $J=3-2$ line we show in Fig.~\ref{fig:general_hcop} the specific
emissivity (per molecule) and the excitation temperature
for the four lowest levels ($J \le 4$) of HCO$^+$ 
computed at a gas kinetic temperature of 200 K, and assuming H$_2$
as the only collision partner. The H$_2$ density at the first
peak is around $10^4$~cm$^{-3}$ falling into the subthermal,
optically thin regime with a linear increase of the emissivity
with density an excitation temperatures well below the kinetic 
temperature. Increasing the density -- or equivalently adding
more collision partners -- can bring us into the maser regime
with $T_{\rm ex} < 0$, seen as a gap in the $T_{\rm ex}$ plots.
However, when considering the emission per molecule, even in 
that regime, the increase is always at most linear with density,
usually sublinear. At very high densities we even see a negative
effect. Consequently, we expect that the contribution
from collisions with electrons can enhace the emission of the
$J=3-2$ line by a factor of at most about $20000/3000\approx 7$.

The SimLine runs actually show that collisions with electrons 
increase the intensity of the HCO$^+ (3–2)$ peak by a factor of
two in S235 A. In S235 C, the additional excitation by electrons
increases the HCO$^+ (3–2)$ intensity by a factor of five. This
is still about 10 and 100 times smaller than the values observed in S235A
and S235C, respectively. We note that the new SimLine calculation 
is probably rather an upper limit, because for higher $J$ transitions the
H$_2$ collisional rates increase somewhat faster than the electron rates
so that the factor 20000 based on the $3-2$ transition is on the
high side when considering the full cascade of transitions including
higher levels.
We therefore conclude that collisions with electrons do
not resolve the problem with the low  HCO$^+ (3–2)$
emission in the model.

\section{Alternative initial abundances}\label{app:highcarbon}

In this section, we present an alternative model for the S235~C PDR with higher elemental carbon abundance, taken from the 'ISM' set of CLOUDY~\citep{Ferland2013}. The carbon abundance relative to hydrogen is $x({\rm C})=2.51\cdot10^{-4}$; the oxygen abundance is $x({\rm O})=5.01\cdot10^{-4}$. These values are about two times higher than those we use in our calculations (see Table~\ref{tab:initial}) and are similar to the abundances from \citet{Schneider2018}.   The general  appearance of the physical and chemical structure of the PDR looks similar to the view in Fig.~\ref{fig:modelphys}, therefore, we only present pv~diagrams for the \thirtCII{}, \CII{} and \OI{} lines. The main qualitative difference between our basic model for S235~C and the model with enhanced carbon abundance is the appearance of the double-peaked \CII{} line profile in the latter. The profile of the \thirtCII{} line remains single-peaked. The \OI{} line profile is double-peaked, as in the basic model. Unfortunately, in the model with enhanced carbon abundance, we could not obtain the double-peaked \CII{} line in the S235~A PDR, due to insufficient depth of the C$^+$-layer in the model. The \hii{} regions are situated in the same molecular cloud at a distance of about 1~pc from each other. Therefore, we assume the same elemental abundances in each of them. As we could not obtain the double-peaked \CII{} line in S235~A using the 'ISM' abundances, we present the model with the 'high metals' elemental abundances \citep{Wakelam2008}, which are considered more conservative according to measurements by \citet{Sofia2004}. 

\begin{figure}
\includegraphics[width=0.48\columnwidth]{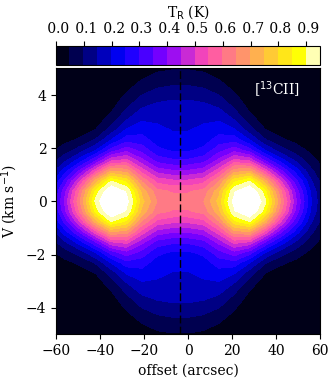}
\includegraphics[width=0.48\columnwidth]{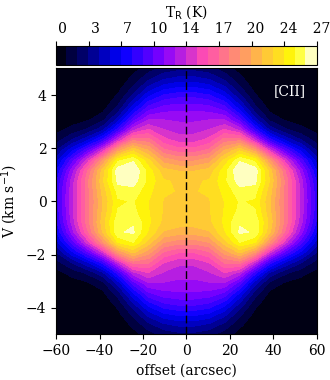}\\
\includegraphics[width=0.48\columnwidth]{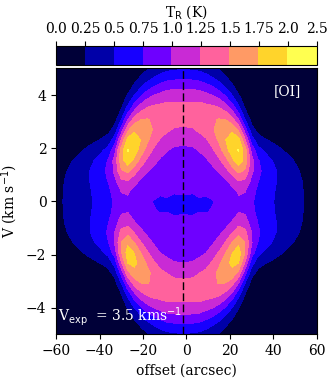}
\caption{Simulated pv~diagrams in the model with elemental abundances taken from the 'ISM' set form CLOUDY.}
\label{fig:altcarbonabund}
\end{figure}

\bsp
\label{lastpage}
\end{document}